\documentclass[lettersize,journal]{IEEEtran}
\usepackage{amsmath,amsfonts}
\usepackage{algorithm}
\usepackage{algpseudocode}
\usepackage{array}
\usepackage{subfig}
\usepackage{textcomp}
\usepackage{stfloats}
\usepackage{dsfont}
\usepackage{url}
\usepackage{verbatim}
\usepackage{multirow}
\usepackage{graphicx}
\usepackage{color}
\usepackage{cite}
\usepackage{amssymb}
\usepackage{balance}
\usepackage{soul, color, xcolor}
\begin{document}
\raggedbottom
\newcommand{\RommaOne}{\uppercase\expandafter{\romannumeral1}}
\newcommand{\RommaTwo}{\uppercase\expandafter{\romannumeral2}}
\newcommand{\RommaThree}{\uppercase\expandafter{\romannumeral3}}
\title{Scalable UAV Multi-Hop Networking via Multi-Agent Reinforcement Learning with Large Language Models}

\author{Yanggang Xu, Jirong Zha, Weijie Hong, Xiangmin Yi, Geng Chen, Jianfeng Zheng, Chen-Chun Hsia, Xinlei Chen
\thanks{This work was supported by the Natural Science Foundation of China under Grant 62371269, Shenzhen Low-Altitude Airspace Strategic Program Portfolio (Grant No. Z25306110), Shenzhen Science and Technology Program ZDCYKCX20250901094203005 and Meituan Academy of Robotics Shenzhen.


\  (Corresponding author: Xinlei Chen. Equal contribution: Yanggang Xu, Jirong Zha.)

Yanggang Xu, Jirong Zha, Xiangmin Yi and Chen-Chun Hsia are with the Shenzhen International Graduate School, Tsinghua University, Shenzhen, China (yanggangxu18@gmail.com; zhajirong23@mails.tsinghua.edu.cn; yxm25@mails.tsinghua.edu.cn; hsiachenchun@gmail.com).

Weijie Hong and Jianfeng Zheng are with the Shenzhen Smartcity Communication, Shenzhen, China (hongweijie@smartcitysz.com; zhengjianfeng@smartcitysz.com).

Geng Chen is with the College of Software, Jilin University, Changchun, China (cg1043148036@outlook.com).

Xinlei Chen is with the Shenzhen International Graduate School, Tsinghua University, Shenzhen, China (chen.xinlei@sz.tsinghua.edu.cn).

}}

\markboth{Journal of \LaTeX\ Class Files,~Vol.~14, No.~8, August~2021}%
{Shell \MakeLowercase{\textit{et al.}}: A Sample Article Using IEEEtran.cls for IEEE Journals}


\maketitle

\begin{abstract}

In disaster scenarios, establishing robust emergency communication networks is critical, and unmanned aerial vehicles (UAVs) offer a promising solution to rapidly restore connectivity. However, organizing UAVs to form multi-hop networks in large-scale dynamic environments presents significant challenges, including limitations in algorithmic scalability and the vast exploration space required for coordinated decision-making.
To address these issues, we propose MRLMN, a novel framework that integrates multi-agent reinforcement learning (MARL) and large language models (LLMs) to jointly optimize UAV agents toward achieving optimal networking performance.
The framework
incorporates a grouping strategy with reward decomposition to enhance algorithmic scalability and balance decision-making across UAVs. In addition, behavioral constraints are applied to selected key UAVs to improve the robustness of the network.
Furthermore, the framework integrates LLM agents, leveraging knowledge distillation to transfer their high-level decision-making capabilities to MARL agents. This enhances both the efficiency of exploration and the overall training process.
In the distillation module, a Hungarian algorithm-based matching scheme is applied to align the decision outputs of the LLM and MARL agents
and define the distillation loss.
Extensive simulation results validate the effectiveness of our approach, demonstrating significant improvements in network performance over the MAPPO baseline and other comparison methods, including enhanced coverage and communication quality.
\end{abstract}


\section{Introduction}
Natural disasters worldwide have devastating impacts, not only causing loss of life and infrastructure damage but also disrupting social and economic systems.
In 2024, the United States experienced 27 climate disasters that caused over \$1 billion in damage each, with total losses reaching \$184.8 billion and 568 fatalities reported \cite{NOAA2025}.
Communication infrastructure, including base stations (BSs) and fiber-optic cables, is highly vulnerable to disasters such as floods and earthquakes.
Communication failures hinder the delivery of humanitarian aid and disrupt life-saving rescue efforts. These breakdowns intensify the challenges faced by disaster-stricken communities. They highlight the urgent need for resilient, rapidly deployable communication systems to support emergency response and recovery.

Unmanned Aerial Vehicles (UAVs) have emerged as a promising solution to address the communication challenges posed by disasters, owing to their exceptional mobility and agility \cite{1intro2}.
UAVs can serve as mobile relay stations, forming multi-hop networks to provide communication services to distant user equipments (UEs).
Furthermore, UAVs, operating in air-to-ground and air-to-air channels with minimal obstructions, are more likely to establish reliable line-of-sight propagation paths, thus ensuring robust communication \cite{3LoS}.
By strategically optimizing UAV flight trajectories and positions, the quality of communication channels can be significantly enhanced, thereby improving the overall performance of UAV-assisted network systems.

\begin{figure}
\centering
\includegraphics[width=\linewidth]{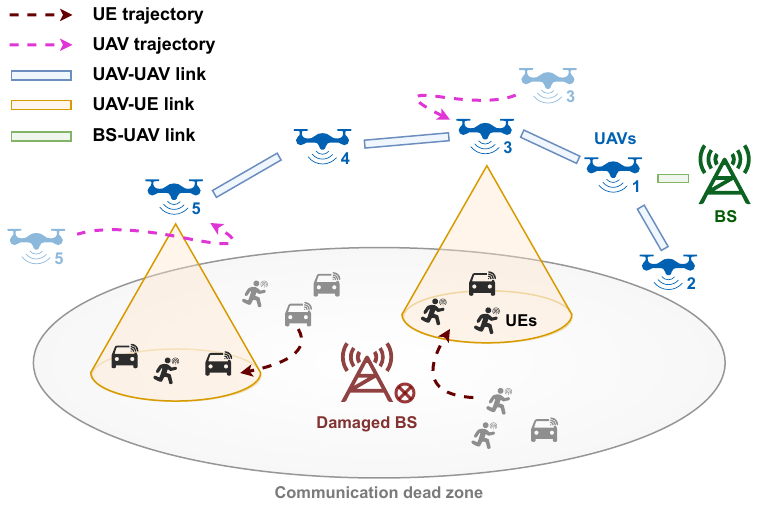}
\caption{In disaster scenarios, UAVs can rapidly establish temporary multi-hop wireless networks in communication dead zone, thereby restoring connectivity for UEs.}
\label{intro figure}
\vspace{-0.3cm}
\end{figure}

This paper leverages UAV swarms to establish expansive multi-hop networks by connecting distant, operational fixed BSs in disaster-stricken regions, shown in Figure \ref{intro figure}.
In situations where disasters create communication dead zones, multiple UAVs are strategically deployed to serve ground UEs. They form a relay network that bridges isolated areas with available BSs, thereby connecting the local network to the core infrastructure.
This configuration not only extends coverage to remote areas where conventional communication is infeasible, but also enhances network reliability through the creation of multiple routing paths. In particular, UAV mobility and multi-hop pathways provide the flexibility needed to adapt to the dynamic movement of ground UEs, enabling adaptive routing.
In a multi-hop network, precise coordination among multiple UAVs is essential. This necessitates sophisticated trajectory planning and deployment strategies to ensure that each UAV maintains optimal positioning for uninterrupted connectivity.

With the rapid advancement of UAV and Internet of Things technologies, extensive research has been devoted to UAV-based communication and networking tasks 
\cite{chen2024soscheduler, zha2025aircopbench, cheng2024multi}.
Numerous studies have sought to optimize the association and coordination between UAVs and UEs through refined transmission strategies \cite{intro_research1}, efficient resource management \cite{intro_research2}, and strategic UAV deployment \cite{intro_research3}.
Several studies have further explored the applications of UAVs in wireless sensor networks 
\cite{intro_research4, wang2025aerial}, data offloading, and edge computing \cite{jian2024lvcp}.
Reinforcement learning (RL) techniques, owing to their exceptional decision-making and planning capabilities, have been applied to optimize UAV trajectory planning in dynamic and uncertain environments \cite{31RL6}. 
These methods aim to improve overall communication performance while addressing the challenges imposed by various operational constraints and environmental uncertainties.
However, existing studies primarily focus on the planning of a limited number of UAVs, without addressing the complex coordination required for large-scale multi-UAV systems.
In large-scale environments, communication constraints often require the use of multi-hop networks, where connectivity is highly dynamic and unstable. Coordinating UAV swarm within such networks remains a significant challenge.

Optimizing dynamic multi-hop networks formed by UAVs presents several key challenges:

\noindent 1) One major difficulty lies in the joint optimization and balancing of UAV-specific strategies, which demands a scalable algorithm capable of managing complex coordination.
Within the swarm, UAVs must coordinate their roles to collectively establish and maintain a stable network topology.
Each UAV's decision-making not only dictates its own relay selection and communication scheduling, but also influences the strategies of other UAVs within the network.
An imbalance in these roles may lead to coverage gaps or network disconnections, triggering cascading effects that undermine network stability and efficiency.
The interdependence of these strategic decisions further amplifies the complexity of coordination, requiring UAVs to continuously adjust their behaviors to achieve a globally balanced and optimized network structure.

\noindent 2) The second challenge arises from the spatial complexity and dynamic nature of multi-hop UAV networks, which significantly expand the exploration space and complicate the training process. In large-scale disaster scenarios, the considerable distances between available BSs and scattered UEs result in an overwhelming number of possible UAV deployment strategies, making it difficult to determine an optimal configuration. This challenge is further exacerbated by the continuously changing network topology and fluctuating channel conditions. Furthermore, the stochastic distribution and unpredictable mobility of UEs introduce further uncertainty into network planning.
These spatial and temporal complexities not only increase the difficulty of optimizing UAV placement and routing, but also require continuous adaptation to maintain stable and efficient communication.

\noindent These challenges highlight the need for scalable algorithms that enable joint optimization of multi-agent strategies to maintain equilibrium while effectively managing the extensive exploration space. Developing such algorithms is essential for achieving efficient coordination and adaptive decision-making in complex, dynamic multi-hop network environments.

To address the aforementioned challenges, this paper investigates the application of UAV swarms in multi-hop networking for disaster emergency scenarios, focusing on optimizing UAV trajectories to maximize both communication coverage and quality under connectivity constraints. To tackle this problem, we propose a novel multi-agent reinforcement learning (MARL) framework, referred to as MRLMN (Multi-agent Reinforcement learning with Large language model in Multi-hop Networking). The key contributions of this paper are summarized as follows:
\begin{itemize}
    \item
    This paper formulates the UAV-enabled multi-hop networking task as a multi-objective optimization problem aimed at maximizing network coverage and communication quality while satisfying connectivity constraints. This problem is modeled as a stochastic game to capture the dynamic interactions among UAVs and between UAVs and the environment, explicitly considering collaboration and coordination challenges in large-scale, multi-UAV, and multi-hop network scenarios.

    \item  
    This paper proposes a task-oriented agent grouping strategy and an information aggregation mechanism within a MARL framework. A reward decomposition model is designed based on the grouping strategy to facilitate coordinated decision-making, mitigate non-stationarity, and enhance scalability for large UAV swarms. Behavioral constraints are further applied to critical UAV groups to improve network robustness and prevent detrimental topology disruptions.

    \item
    This paper designs a knowledge distillation framework that combines Large Language Models (LLMs) with MARL for UAV networking. In our framework, the LLM serves as an offline advisor that provides high-level strategic insights, which are distilled into the MARL agents to guide policy learning, without being deployed for real-time UAV control. LLM outputs are aligned with MARL actions through a per-agent decision matching scheme. The knowledge is then transferred via a tailored distillation loss, facilitating more efficient policy exploration, addressing the cold-start problem, and enhancing the convergence of MARL training.

    \item Extensive simulations are conducted to evaluate the proposed approach across different environment scales, UAV swarm sizes, and baseline comparisons. The results demonstrate that MRLMN consistently outperforms existing methods in terms of network coverage, communication quality, and robustness, validating the effectiveness of the proposed mechanisms.
\end{itemize}

The remainder of this paper is structured as follows. Section \ref{Related Work} reviews related work on UAV-based networking. Section \ref{System Model} formulates the problem and introduces the system model. Section \ref{Methodology} presents the proposed MRLMN framework. Section \ref{Performance Evaluation} describes the simulation setup and evaluates performance by comparing the proposed method with other approaches. Finally, Section \ref{Conclusion} concludes the paper and outlines potential directions for future research.

\section{Related Work}
\label{Related Work}

This section reviews existing research relevant to UAV-assisted networking, covering optimization-based, RL-based, and LLM-assisted methods. The discussion summarizes key developments in each category, identifies their limitations, and emphasizes the advantages of the proposed approach relative to prior work.

\subsection{Optimization-Based Methods}
In the problem of employing UAVs to provide communication networks for UEs, designing trajectory and resource planning algorithms is complex due to the coupling of various factors. Several studies propose algorithms based on optimization theory to tackle this intricate problem
\cite{7ML3}.
The majority of these works model the trajectory optimization problem for UAVs within the context of combinatorial optimization problems such as the Traveling Salesman Problem (TSP) and Vehicle Routing Problem (VRP). Building on these well-established frameworks, researchers have developed specialized algorithms to effectively address the optimization challenges present in different UAV-assisted networking scenarios.
In \cite{11ML1}, block coordinate descent and successive convex optimization techniques are applied to optimize UE communication scheduling and UAV trajectory planning. The work in \cite{12ML2} decomposes the optimization problem into different sub-problems and employs fast global K-means along with an interior-point method to optimize the locations of UAVs. In the context of UAV networking problems, some studies focus on optimizing the uplink and downlink communication rates \cite{11ML1}, UAV energy consumption \cite{13ML11}, quality of service \cite{15ML7} and the number of UEs covered \cite{18ML9}.
Under controlled conditions, optimization-based approaches for UAV deployment and planning have demonstrated strong performance and robustness. 

However, these methods are significantly hindered by high execution complexity in practical applications \cite{intro_research4}. In large-scale dynamic environments, the inherent non-convexity of the optimization problems impedes the attainment of global optimality, thus limiting their overall effectiveness.
Furthermore, these algorithms struggle to handle the vast state and action spaces in such networking scenarios, where the high dimensionality complicates accurate modeling and further increases computational complexity.
Collectively, these factors constrain the ability of optimization-based methods to rapidly generate effective decisions in rapidly changing and unpredictable large-scale environments.

To address these limitations, the proposed approach leverages MARL to efficiently explore high-dimensional state and action spaces, reducing reliance on precise modeling of the networking task. Task-oriented grouping and reward decomposition simplify the coordination among multiple UAVs, mitigating the computational complexity associated with large-scale deployments. These combined mechanisms allow the framework to generate effective and robust UAV deployment decisions in dynamic, large-scale multi-hop networking scenarios.

\subsection{RL-Based Methods}
In recent years, reinforcement learning (RL) has emerged as a prominent approach for addressing complex optimization challenges in UAV-assisted networks
\cite{5RLintro1},
demonstrating remarkable capabilities in sequential decision-making and scheduling tasks.
In \cite{21RL3}, the authors introduce a constrained deep Q-network to maximize downlink capacity while ensuring comprehensive coverage for all UEs. \cite{22RL4} proposes a dual-attention RL technique to address the time-varying UE traffic and mobility challenges in the environment. Additionally, several studies integrate UAV-assisted networking with data offloading and edge computing. These works employ RL algorithms to achieve fair throughput among UAVs
\cite{20RL2}
or to minimize the content acquisition delay for UEs \cite{intro_research7}. Song et al. \cite{26RL13} propose an improved evolutionary multi-objective RL algorithm to address both trajectory control and task offloading problems of the UAVs.

In this context, MARL \cite{27MARL1} algorithms enable each UAV to act as an individual decision-maker while leveraging interactions with other UAVs, making them well-suited for multi-hop wireless networks.
\cite{29RL8} utilizes a multi-agent deep deterministic policy gradient (MADDPG) approach to optimize task offloading strategies between UEs and UAVs. Similarly, \cite{30RL12} proposes a decentralized multi-agent soft actor-critic algorithm to optimize spectral efficiency among UAVs. To capture the complex relationships between UAVs and UEs, \cite{31RL6} introduces a heterogeneous graph-based formulation that is integrated into MARL frameworks to facilitate the learning of distributed policies for UAVs. And \cite{32GNN1} develops an attention-based heterogeneous graph neural network combined with model-based RL to optimize the UAVs' resource allocation.

Despite significant advancements in MARL, some issues still persist. Many studies assume that the backhaul network interfacing with the core network is fully configured, thereby neglecting the optimization of relay nodes. In large-scale dynamic environments, such assumptions overlook the scalability challenges of coordinating dozens of UAVs for multi-hop relaying. As the number of UAVs increases, RL techniques encounter convergence issues due to the exponential growth of state and action spaces. The intrinsic randomness of RL further complicates long-term link maintenance in UAV networks. Another critical challenge in MARL is the credit assignment, which complicates the evaluation of each UAV's contribution when only a global reward signal is available. For instance, the suboptimal performance of a single relay UAV can trigger network disconnections and a rapid decline in the global reward, potentially misleading other agents regarding the efficacy of their policies. Similarly, when overall performance gains are driven by a few critical UAVs, the remaining agents may overestimate their contributions.
Moreover, the wide range of environmental states in large-scale scenarios results in a vast exploration space, thereby reducing algorithmic robustness. In the early stages of RL training, random initialization of model parameters leads to counterintuitive decisions that prompt unproductive exploration, ultimately impeding the training process.

In response to these challenges, in the proposed framework, UAVs are coordinated through a task-oriented grouping strategy that assigns specific roles, ensuring that relay responsibilities are explicitly considered.   To mitigate the scalability and convergence issues arising from high-dimensional state and action spaces, a reward decomposition scheme is applied, which distributes global feedback into more localized signals, thereby improving credit assignment for individual UAVs and stabilizing learning. Behavioral constraints are enforced on critical UAVs to prevent disruptive topology changes, ensuring network robustness despite the intrinsic randomness of RL.   Finally, to reduce the burden of exploration in vast and complex environments, large language models are employed during offline training to provide high-level strategic guidance, which is distilled into MARL policies to accelerate learning, guide effective UAV deployment, and maintain decentralized decision-making suitable for real-time operation.

\subsection{LLM-Based Methods}

The development of LLMs has facilitated their application in robotic planning, where they exhibit strong reasoning and decision-making capabilities
\cite{voyager,zhou2024large}.
Building on this progress, recent studies have extended the use of LLM agents to UAV planning \cite{llm_uav, xu2024emergency}, enhancing the efficiency of the planning process in complex environments. Furthermore, integrating LLMs with RL has emerged as a promising research direction. LLMs have been utilized to shape reward functions \cite{Guiding}, enabling more nuanced and context-aware feedback for RL agents. They have also been applied to improve state representation \cite{llm_state} by capturing intricate relationships within the environment, which enhances the agent's understanding of the environmental states. Additionally, LLMs have been employed to refine action selection \cite{llm_action}, providing guidance that allows RL agents to make more informed and strategically aligned decisions, ultimately improving the performance of the system.
Despite their potential, the application of LLMs in robotics and UAV planning faces notable challenges. In particular, LLMs' sensitivity to input prompts can lead to inconsistent decision-making, especially in complex and dynamic tasks that require precise control. Moreover, the gap between the generalized knowledge of LLMs and specific domain requirements further complicates their practical integration, with the issue being more pronounced in problems with complex constraints.
Existing methods that integrate LLMs with RL often assume low-dimensional or structured action and state spaces and rely on direct translation of LLM outputs into agent actions or hierarchical abstractions. These assumptions are unsuitable for multi-UAV multi-hop networking, where action and state spaces are high-dimensional, strongly coupled, and constrained by connectivity and communication requirements. 

In the proposed approach, the semantic-level reasoning of the LLM is combined with the robustness and coordination capability of MARL to address these challenges. During MARL training, the LLM provides high-level strategic guidance for UAV deployment and networking. MRLMN then aligns these strategies with decentralized agent behaviors by assigning LLM-suggested roles via the Hungarian algorithm and distilling the resulting priors into the MARL policies. At deployment, MARL agents execute decisions independently, ensuring scalable and distributed control without requiring online interaction with the LLM.

\section{System Model}
\label{System Model}

In this section, the system components of the UAV-enabled emergency network and their spatial dynamics within the disaster environment are defined.  The communication model is then introduced together with the connectivity constraints required to maintain a feasible multi-hop backhaul link to the core network.  Building on these elements, the networking task is formulated as a joint trajectory optimization problem that aims to maximize user coverage and communication quality under practical operational constraints.

\subsection{Problem Formulation}

In the networking system, we consider a setup with $U$ relay UAVs, $M$ mobile UEs, and $G$ BSs, represented by the sets $\boldsymbol{\mathcal{U}} = \{1, 2, \dots, U\}$, $\boldsymbol{\mathcal{M}} = \{1, 2, \dots, M\}$, and $\boldsymbol{\mathcal{G}} = \{1, 2, \dots, G\}$, respectively. The system operates in time slots, indexed as $t \in \boldsymbol{\mathcal{T}}$, where $\boldsymbol{\mathcal{T}} = \{1, 2, \dots, T\}$. Treating each entity from the UAV, UE, and BS groups as a node, the overall node set is expressed as $\boldsymbol{\mathcal{N}} = \boldsymbol{\mathcal{U}} \cup \boldsymbol{\mathcal{M}} \cup \boldsymbol{\mathcal{G}}$. The interaction between these nodes facilitates the establishment of a multi-hop communication network, enabling UEs to maintain consistent connections with BSs. The position of each node $n\in\boldsymbol{\mathcal{N}}$ at a specific time slot $t$ is described in three-dimensional space by $\boldsymbol{l}_n(t) = (x_n(t), y_n(t), z_n(t))$. The Euclidean distance between any two nodes $i, j \in \boldsymbol{\mathcal{N}}$ at time $t$ is given by $d_{i,j}(t) = ||\boldsymbol{l}_i(t)-\boldsymbol{l}_j(t)||_2$.
Prior studies on UAV-enabled communication have established modeling principles for node interactions and relay-assisted communication, offering validated abstractions for UAV modeling \cite{zhao2025flight}, link quality characterization and connectivity constraints
\cite{hussain2024computing}.
Complementary work on UAV-supported UE connection and communication adaptation \cite{p1} 
further provides mechanisms for describing how UAVs maintain stable communication connectivity while serving UEs 
\cite{chen2025survey,zha2024diffusion, wang2025novel}. 
Building on these foundations, the communication model, connectivity constraints, and optimization objective in this study can be formally established, thereby constructing a complete and consistent formulation of the UAV multi-hop networking task.

\subsection{Communication Model}
In the proposed networking scenario, the communication dynamics are modeled across three types of links: UAV-UE links that facilitate data transmission between UAVs and UEs, UAV-UAV links that enable communication and coordination among UAVs, and BS-UAV links that support connectivity between BSs and UAVs. 

\begin{itemize}
    \item \textbf{UAV-UE links:}
    For the UAV-UE links, a probabilistic path loss framework is employed to model the distinct characteristics and occurrence probabilities of Line-of-Sight (LoS) and Non-Line-of-Sight (NLoS) conditions. This approach captures the additional path loss caused by environmental factors such as shadowing and scattering, which significantly affect air-to-ground communication in realistic propagation environments. The LoS and NLoS path loss model for UAV $u \in \boldsymbol{\mathcal{U}}$ and UE $m \in \boldsymbol{\mathcal{M}}$ is defined as
    \begin{equation}
        \begin{aligned}
            PL_{u,m}^\text{LoS}(t) = 20\log(\frac{4\pi f_c}{c}) + 20\log(d_{u,m}(t)) 
            + \eta_\text{LoS},
        \end{aligned}
    \end{equation}
    \begin{equation}
    \begin{aligned} 
    PL_{u,m}^\text{NLoS}(t) = 20\log(\frac{4\pi f_c}{c}) + 20\log(d_{u,m}(t)) 
    + \eta_\text{NLoS},
    \end{aligned} 
    \end{equation}
    where $f_c$ is the carrier frequency of the channel, $c$ is the speed of light,
    $\eta_{\text{LoS}}$ and $\eta_{\text{NLoS}}$ are constant values representing the excessive path loss for LoS and NLoS links, respectively.
    The occurrence probability of the LoS channel follows
    \begin{equation}
    P_{u,m}^\text{LoS}(t) = \frac{1}{1+a\exp{[-b(\frac{180}{\pi}\arcsin(\frac{z_{u}(t)}{d_{u,m}(t)})-a)]}},
    \end{equation}
    where $z_{u}(t)$ is the height of the UAV, $a$ and $b$ are environmental constants. The occurrence probability of the NLoS channel is given by $P_{u,m}^\text{NLoS}(t)=1-P_{u,m}^\text{LoS}(t)$. Therefore, the path loss of the UAV-UE links is modeled as
    \begin{equation}
    \begin{aligned} 
    PL^{\text{UAV-UE}}_{u,m}(t) = P_{u,m}^\text{LoS}(t)PL_{u,m}^\text{LoS}(t)
    + P_{u,m}^\text{NLoS}(t)PL_{u,m}^\text{NLoS}(t).
    \end{aligned} 
    \end{equation}

    \item \textbf{UAV-UAV links:} 
    For UAV-UAV links, where signal propagation occurs in unobstructed airspace with minimal interference from environmental obstacles and LoS link is the dominant mode of communication, the free-space path loss (FSPL) model is adopted. Thus, the path loss for UAV-UAV links between $u, v \in \boldsymbol{\mathcal{U}}$ is given by
    \begin{equation}
    PL^{\text{UAV-UAV}}_{u,v}(t) = 20\log(\frac{4\pi f_c}{c}) + 20\log(d_{u,v}(t))+ \eta_\text{LoS}.
    \label{FSPL eq}
    \end{equation}
    
    \item \textbf{BS-UAV links:} 
    In this paper, the BS antennas are assumed to be mounted at a high elevation, consistent with real-world deployments. This positioning ensures a largely unobstructed communication environment with UAVs, allowing the channel to be approximated as free space. Under this assumption, the path loss for the link between UAV $u\in\boldsymbol{\mathcal{U}}$ and BS $g\in\boldsymbol{\mathcal{G}}$ is modeled using the FSPL framework, 
    \begin{equation}
    PL^{\text{BS-UAV}}_{g,u}(t) = 20\log(\frac{4\pi f_c}{c}) + 20\log(d_{g,u}(t))+ \eta_\text{LoS}.
    \end{equation}
\end{itemize}

Therefore, the received signal power from node $i\in\boldsymbol{\mathcal{N}}$ to $j\in\boldsymbol{\mathcal{N}}$ is given by
\begin{equation}
P_{i,j}^{\text{RX}}(t) = 
P^{\text{TX}}_iG^{\text{TX}}_iG^{\text{RX}}_j10^{-PL_{i,j}(t)/10},
\end{equation}
where 
$P_i^{\text{TX}}$ and $P_{i,j}^{\text{RX}}$ are the transmitted signal power and received signal power, $G_i^{\text{TX}}$ and $G_j^{\text{RX}}$ are the gains of the transmitter and receiver antennas, respectively. Accordingly, the signal-to-noise ratio (SNR) from node $i$ to $j$ is denoted as
\begin{equation}
\rho_{i,j}(t) = \frac{P_{i,j}^{\text{RX}}(t)}{10^{N_A/10}},
\end{equation}
where $N_A$ represents the noise power in dB form.
In the environment, the noise power is computed according to
\begin{equation}
    N_A=-174 + 10\log B + \textit{NF},
\end{equation}
where -174 dBm/Hz represents the thermal noise power spectral density at room temperature, $B$ represents the system bandwidth in Hz and $\textit{NF}$ denotes the noise figure. 
The available data rate from node $i$ to $j$ is determined based on the Shannon capacity formula as
\begin{equation}
r_{i,j}(t) = B_{i,j}\log_2(1+\rho_{i,j}),
\end{equation}
where $B_{i,j}$ denotes the bandwidth. 

\subsection{Connectivity Constraint}
\label{Connectivity}

In multi-hop networks, the connectivity between nodes is crucial in determining whether a UE can successfully connect to the core network. In this paper, binary variables $c^\text{UE}_m$ and $c^\text{UAV}_u$ are denoted to indicate whether UE $m$ or UAV $u$ can establish a connection to an available BS through the multi-hop network.
The connectivity of each link is determined based on a predefined SNR threshold $\rho_{\text{th}}$, where a link is considered disconnected if its SNR falls below the threshold.
Moreover, UAVs can serve as relay nodes for UEs or other UAVs, constituting integral components of the multi-hop network that connects to the BS.
A UAV can connect to the BS either directly or through multiple hops, while a UE accesses the core network via UAVs linked to the BS.
Consequently, we have
\begin{equation}
    c_u^\text{UAV}(t)=\left\{
            \begin{aligned}
                1,&\  \text{if } \exists g\in\boldsymbol{\mathcal{G}}, \rho_{g,u}(t)\geq\rho_{\text{th}},\\
                 &\ \text{or } \exists v\in\boldsymbol{\mathcal{U}}\backslash\{u\}, 
                  c_v^\text{UAV}(t)=1,\rho_{u,v}(t)\geq\rho_{\text{th}}\\
                0,&\ \text{otherwise}
            \end{aligned}
            \right.
    \label{constraint_1}
\end{equation}
and
\begin{equation}
    c_m^\text{UE}(t)=\left\{
        \begin{aligned}
            1,&\  \text{if }\exists u\in\boldsymbol{\mathcal{U}}, 
              c_u^\text{UAV}(t)=1,\rho_{u,m}(t)\geq\rho_{\text{th}}\\
            0,&\  \text{otherwise}.
        \end{aligned}
        \right.
    \label{constraint_2}
\end{equation}

Within the network, UAVs select the backhaul path with the fewest relay hops, whereas UEs establish only the connection that yields the maximum data rate. Accordingly, the data rate available to UE $m$ is defined as
\begin{equation}
r^\text{UE}_m(t) = \max \{ r_{u,m}(t) | c_u^\text{UAV}(t)=1,\rho_{u,m}(t)\geq\rho_{\text{th}},u \in \boldsymbol{\mathcal{U}} \}.
\label{constraint_3}
\end{equation}

\subsection{Objective Model}
In this paper, the objective is to maximize the number of connected UEs and the accessible data rate under a set of constraints by optimizing UAV trajectories. To formulate the networking task, we define all UAVs' trajectories as $\boldsymbol{\tau} = \{\boldsymbol{l}_u(t)|u\in\boldsymbol{\mathcal{U}},t\in\boldsymbol{\mathcal{T}}\}$, which records the coordinates of each UAV over time. The corresponding optimization problem is then defined as
\begin{align}
    \boldsymbol{\tau}^*=&\ \mathop{\arg\max}\limits_{\boldsymbol{\tau}}  \;(\frac{1}{T}\sum_{t=1} ^T(\frac{1}{M}\sum_{m=1} ^M c_m^\text{UE}(t) + \frac{\kappa}{M}\sum_{m=1} ^M r^\text{UE}_m(t))) \tag{P1}\\
    s.t.\ \ 
    &
    ||\boldsymbol{l}_u(t+1)-\boldsymbol{l}_u(t)||_2\leq\Delta \boldsymbol{l}_u, \forall u\in\boldsymbol{\mathcal{U}} \tag{C1}\\
    &
    \boldsymbol{l}_u(t)\in \boldsymbol{L}^E, \forall u\in\boldsymbol{\mathcal{U}} \tag{C2}\\
    &
    \text{\eqref{constraint_1} and \eqref{constraint_2}} \tag{C3}
\end{align}
where $\kappa$ serves as a weighting factor that balances the objective function.
The objective function, as shown in equation (P1), aims to maximize the coverage and data rate of all UEs over the optimization period. The first term represents the number of connected UEs, while the second term captures the data rate achieved by UEs.
Constraint (C1) limits the movement of each UAV such that the distance it can travel in a single time slot is $\Delta \boldsymbol{l}_u = \omega \Delta t$, where $\omega$ denotes the UAV speed and $\Delta t$ represents the duration of one time slot.
Constraint (C2) restricts UAV trajectories to the spatial confines of the environment, denoted by $\boldsymbol{L}^E$. Moreover, the optimization problem is further bounded by communication and connectivity constraints, as delineated in equations \eqref{constraint_1} and \eqref{constraint_2}.

\section{Methodology}
\label{Methodology}

To address the scalability challenges and extensive exploration space in large-scale UAV networking, this section introduces the MRLMN framework.  The networking objective is reformulated as a stochastic game, and a decentralized learning architecture based on the IPPO is adopted.  Coordination is facilitated through an information aggregation mechanism and a role-based grouping strategy with decomposed rewards, while behavioral constraints ensure feasible motion and communication.  A knowledge distillation module is employed to integrate LLM guidance, significantly improving exploration efficiency and guiding the learning of agent policies in complex decision scenarios.

\subsection{Stochastic Game Formulation}
To capture the dynamic and uncertain nature of the operational environment and the interdependent decision-making processes of individual UAVs, the multi-UAV networking problem is modeled as a stochastic game defined by the tuple $(\boldsymbol{\mathcal{U}}, \boldsymbol{\mathcal{S}}, \{\boldsymbol{\mathcal{A}}^u\}_{u\in\boldsymbol{\mathcal{U}}}, P, \{\boldsymbol{\mathcal{R}}^u\}_{u\in\boldsymbol{\mathcal{U}}}, \gamma)$.

\subsubsection{State space $\boldsymbol{\mathcal{S}}$}
The state space of the networking environment is defined to encompass a) the spatial coordinates of all nodes within the network, b) the status of communication links, including SNR and data rate, c) the connectivity status of each UAV to the available BS via the multi-hop network, denoted by $\{c_u^\text{UAV}(t)\}_{u\in\boldsymbol{\mathcal{U}}}$.

\subsubsection{Action space $\boldsymbol{\mathcal{A}}=\{\boldsymbol{\mathcal{A}}^u\}_{u\in\boldsymbol{\mathcal{U}}}$}
The action space specifies the set of movement decisions available to each UAV. In this paper, the UAVs move at a fixed cruise speed $\omega$, which is treated as a constant system parameter. Accordingly, the action space includes only the selection of movement direction within the planar domain, modeled as eight discrete directional options together with a hovering action. This formulation captures the essential trajectory control while simplifying the decision-making complexity for MARL agents.

\subsubsection{State transition probability $P$}
The state transition defines how the system evolves based on UAV actions and environmental dynamics. Given the current state $s_t$, each UAV $u$ selects an action $a_t^u$, leading to a new state $s_{t+1}$ with probability $P(s_{t+1}|s_t,\{a_t^u\}_{u\in\boldsymbol{\mathcal{U}}})$.

\subsubsection{Reward model $\boldsymbol{\mathcal{R}}=\{\boldsymbol{\mathcal{R}}^u\}_{u\in\boldsymbol{\mathcal{U}}}$}
The reward model $\boldsymbol{\mathcal{R}}$ quantifies UAV actions by evaluating network connectivity and data rate. Accordingly, the team reward, representing the environmental feedback on the effectiveness of UAV networking, is defined as
\begin{equation}
    R_t=\frac{1}{M}\sum_{m=1} ^M c_m^\text{UE}(t) + \frac{\kappa}{M}\sum_{m=1} ^M r^\text{UE}_m(t).
    \label{team_rew}
\end{equation}
Guided by this formulation, a novel reward model is proposed to support UAV decision-making, with detailed formulation provided in Section \ref{Grouping and reward}. The discount factor $\gamma$ is incorporated to balance immediate and future rewards.

\subsection{MARL Framework and Algorithm Overview}
In this paper, the proposed model is designed based on the MARL algorithm, where multiple agents interact within a shared environment and iteratively refine their behaviors through trial-and-error. Within the model, each UAV $u$ operates based on its own policy $\pi^u(a_t^u|o_t^u;\theta^u)$, which dictates the action it should take given its observation $o_t^u$ and learnable parameters of the policy network $\theta^u$. 
The objective is to find a set of UAV policies that maximizes the joint discounted cumulative reward
$\sum_{k=0}^{T-t}\gamma^kR_{t+k}$
under the joint policy $\pi(\boldsymbol{a}_t|s_t)=\prod_{u\in\boldsymbol{\mathcal{U}}} \pi^u(a_t^u|o_t^u)$, where $\boldsymbol{a}_t$ represents the joint action of all UAVs at time $t$.
To train each UAV with a decentralized policy $\pi^u$, the training approach is designed based on the Proximal Policy Optimization (PPO) algorithm \cite{PPO}.
The policy objective for each UAV $u$ is initially defined as 
\begin{equation}
    \mathcal{L}_u^\text{PPO}(t,\theta^u) = \mathds E_{o_t^u,a_t^u}[\min(\zeta_t^{\theta^u} A_t^u,\text{clip}(\zeta_t^{\theta^u}, 1-\epsilon, 1+\epsilon)A_t^u)],
\end{equation}
where $\zeta_t^{\theta^u}=\pi^u(a_t^u|o_t^u;\theta^u)/\pi^u(a_t^u|o_t^u;\theta^u_\text{old})$ is the probability ratio between the updated and old policies, $A_t^u$ is the advantage estimation, $\epsilon$ is a hyperparameter that controls the clipping range, restricting policy updates to ensure stable training.
The PPO entropy loss and critic loss \cite{PPO} are also incorporated into the training process. The entropy loss encourages exploration by penalizing deterministic behavior and promoting action diversity. The critic loss is designed to improve value estimation accuracy, providing a more reliable baseline for policy updates and enhancing training stability.
In this work, PPO is adopted due to its stable training behavior, where the clipped policy-update constraint prevents abrupt policy shifts that could otherwise induce disruptive topology changes or unexpected UE disconnections. Building upon PPO, the Independent PPO (IPPO) \cite{IPPO} formulation is adopted as the base framework because its decentralized policy and critic design allow each UAV to make decisions independently from local observations. This structure reduces the coordination complexity that arises when controlling large UAV swarms, mitigates the curse of dimensionality associated with multi-agent state and action spaces, and supports efficient parallel learning. As a result, IPPO offers natural scalability and robustness in dynamic multi-UAV multi-hop networking environments, making it well-suited for the proposed MRLMN architecture.

To address the challenges associated with UAV networking, several key modules are proposed based on the IPPO framework.
To enhance inter-agent cooperation and improve algorithm scalability, an information aggregation module (see Section \ref{Information Aggregation}) and a grouping strategy (see Section \ref{Grouping and reward}) are introduced as core components. Meanwhile, the reward mechanism is adjusted to deliver clearer feedback tailored to each agent's training process (see Section \ref{Grouping and reward}). During MARL training, the loss function is further augmented with a behavioral constraint term $\mathcal{L}^{\text{BC}}$ to encourage the maintenance of robust multi-hop connectivity (see Section \ref{Behavioral_Constraint}). Additionally, a knowledge distillation loss $\mathcal L^\text{KD}$ based on LLMs is designed to promote structured exploration within the MARL paradigm (see Section \ref{LLM_KD_section}).
Consequently, the overall objective for each agent $u$ is expressed as
\begin{equation}
    \max_{\pi^u} \; \mathcal{L}_u^\text{PPO}(t) - \beta_1 \mathcal{L}^{\text{KD}}_u(t) -  \beta_2\mathcal{L}^{\text{BC}}_u(t),
\end{equation}
where $\beta_1$ and $\beta_2$ are weighting coefficients that balance the relative importance of the primary objective with other terms.
Through the integration of these components, the proposed framework effectively balances the optimization of network performance, the enforcement of connectivity constraints, and the incorporation of task-specific insights.

\subsection{Information Aggregation for Coordination}
\label{Information Aggregation}
To enhance inter-agent cooperation, an information aggregation module is designed for each agent, and its output, together with the agent's local observation, serves as the input to both the policy and critic networks.
Effective cooperation among UAVs in a multi-agent networking environment relies on the exchange of local observations, enabling more informed decision-making and improved coordination.
In the networking problem, UAVs are assumed to be able to share their local observations through the established communication network, enabling each UAV to acquire a more comprehensive view of the environment. The shared information at time $t$ is considered as
\begin{equation}
\begin{aligned} 
\xi(t)=\{&\{\boldsymbol{l}_n(t)\}_{n\in\boldsymbol{\mathcal{N}}},\{\rho_{i,j}(t)\}_{i,j\in\boldsymbol{\mathcal{N}}},\\&\{r^\text{UE}_m(t)\}_{m\in\boldsymbol{\mathcal{M}}},
\{c_u^\text{UAV}(t)\}_{u\in\boldsymbol{\mathcal{U}}}\}.
\end{aligned} 
\label{state_equation}
\end{equation}
$\xi(t)$ comprises four components: (i) the spatial coordinates $\boldsymbol{l}$ of UAVs, (ii) the quality of multi-hop links, measured by SNR $\rho$, (iii) the communication quality $r^\text{UE}$ received by UEs, and (iv) the link status $c^\text{UAV}$ of UAVs.
The local observations are concatenated into $\xi(t)$ based on the four distinct categories rather than simply aggregating individual agents' local observations, making the aggregation more structured and organized.
To establish information sharing among agents, the total observation for each UAV is defined as the concatenation of $\xi(t)$ and a subset of the UAV's local observation,
\begin{equation}
o_t^u=\text{concat}(\xi(t),\boldsymbol{l}_u(t),\{\rho_{u,i}(t)\}_{i\in\boldsymbol{\mathcal{N}}}),
\end{equation}
where $\text{concat}(\cdot)$ denotes the concatenation operation, with all vectors flattened before being combined.
The local observation of UAV $u$ includes its position $\boldsymbol{l}_u$ and the link qualities ${\rho_{u,i}}$ of all links associated with it.
The inclusion of the local observation is essential to ensure that each observation is uniquely associated with a specific UAV, enabling the policy network to map outputs unambiguously to the corresponding decentralized agent. This structure supports stable policy learning and enables agents to have a more comprehensive understanding of the environment, promoting better coordination and decision-making.
In practical deployments, this design is feasible as UAVs exchange only compact coordination messages over the multi-hop network. The information aggregation module can be physically separated from the UE service module, thereby reducing interference and improving system reliability.

\subsection{Task-Based Agent Grouping and Reward Decomposition}
\label{Grouping and reward}

\begin{figure}
\centering
\includegraphics[width=\linewidth]{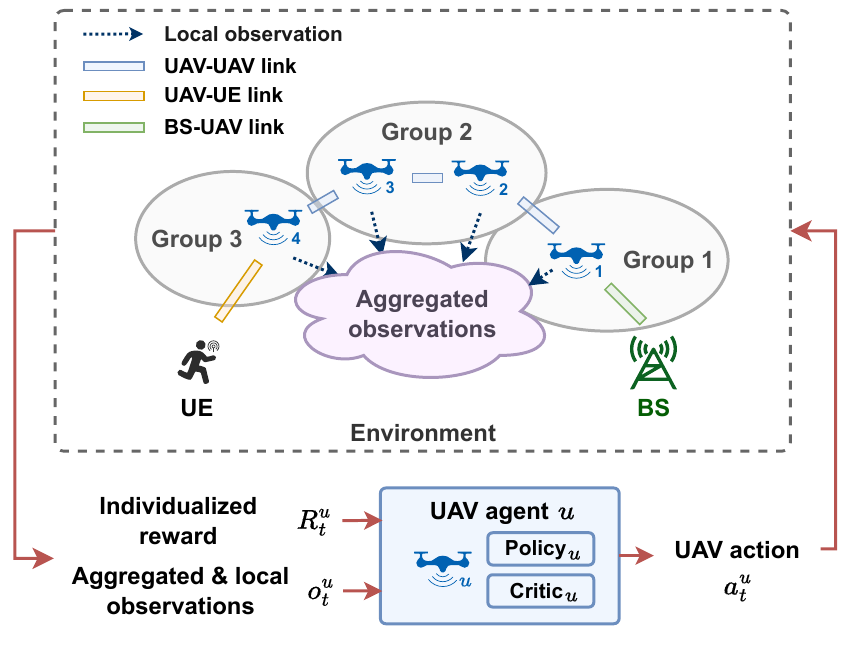}
\caption{UAVs are grouped by role, with each UAV deploying an independent PPO-based policy and critic network. Local observations are shared among agents, and each UAV receives individualized reward components.}
\label{group method}
\vspace{-0.3cm}
\end{figure}

To ensure efficient policy training and model scalability, UAVs are grouped according to their specific roles in the networking task, shown in Figure \ref{group method}. Within the networking environment, some UAVs are initially positioned closer to the BSs, making them well-suited for data relay functions. Conversely, some UAVs are located nearer to the UEs yet remain relatively distant from available BSs in emergency conditions, thereby necessitating robust communication links with users. Additionally, certain UAVs may be required to balance both responsibilities. Consequently, the responsibilities and training objectives of different UAV agents vary. To address this heterogeneity, the UAVs are partitioned into different groups $\mathds{G}_i,i\in\{1,2,\dots,N_\mathds{G}\}$.
Specifically, each UAV $u$ is assigned to a group based on its distance to the nearest BS at time step 0, defined as
\begin{equation}
d_u^{\boldsymbol{\mathcal{G}}}=\text{min}\{d_{u,g}(t=0)|g\in\boldsymbol{\mathcal{G}}\}.
\label{grouping_dist}
\end{equation}
To enable efficient group partitioning,
a quantile-based segmentation strategy is applied to divide them into multiple groups.
In this strategy, the UAVs are sorted based on $d_u^{\boldsymbol{\mathcal{G}}}$ and then partitioned into groups by dividing the ordered list into quantiles.
The number of groups and the size of each group are jointly determined by the total number of UAVs and the spatial scale of the environment.
UAVs with smaller $ d_u^{\boldsymbol{\mathcal{G}}} $, typically located closer to BSs, are assigned to smaller groups, with group sizes approximately matching the number of BSs. In contrast, UAVs with larger $ d_u^{\boldsymbol{\mathcal{G}}} $ values are grouped into larger subsets to improve communication coverage for UEs.
Thus, this grouping strategy ensures that UAVs within the same group exhibit similar proximities to the BSs, while UAVs in different groups display distinct distance ranges. By aligning the training objectives with the distinct roles and responsibilities of UAVs in network tasks, this approach enables the development of customized training strategies. 

Considering the distinct UAV groups $\mathds{G}_i$, a reward decomposition module is developed to assign differentiated rewards to each agent, thereby ensuring tailored feedback. In the networking task, the overall team reward is formulated to capture both the connectivity and quality of communication service, as defined in equation \eqref{team_rew}.
To ensure that each UAV receives targeted feedback, the networking task is subdivided into distinct components that delineate different agent responsibilities. The reward decomposition is formulated based on the following metrics:

\subsubsection{UAV-UE connection}
This component evaluates the number of UEs connected directly to a given UAV $u$. Initially, a criterion for determining whether UE $m$ is directly connected to UAV $u$ is defined as
\begin{equation}
\begin{aligned} 
I_{u,m}(t)=\mathds 1(u=&\mathop{\arg\max}\limits_{v \in \boldsymbol{\mathcal{U}}} \; \{ r_{v,m}(t) | \\ 
&c_v^\text{UAV}(t)=1,\rho_{v,m}(t)\geq\rho_{\text{th}} \}).
\end{aligned} 
\end{equation}
where $\mathds 1(\cdot)$ is the indicator function that returns 1 if UAV $u$ can provide the highest communication quality to UE $m$ among all UAVs that $m$ can directly connect to, and UAV $u$ itself maintains a connection to the BSs. Therefore, to encourage UAVs to maintain high-quality communication links, the total data rate provided by UAV $u$ to its connected UEs is measured by
\begin{equation}
    R^{\text{Conn}}_u(t)=\sum_{m=1}^{M}I_{u,m}(t)r_{m,u}(t).
\end{equation}
The reward $R^{\text{Conn}}_u$ incentivizes UAVs based on their direct connections with UEs, promoting efficient connectivity management within the network.

\subsubsection{Relay responsibility}
In a multi-hop network, without adequate and direct incentives, UAVs may not optimally perform their relay duties, potentially disrupting network robustness, stability and overall performance. To quantify the contribution of UAV $u$ in relaying data for other UAVs and UEs, $R^{\text{RE}}_u$ is defined to measure the total number of UEs that relay through UAV $u$ when $u$ serves as a relay among other UAVs.
Primarily, the communication path through which a UAV connects to the BSs via a multi-hop network is defined as
\begin{equation}
    \text{path}_u(t)=\left\{
            \begin{aligned}
                &(u \rightarrow g),\  \text{if } \exists g\in\boldsymbol{\mathcal{G}}, \rho_{g,u}(t)\geq\rho_\text{th}\\
                &(u \rightarrow \text{path}_v(t)),\ \text{if condition (\ref{path}.1)} \\
                &\emptyset,\ \text{otherwise},
            \end{aligned}
            \right.
    \label{path}
\end{equation}
where $\rightarrow$ indicates
the backhaul data transmission between network nodes.
And condition (\ref{path}.1) states
\begin{align}
    \nexists g\in\boldsymbol{\mathcal{G}}, \rho_{g,u}(t)\geq\rho_\text{th}
    \text{ and }
    v=\mathop{\arg\min}\limits_{v\in\boldsymbol{\mathcal{V}}} \;\text{len}(\text{path}_v(t)),\tag{\ref{path}.1}
\end{align}
where $\text{len}(\cdot)$ calculates the number of communication hops in a path.
Additionally, $\boldsymbol{\mathcal{V}}=\{v|v\in\boldsymbol{\mathcal{U}}\backslash\{u\},\rho_{u,v}(t)\geq\rho_\text{th},\text{path}_v(t)\neq\emptyset\}$ represents the set of UAVs that UAV $u$ can directly connect to, where each UAV in this set can establish a multi-hop backhaul link to the BSs.
Formula (\ref{path}) ensures that UAVs select the link with the fewest hops for data relaying. Based on all communication paths in the multi-hop network, the set of UAVs relayed through UAV $v$ can be defined by
\begin{equation}
\boldsymbol{\mathcal{U}}^{\text{RE}}_u(t)=\{v|v\in\boldsymbol{\mathcal{U}}\backslash\{u\},u\in\text{path}_v(t)\}.
\end{equation}
Therefore, the relay reward for UAV $u$ is designed as
\begin{equation}
R_u^{\text{RE}}(t) = \sum_{v \in \boldsymbol{\mathcal{U}}^{\text{RE}}_u(t)}(\sum_{m=1}^{M}I_{v,m}(t)r_{m,v}(t) ).
\end{equation}

Based on the two reward components, the rewards for UAVs are composed of the overall team reward and individual contributions related to connecting with UEs and relaying data.
These components are aggregated according to specific weights $\alpha$ assigned based on the group $\mathds{G}$ that each UAV belongs to. The aggregated result serves as the UAV's individual reward throughout the training process,
\begin{equation}
R^u_t=R_t+\alpha_1^uR^{\text{Conn}}_u(t)+\alpha_2^uR^{\text{RE}}_u(t).
\end{equation}
As UAVs are categorized by role, the weights $\alpha$ are assigned based on group affiliation. For UAVs in groups that prioritize relaying, $\alpha_2^u$ is relatively larger, emphasizing the relay reward $R^{\text{RE}}_u(t)$. Conversely, UAVs in groups that focus on UE connectivity have a higher $\alpha_1^u$, reinforcing the importance of $R^{\text{Conn}}_u(t)$.
Based on this method, UAV agents prioritize tasks aligned with their designated roles during training, thereby enhancing the efficiency of policy learning.

\subsection{Behavioral Constraint for Robustness}
\label{Behavioral_Constraint}
In this paper, behavioral constraints are introduced to regulate the actions of UAVs directly connected to the BS within multi-hop networks. Notably, disconnections within this group, denoted as $\mathds{G}_\text{BS}$, present a significant risk, as they can trigger cascading failures along subsequent relay paths and ultimately lead to widespread network disruption.
Thus, the UAVs in group $\mathds{G}_\text{BS}$ serve as vital intermediaries, facilitating communication among nodes located beyond the BS's direct transmission range, thereby sustaining overall connectivity and optimizing objectives.
To address this issue, if the SNR of the links between UAV $u\in\mathds{G}_\text{BS}$ and all BSs falls below $\rho_\text{th}$, the UAV should be guided toward the BS with the highest SNR, defined as
\begin{equation}
g^*=\mathop{\arg\max}\limits_{g\in\boldsymbol{\mathcal{G}}} \;\rho_{u,g}.
\end{equation}
The expected directional guidance for the UAV is subsequently computed as
\begin{equation}
\boldsymbol{z}_u^\text{BC}(t)=
\frac{\boldsymbol{l}_{g^*}(t)-\boldsymbol{l}_u(t)}{||\boldsymbol{l}_{g^*}(t)-\boldsymbol{l}_u(t)||_2}.
\end{equation}
To maintain consistency between the computed guidance direction and the MARL action space, each action $a_i\in\boldsymbol{\mathcal{A}}$ is mapped to a direction vector as $\boldsymbol{z}_i=\text{map}(a_i)$. For an agent $u$ with action space $\boldsymbol{\mathcal{A}}^u$ , the corresponding set of directional vectors is formulated as
\begin{equation}
    \boldsymbol{Z}^u = \{ \text{map}(a_i) \}_{a_i \in \boldsymbol{\mathcal{A}}^u}
    \label{actionMapping}
\end{equation}
To ensure alignment between the derived guidance and the predefined discrete action space, the desired movement direction toward $g^*$ is mapped to the closest available action in the action space $\boldsymbol{\mathcal{A}}$, as defined by
\begin{equation}
\boldsymbol{z}^*_u(t)=\mathop{\arg\max}\limits_{\boldsymbol{z}_i\in\boldsymbol{Z}^u} \;\cos(\boldsymbol{z}_u^\text{BC}(t), \boldsymbol{z}_i).
\end{equation}
To mitigate the risk of large-scale disconnection, a supplementary loss term is introduced for each UAV $u\in\mathds{G}_\text{BS}$ at time $t$, defined as
\begin{equation}
\begin{aligned}
\mathcal{L}^{\text{BC}}_u(t) = -\mathds{1} ( (\max_{g\in\boldsymbol{\mathcal{G}}}&  \;\rho_{u,g}(t)) < \rho_{\text{th}},u\in\mathds{G}_\text{BS})\\&w_\text{BC} \log \pi^u(\text{map}^{-1}(\boldsymbol{z}_u^*(t))|o_t^u),
\end{aligned}
\end{equation}
where $\text{map}^{-1}(\cdot)$ represents the inverse of the mapping function $\text{map}(\cdot)$, and
$w_\text{BC}=||\boldsymbol{l}_{g^*}(t)-\boldsymbol{l}_u(t)||_2$ scales the loss according to the UAV-BS distance.
The imposed constraint on UAV behavior ensures stable connections with BSs, preventing erratic movements that may lead to large-scale network disconnections. This regulation is necessary for UAVs directly connected to BSs, as their movement patterns are more deterministic.
In contrast, other UAV groups place greater emphasis on non-myopic planning, requiring increased flexibility to adapt to dynamic network conditions, making strict constraints unnecessary.

\subsection{LLM Agent and Knowledge Distillation}
\label{LLM_KD_section}

\begin{figure}
\centering
\includegraphics[width=\linewidth]{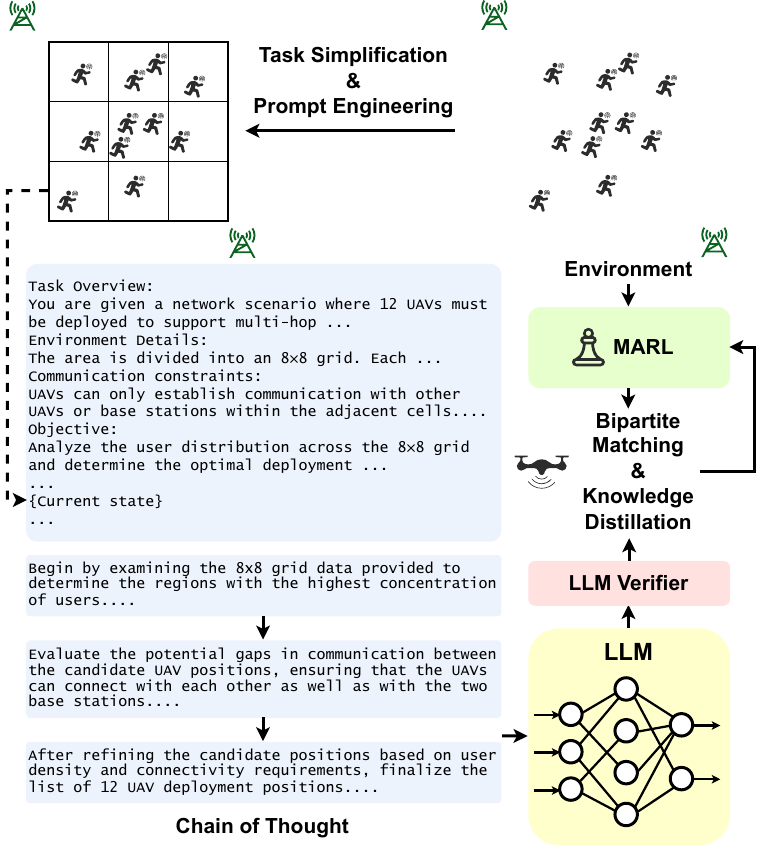}
\caption{A knowledge distillation mechanism is proposed to transfer the LLM's decision-making capabilities to MARL agents. This design incorporates a bipartite matching strategy and a distillation loss function to align the decisions of the LLM and MARL agents, while simultaneously utilizing LLM's chain-of-thought reasoning.}
\label{LLM method}
\vspace{-0.3cm}
\end{figure}

Given that LLMs possess knowledge and capabilities that are aligned with human preferences, they can be leveraged to guide and enhance the RL training process, thereby reducing unproductive exploration. LLMs do not aim to achieve globally optimal UAV networking solutions, but instead provide high-level strategic guidance that narrows the MARL exploration space.  This guidance effectively mitigates the cold-start problem during the initial stages of MARL training, improving learning efficiency and stability.
Pretrained LLMs, although primarily trained on natural language data, can understand high-level semantic information about tasks and scenarios. By embedding chain-of-thought (CoT) reasoning into the inference process, LLM-based agents can interpret tasks, extract key features of networking tasks and environmental states, and make informed decisions.
Additionally, random initialization in MARL often leads to inefficient exploration in the early stages, heightening the risk of converging to a local optimum. Given the scarcity of feasible network configurations within the vast search space, we propose utilizing LLM-driven decision-making to enhance the MARL training process.
Therefore, an LLM agent is designed to offer an initial attempt at tackling the UAV multi-hop networking task, as shown in Figure \ref{LLM method}.
To ensure that the LLM fully comprehends the networking task and the environmental state for better reasoning, the networking scenario is simplified to preclude the analysis of thousands of precise numerical values, such as location coordinates.
And LLM agents generally do not perform at the level of models that have undergone domain-specific training. In this paper, the LLM module is designed primarily to supply RL agents with decisions that align with common sense, thereby aiding their exploration and training processes. Thus, the adoption of an appropriately simplified environment is considered both acceptable and beneficial.
Notably, a single UAV can provide communication services over a contiguous region. In line with this regional characteristic of network service delivery, the entire environment is partitioned into grid cells.
The grid width is configured to ensure that the UAV positioned at the center of each grid cell can communicate with the UAVs in the adjacent cells and establish connections with the majority of UEs within the grid, i.e. meeting the SNR requirements.
Thus, the grid side length is given by $d^\text{grid}=\frac{1}{\sqrt{2}}(\max\{d_{u,v}|\rho_{u,v}\geq\rho_{\text{th}}\})$, where $u, v\in\boldsymbol{\mathcal{U}}$ and the SNR $\rho_{u,v}$ is determined by the distance $d_{u,v}$ between UAVs.
Consequently, the distribution of UEs is quantified by counting the number of UEs in each grid cell. Moreover, the UAV positions generated by the LLM are constrained to the centers of these grid cells, reducing the complexity of considering connectivity constraints.

To effectively harness the LLM's capabilities, a structured prompt is designed to encapsulate the key aspects of the task, enabling the LLM to generate precise, structured outputs. The input prompt to the LLM consists of several components:
\begin{itemize}
    \item \textbf{Scenario description:}
    Detail the UAV networking tasks, communication constraints, and emphasize relaying in multi-hop network organization.
    \item \textbf{Model behavior and objective:}
    Clarify that the LLM should analyze UE distribution to determine optimal UAV deployment, ensure relay connections, and maximize the objective function.
    \item \textbf{Output constraints:} Specify the desired output format, e.g. ``\texttt{... using the following format: $\backslash$"[(UAV 3D coordinates), (UAV 3D coordinates),...]$\backslash$"...}"
    \item \textbf{Few-shot examples:} Include representative input-output examples to guide the LLM in generating structured and accurate responses.
    \item \textbf{Current states:} Provide key environmental information, including locations of BSs and UE distribution.
\end{itemize}

Directly mapping current states to UAV positions poses a challenge for the LLM agent \cite{CoT}, as it requires accounting for UE distribution, organization of multi-hop networks and ensuring UAV connectivity constraints for networking. Inspired by the CoT approach \cite{CoT}, the LLM reasoning process for the problem is partitioned into three sequential steps: a) Analyze UE distribution. Identify densely populated areas and determine candidate UAV deployment locations. b) Address connectivity gaps. Evaluate potential UAV connectivity gaps and reorganize UAV placements to ensure robust network connectivity. c) Determine the final UAV deployment based on the analysis of connectivity constraints and UE distribution.
This stepwise approach enables the LLM to use more intermediate tokens to sequentially analyze the scenario and requirements, thereby enhancing the overall quality of the response.

To ensure reliable integration of LLM-generated guidance into MARL training, a rule-based verifier is used to validate UAV deployment plans and filter out infeasible outputs. CoT reasoning by the LLM can produce incorrect or inconsistent intermediate steps when generating high-level strategic plans for complex multi-UAV multi-hop networking tasks. The verifier checks that UAV positions are within operational boundaries and physically reachable. It verifies network connectivity by confirming that multi-hop paths remain connected, only a small number of UAVs are isolated, and sufficient UE coverage is maintained according to predefined thresholds. Deployment plans that fail these checks are discarded, preventing infeasible guidance from influencing MARL policy learning. Other methods, such as Retrieval-Augmented Generation \cite{pmlr-v235-sun24e}, Reasoning and Acting \cite{react}, or Supervised Fine-Tuning \cite{10.5555/3692070.3693945}, can also improve the reliability of LLM outputs. However, they introduce substantial computational overhead during training. Discarding invalid outputs using a lightweight, rule-based verifier ensures the quality and reliability of LLM-generated guidance while providing an efficient and practical solution for the task.

Building on the designed LLM agent, a knowledge distillation mechanism is proposed to further enhance MARL performance and guide its training. Within this framework, the LLM is designated as the teacher and the MARL agents as the students. A distillation loss is designed to transfer the LLM's decision-making capabilities to the MARL agents by quantifying the similarity in network outputs between the LLM and the MARL policies.
To enable the current UAVs to efficiently approach the positions provided by the LLM, an optimal bipartite matching is computed, establishing a one-to-one correspondence between the LLM-inferred positions $\{\boldsymbol{l}^{\text{LLM}}_u(t)\}_{u\in\boldsymbol{\mathcal{U}}}$ and the current UAV positions $\{\boldsymbol{l}_u(t)\}_{u\in\boldsymbol{\mathcal{U}}}$.
Thus, a permutation function $\sigma\in\mathfrak{S}_U$
is introduced, where $\mathfrak{S}_U$ denotes the symmetric group of all permutations over $U$ elements.
The goal is to determine $\sigma$ that minimizes the pairwise cost
\begin{equation}
\label{match_loss}
    \mathcal{L}^{\text{MATCH}}(\boldsymbol{l}_u,\boldsymbol{l}^{\text{LLM}}_{\sigma(u)})=||\boldsymbol{l}_u-\boldsymbol{l}^{\text{LLM}}_{\sigma(u)}||_2,
\end{equation}
defined as the Euclidean distance between the UAV indexed by $u$ in the environment and its matched counterpart indexed by $\sigma(u)$ given by the LLM.
To obtain the optimal permutation, the objective is formulated as
\begin{equation}
\sigma^*=\mathop{\arg\min}_{\sigma\in\mathfrak{S}_U} \;\sum_{u\in\boldsymbol{\mathcal{U}}}\mathcal{L}^{\text{MATCH}}(\boldsymbol{l}_u,\boldsymbol{l}^{\text{LLM}}_{\sigma(u)}).
\end{equation}
In this paper, the optimization of the permutation is achieved using the Hungarian algorithm \cite{DETR}.
Under the optimal permutation, the LLM-expected action for each UAV $u$ in the current state is derived as 
\begin{equation}
\boldsymbol{z}_u^\text{LLM}(t)=\boldsymbol{l}^{\text{LLM}}_{\sigma^*(u)}(t)-\boldsymbol{l}_u(t).
\end{equation}
To align the LLM-inferred action with the MARL agent's discrete action space, a soft target distribution is constructed based on the cosine similarity between the inferred action and each candidate action in the predefined action space.
Specifically, for each discrete action in the MARL action space, the soft target probability is formulated as
\begin{equation}
\widetilde p_u(\boldsymbol{z}_i,t)=\frac{\exp(\cos(\boldsymbol{z}_u^\text{LLM}(t),\boldsymbol{z}_i)/\Omega)}{\sum_{\boldsymbol{z}_j\in\boldsymbol{Z}^u}\exp(\cos(\boldsymbol{z}_u^\text{LLM}(t),\boldsymbol{z}_j)/\Omega)},\boldsymbol{z}_i\in\boldsymbol{Z}^u
\end{equation}
where $\cos(\cdot)$ ensures that actions more aligned with the LLM-inferred direction receive higher probabilities, $\boldsymbol{Z}^u$ represents the set of directional vectors corresponding to the action space, as defined in equation \eqref{actionMapping} and $\Omega$ is a temperature parameter that controls the smoothness of the probability distribution. The soft target distribution encapsulates information from LLM, offering nuanced guidance during training \cite{KD}. The distillation loss is then defined as the cross-entropy between the MARL agent's policy and the soft target distribution
\begin{equation}
\mathcal L_u^\text{KD}(t)=-\sum_{\boldsymbol{z}_i\in\boldsymbol{Z}^u}\widetilde  p_u(\boldsymbol{z}_i,t)\log\pi^u(\text{map}^{-1}(\boldsymbol{z}_i)|o_t^u).
\end{equation}
This formulation enables MARL agents to effectively leverage the LLM's inferred actions as supervisory signals, allowing them to learn the LLM's decision-making capabilities. In the proposed framework, the LLM participates exclusively during the offline training phase, providing high-level strategic guidance for UAV deployment and multi-hop network coordination.   Its outputs are incorporated into MARL policies through a soft target-based distillation loss, ensuring that strategic priors guide the learning process.   This offline integration prevents the LLM from participating in real-time UAV control, avoiding additional latency and computational overhead, and making the method feasible for practical network deployments.   During online deployment, each UAV executes its MARL policy independently, without LLM inference, ensuring that real-time control is decentralized, incurs no extra computational burden, and remains practical and reproducible in large-scale multi-hop UAV networks. This technique promotes structured exploration and enhances both the training and inference processes of MARL.

\subsection{Overall Algorithm and Complexity Analysis}

\begin{algorithm}[H]
	\caption{Training procedure of MRLMN}
	\label{alg:mrlmn}
	\begin{algorithmic}
		\State \textbf{Initialize:} Policy networks $\pi_{\theta}$, PPO replay buffer $\mathcal{D}$
		\State \textbf{Parameters:} $N_{\text{ep}}$ (episodes), $T$ (episode horizon), $K$ (training epochs), $Q_{\text{LLM}}$ (LLM guidance interval)
		\State \textbf{Cache:} $\mathcal{T}^{\text{LLM}}_{\text{cached}} \gets \emptyset$  \Comment{Cached LLM-suggested UAV deployment}
		
		\For{episode $k = 1 \dots N_{\text{ep}}$}
		\State $s_0 \sim p_{\text{init}}(s)$ \Comment{Reset environment}
		\State Initialize agent grouping for the episode
		
		\For{step $t = 1 \dots T$}
		\State $\mathbf{o}_t^{\text{ind}} \gets \text{GetEnvironmentObservation}(s_t)$
		\State $\mathbf{o}_t \gets \text{AggregateInfo}(\mathbf{o}_t^{\text{ind}}, \text{grouping})$ 
		\State \Comment{Information aggregation}
		
		\If{$t \bmod Q_{\text{LLM}} == 0$} 
		\State $\mathcal{T}^{\text{LLM}}_{\text{cached}} \gets \text{LLM\_Inference}(\mathbf{o}_t)$ 
		\State \Comment{Update and cache LLM guidance}
		\EndIf
		
		\State $\mathbf{a}_t \sim \pi_{\theta}(\cdot \mid \mathbf{o}_t)$ 
		\State $s_{t+1}, \mathbf{r}_t \gets \text{Env.Step}(\mathbf{a}_t)$ \Comment{Environment transition}
		
		\State  
		$\sigma^*=\mathop{\arg\min}_{\sigma\in\mathfrak{S}_U} \;\sum_{u\in\boldsymbol{\mathcal{U}}}\mathcal{L}^{\text{MATCH}}(\boldsymbol{l}_u,\boldsymbol{l}^{\text{LLM}}_{\text{cached}, \sigma(u)})$
		\State \Comment{Bipartite matching}
		
		\State $\mathbf{R}_t \gets \text{Reward\_Decomp}(\mathbf{r}_t, \text{grouping})$ 
		\State \Comment{Reward Decomposition}
		\State $\mathcal{D} \gets \mathcal{D} \cup \{ (\mathbf{o}_t, \mathbf{a}_t, \mathbf{R}_t, \mathbf{o}_{t+1}, \sigma^*) \}$
		
		\EndFor
		
		\For{epoch $1 \dots K$}
		\State $\max_{\pi_{\theta}} \; \mathcal{L}^\text{PPO} - \beta_1 \mathcal{L}^{\text{KD}} -  \beta_2\mathcal{L}^{\text{BC}}$
		\State \Comment{MARL Optimization with knowledge distillation and behavioral constraints}
		\EndFor
		
		\EndFor
	\end{algorithmic}
\end{algorithm}

The complete training procedure of MRLMN is summarized in Algorithm \ref{alg:mrlmn}. The framework operates in a hybrid manner. UAVs continuously execute decentralized control using MLP-based policy networks $\pi_{\theta}$ under the IPPO paradigm, while the LLM is queried only once every $Q_{\text{LLM}}$ steps to provide high-level deployment guidance. At each step, UAVs first aggregate observations from other UAVs, take actions through the learned policies, and collect transitions into the replay buffer $\mathcal{D}$. When LLM guidance is triggered, the suggested deployment topology is cached using $\mathcal{T}^{\text{LLM}}_{\text{cached}}$ and aligned with the UAVs' current configuration through bipartite matching, enabling knowledge distillation to shape subsequent policy updates. IPPO optimization is then performed at the end of each episode, incorporating both the distillation objective and the behavioral constraint. 
From a computational perspective, each UAV is equipped with an MLP consisting of $L$ hidden layers of width $H$. Thus, with $U$ UAVs executing their policies in parallel, the per-step inference cost scales as $\mathcal{O}(U \cdot L \cdot H^2)$. During training, the same order applies to forward and backward passes of both the policy and critic networks. LLM reasoning is executed once every $Q_{\text{LLM}}$ steps, with a single inference cost $\mathcal{C}_{\text{LLM}}$ contributing an amortized $\mathcal{C}_{\text{LLM}} / Q_{\text{LLM}}$ per step. The bipartite matching needed for distillation incurs $\mathcal{O}(U^{3})$ complexity, which is lower than the neural network computations and is invoked once per step. Additionally, the overhead from information aggregation, reward decomposition, and behavioral constraints is minimal. Consequently, MLP-based networks dominate inference complexity, whereas during training, both the neural networks and the amortized LLM guidance jointly determine the overall computational cost.

\section{Performance Evaluation}
\label{Performance Evaluation}

This section evaluates the proposed MRLMN framework in multi-UAV multi-hop networking scenarios under diverse conditions, including varying environment scales, UAV swarm sizes, baseline comparisons, training dynamics, and ablation studies.  The experiments demonstrate the framework's stability, scalability, adaptability, and coordination effectiveness, while also examining the impact of parameter sharing within agent groups on training efficiency and policy quality.  The results highlight the overall superior performance of MRLMN across these settings.

\subsection{Experimental Setup}

\begin{table}[h]
\centering
\caption{Parameter settings}
\label{Main_parameters}
    \begin{tabular}{cc}
    \hline
    Parameters    &Value                 \\
    \hline
    System frequency $f_c$            &Around 2.4 GHz \\
    Bandwidth $B$ for each BS-UAV link    &7 MHz \\
    Bandwidth $B$ for each UAV-UAV link   &5 MHz \\
    Bandwidth $B$ for each UAV-UE link    &1 MHz \\
    Transmitted signal power $P^{\text{TX}}$ of UAV, BS, UE         &1, 10, 0.4 W \\
    The excessive path loss for NLoS links $\eta_\text{NLoS}$   &20 dB \\
    The excessive path loss for LoS links $\eta_\text{LoS}$   &1 dB \\
    Environmental constant $a$, $b$   &9.61, 0.16 \\
    Speed of light $c$ & 3$\times$10$^\text{8}$ m/s\\
    SNR threshold $\rho_\text{th}$ & 25 dB\\
    Noise figure $\textit{NF}$&15 dB \\
    Reward decomposition related weights $\alpha_1$, $\alpha_2$&1, 3 \\
    Objective related weights $\beta_2$&0.3 \\
    Temperature parameter $\Omega$&1 \\
    \hline
    \end{tabular}
\end{table}

\subsubsection{Environment setup}
To assess the performance of our proposed method, we employ a simulation environment that spans an area of approximately 3.5 km $\times$ 3.5 km, within which around 150 UEs and 18 UAVs are deployed. The distribution of UEs follows either a uniform pattern or a two-dimensional Gaussian mixture with multiple centers, and their motion follows a Brownian process at a constant velocity. The BSs are strategically positioned at three of the four corners of the area to preclude direct connections with the UEs, thereby simulating communication conditions typically encountered in disaster scenarios. The UAV speed $\omega$ is limited to a displacement of 30 meters within each time slot.

\subsubsection{Training configuration}
For the training process, a total of 25,000 episodes are conducted, with each episode consisting of 400 time slots. The PPO policy and critic networks are implemented as multi-layer perceptrons with five hidden layers, each employing the tanh activation function. And the learning rate is initially set to $3\times 10^{-4}$ and is progressively lowered to $1\times 10^{-4}$ toward the end of the training process. The LLM guidance is provided using the GPT-4o model \cite{openai_gpt4o}, which generates high-level strategic deployment suggestions during MARL offline training. In the networking task, we assume identical transmitter and receiver antenna gains, with UAV, UE, and BS antenna gains set to 0, 0, and 5 dBi, respectively. In equation (P1), the weighting factor $\kappa$ is defined as 0.025 when the data rate $r^\text{UE}$ is measured in Mbps. The coefficient $\beta_1$, which controls the weight of the knowledge distillation component, is initially set to 0.5 and is gradually reduced to 0.1 during the final phase of training. It is designed to help improve the effectiveness of MARL by gradually shifting the focus from distillation to the training objective. In MARL training, the UAVs are divided into four groups based solely on the distance relationship between each UAV and the set of BS nodes $\boldsymbol{\mathcal{G}}$, as defined by equation (\ref{grouping_dist}). The network parameters of each agent remain independent and are not influenced by the grouping strategy.
Furthermore, a detailed discussion on the relationship between parameter sharing in groups and training performance can be found in Section \ref{pareto discussion}.
Due to the substantial inference latency of the LLM, which adversely affects the training efficiency of the MARL agents, LLM inference is performed only at selected time slots within each episode, specifically once every $Q_{\text{LLM}}$ steps to obtain the required guidance output. For subsequent time slots, the validated LLM outputs are cached and subsequently reused in the computation of the knowledge distillation loss, $\mathcal{L}^{\text{KD}}$.
Since the UE distribution typically experiences only minor changes over short time intervals, the retained outputs perform well during these periods.
Additionally, given that the LLM agent is primarily tasked with guiding MARL agent training and managing an extensive exploration space, even suboptimal network decisions from the LLM are acceptable, ensuring that this intermittent LLM reasoning strategy does not lead to a significant negative impact on knowledge distillation.
Additional critical parameter settings are summarized in Table \ref{Main_parameters}.
To focus on evaluating MARL coordination and policy performance rather than detailed physical-layer optimization, the bandwidth allocation across BS-UAV, UAV-UAV, and UAV-UE links is assumed to be fixed, which serves to mitigate cross-link interference.

\subsubsection{Performance metrics}
The performance of the algorithm is assessed using two primary metrics: the quality of service, defined as the average data rate per UE, calculated as $({1}/{M})\sum_{m=1} ^M r^\text{UE}_m$ and the connected UE proportion $(1/{M})\sum_{m=1} ^M c_m^\text{UE}$, i.e. the proportion of UEs that are successfully connected to the core network relative to the total number of UEs. To evaluate the robustness of each model in the networking task, the number of available UAVs is introduced as another metric. This metric quantifies the number of UAVs that successfully establish and maintain connectivity with the BSs and core network through the multi-hop UAV network. A higher value indicates fewer disconnections, reflecting a more resilient and stable network. Formally, it is measured as $(1/U)\sum_{u\in\boldsymbol{\mathcal{U}}}c_u^\text{UAV}$, where $c_u^\text{UAV}$ denotes the connectivity status of UAV $u$, defined in equation \eqref{constraint_1}. This metric serves as a key indicator of the stability and effectiveness of each algorithm.

\subsection{Comparative Experiments}
\begin{figure}
\centering
\includegraphics[width=0.9\linewidth]{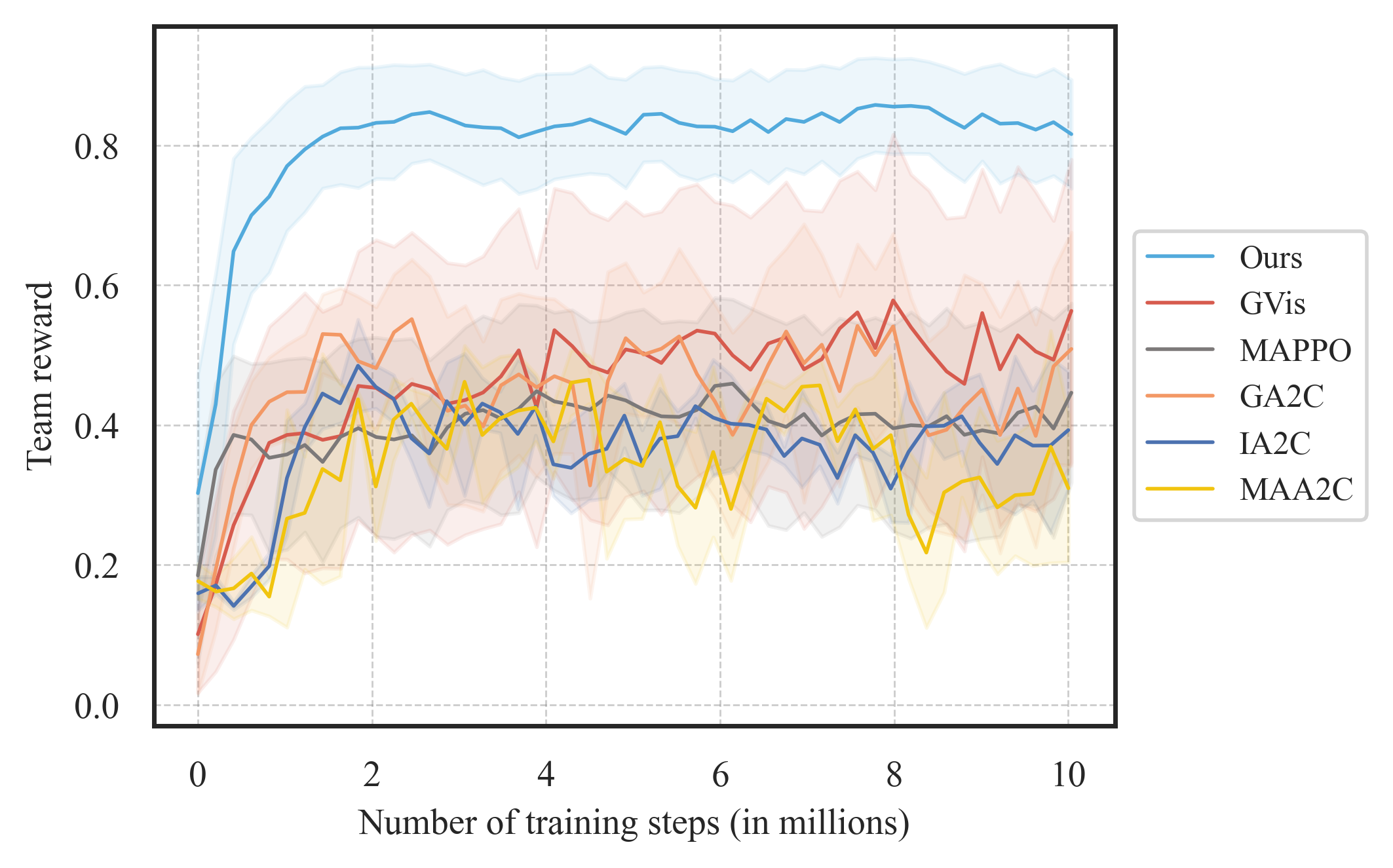}
\caption{Training curves of different models over training steps, with the team reward computed according to equation \eqref{team_rew}. Shaded regions represent the standard deviation.}
\label{training curve}
\vspace{-0.3cm}
\end{figure}

In this section, we present a comprehensive performance evaluation of the proposed MRLMN framework by benchmarking it against five representative baselines that include both independently trained agents and cooperative MARL approaches:
\begin{itemize}
    \item GVis: A MARL framework \cite{31RL6} based on heterogeneous graphs, designed to enhance cooperation among UAVs. This method optimizes local observation management and enables cooperation through explicit information exchange among nodes in the network. Slight modifications are made to adapt this framework for application in the UAV networking scenario.
    \item GA2C: A RL framework \cite{compare2} that integrates graph neural network (GNN) to guide action training within an Advantage Actor-Critic (A2C) architecture. By leveraging hidden representations from network feature correlations, this method captures intricate environmental relationships, thereby enhancing the efficiency of policy training.
    \item MAPPO: A MARL algorithm \cite{28MARL2} introduced for coordinating agents by utilizing a centralized value function during training, while allowing each agent to act independently during execution. This strategy has demonstrated competitive performance on a range of cooperative benchmark tasks.
    \item IA2C: An independent RL method based on the A2C architecture \cite{mnih2016asynchronous}. In this approach, each UAV agent learns its policy independently without explicitly considering the joint state or actions of other agents, serving as a baseline for non-cooperative learning.
    \item MAA2C: A multi-agent variant of A2C that employs a centralized training and decentralized execution paradigm similar to MAPPO but relies on the A2C optimization objective.
\end{itemize}

Figure \ref{training curve} shows the training curves over 10 million steps, depicting the performance trends of different methods in terms of the objective function defined in equation (P1). The experiments are set in a 3.5 km $\times$ 3.5 km environment with 18 UAVs, evaluating the proposed framework under large-scale networking conditions.
The results show that the proposed MRLMN method significantly outperforms the baseline methods throughout the training process. MRLMN exhibits a rapid increase in training performance, stabilizing above 0.8, while the competing methods plateau at much lower values. In detail, GVis performs relatively better among the baselines but still falls short of MRLMN. MAPPO and GA2C stabilize between 0.4 and 0.6, whereas the A2C-based methods converge more slowly and achieve lower overall performance, fluctuating around 0.4. The curves indicate that the proposed algorithm, by integrating multiple modules, achieves superior networking performance.
Specifically, the knowledge distillation module enables reinforcement learning agents to rapidly acquire generalized decision-making capabilities, accelerating the overall training process. In parallel, the reward decomposition and behavioral constraints allow for fine-tuning of agent behaviors, ensuring a better adaptation to the unique challenges of UAV multi-hop networking.

Furthermore, to provide a comprehensive view of scalability and network robustness, we conduct a detailed comparison against the key baselines (GVis, GA2C, and MAPPO) across several metrics under varying network conditions, including changes in geographical area size and the number of deployed UAVs.

\begin{figure}
\centering
\subfloat[]{\includegraphics[width=.49\columnwidth]{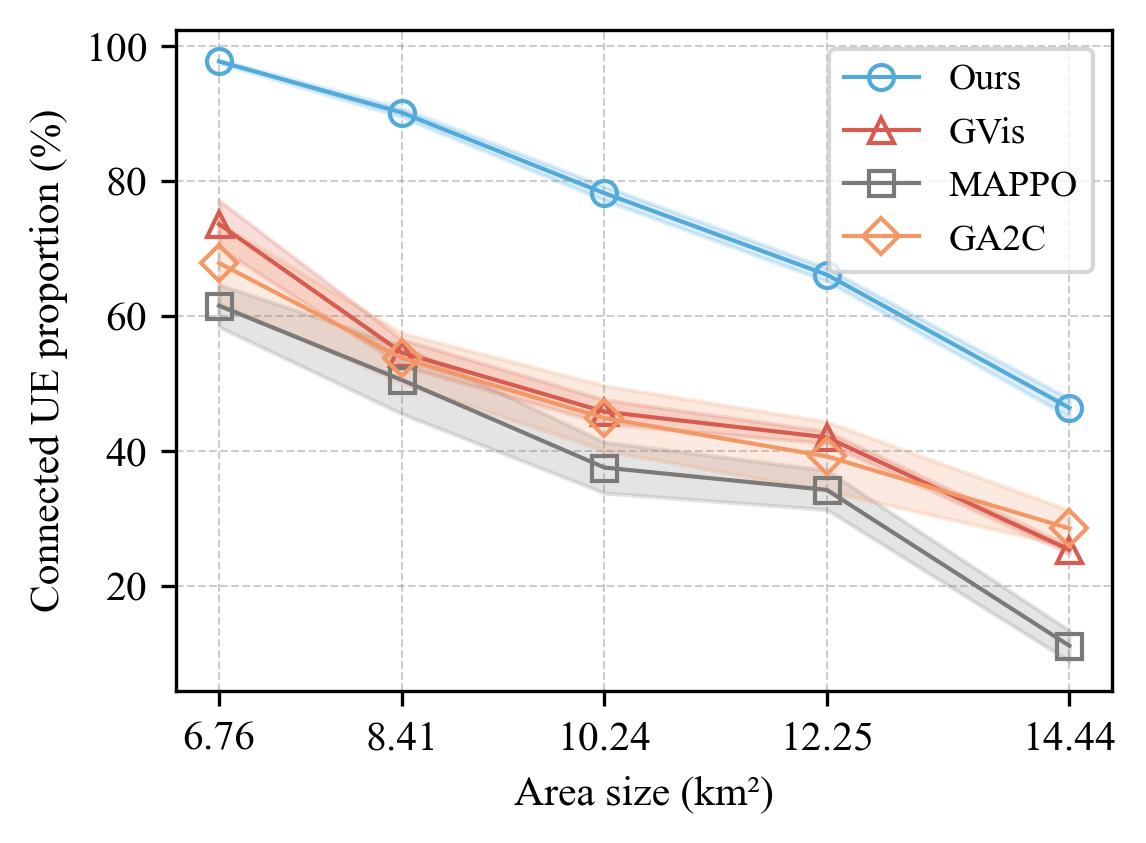}}
\subfloat[]{\includegraphics[width=.49\columnwidth]{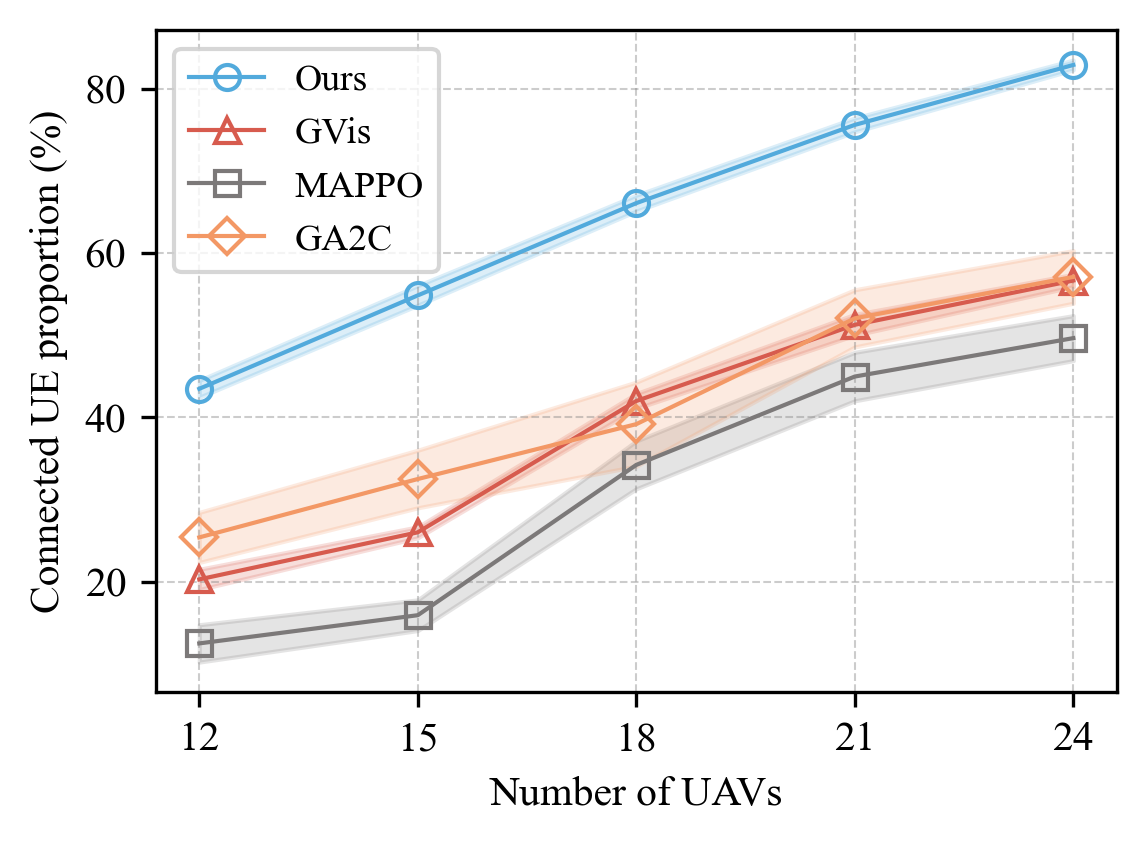}}\\
\vspace{-0.2cm}
\subfloat[]{\includegraphics[width=.49\columnwidth]{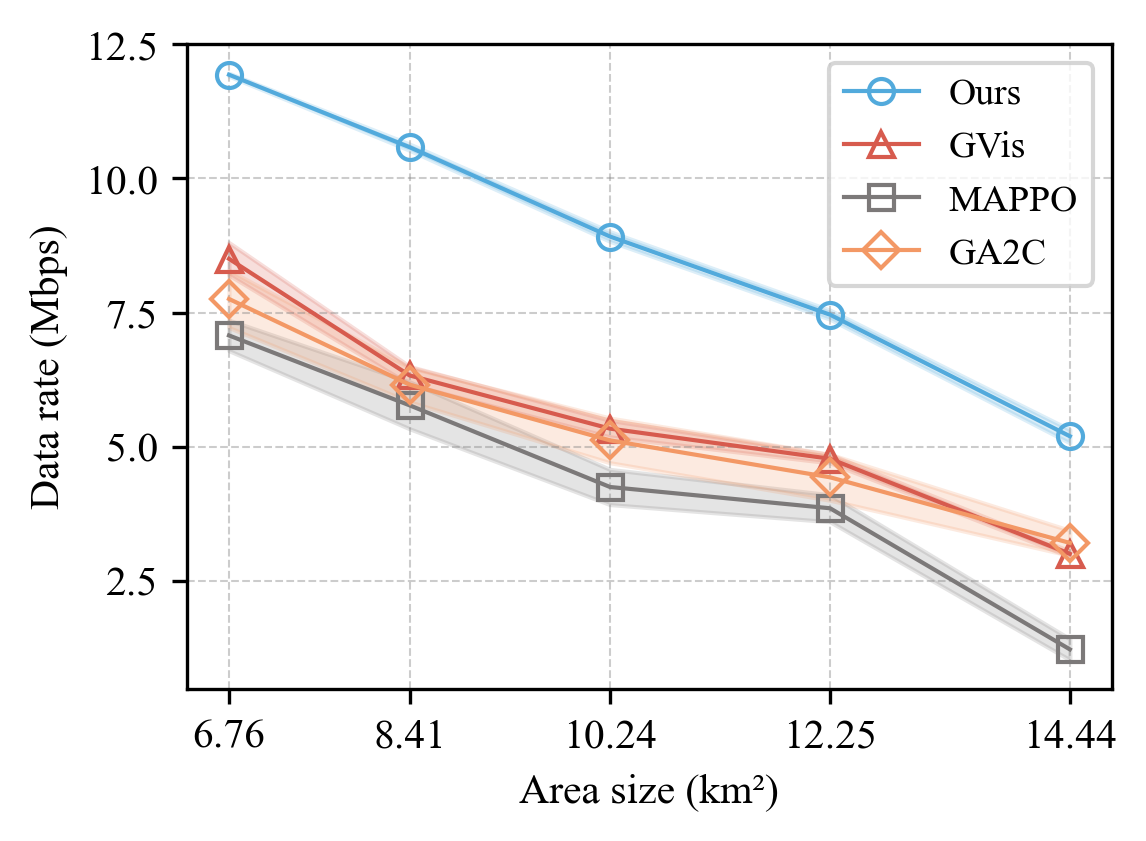}}
\subfloat[]{\includegraphics[width=.49\columnwidth]{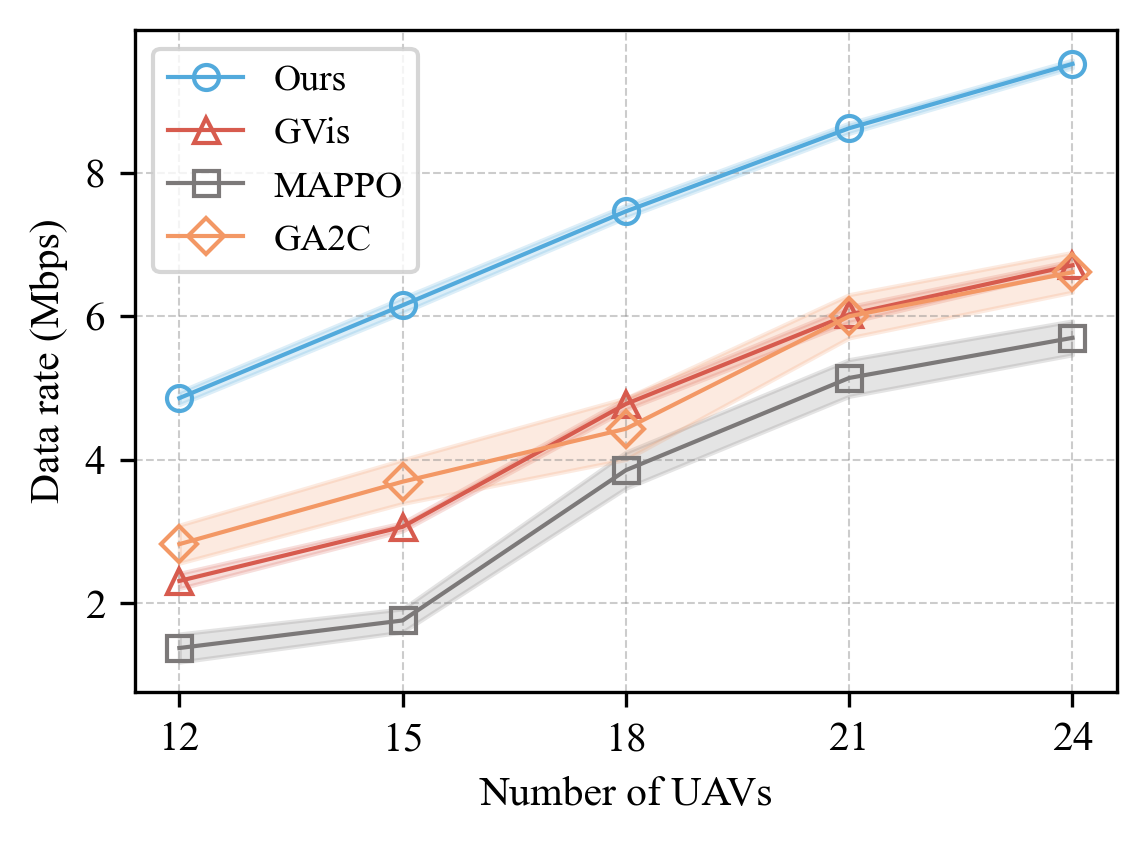}}\\
\vspace{-0.2cm}
\subfloat[]{\includegraphics[width=.49\columnwidth]{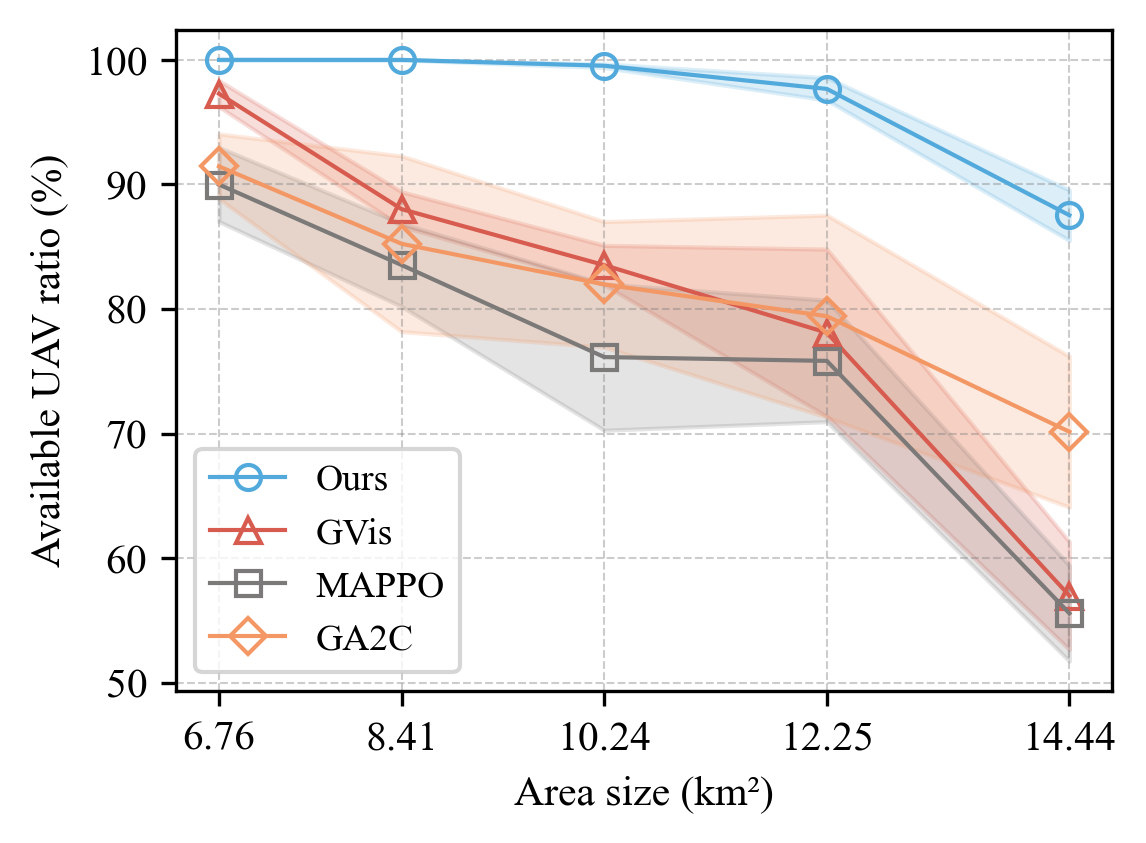}}
\subfloat[]{\includegraphics[width=.49\columnwidth]{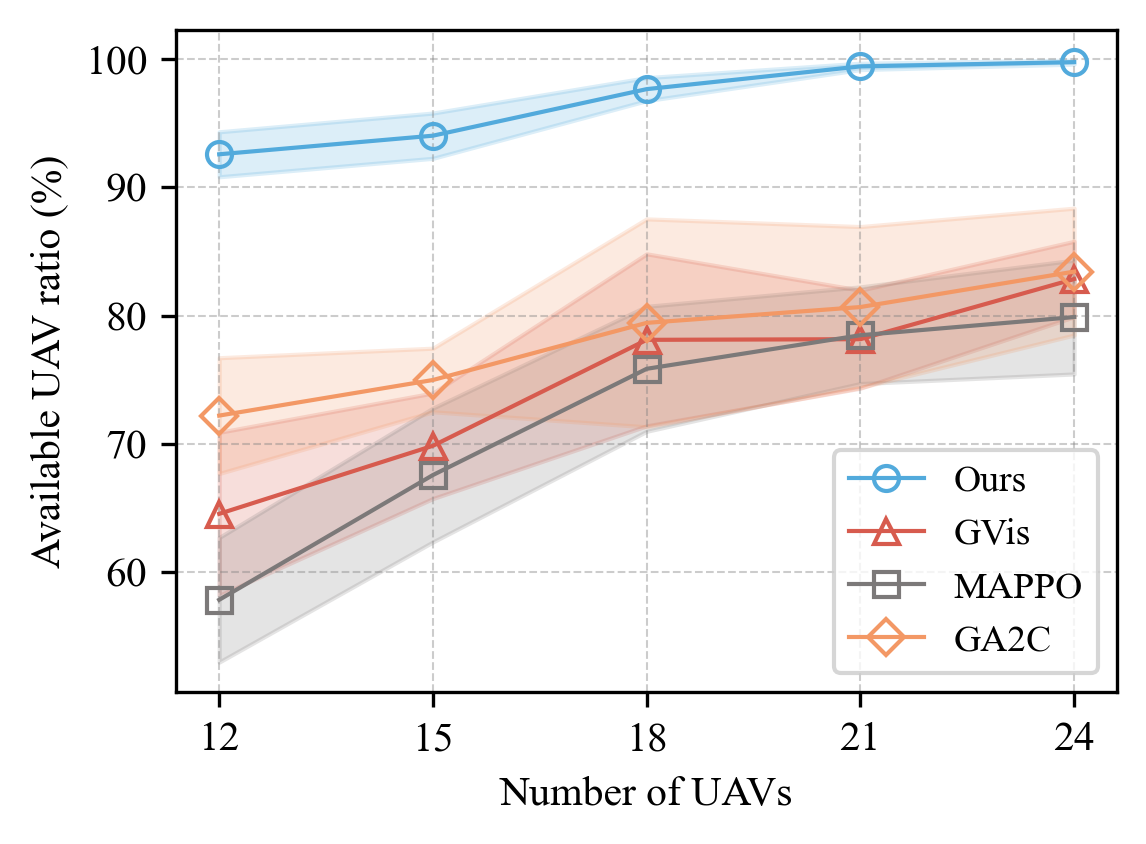}}\\
\vspace{-0.1cm}
\caption{Comparison of model performance across three key performance metrics: connected UE proportion, average data rate, and available UAV ratio. The experiments are conducted under varying environment areas and the number of UAVs. Shaded regions represent the standard deviation.}
\vspace{-0.3cm}
\label{compare}
\end{figure}

\subsubsection{Impact of environment size}
Figure \ref{compare}.(a), (c), (e) examine the influence of expanding the square environment area from 6.76 km$^2$ to 14.44 km$^2$, with the number of UAVs fixed at 18. The shaded regions indicate the standard deviation of each method, providing insight into algorithmic stability. In Figure \ref{compare}.(a), the number of connected UEs for all approaches declines as the area expands, reflecting the increased difficulty of maintaining reliable connections over larger regions populated with numerous UEs. Nevertheless, compared to GVis, GA2C, and MAPPO, MRLMN achieves an average UE coverage improvement of approximately 27\% and maintains greater performance stability across the entire range of area sizes. A similar trend is observed in Figure \ref{compare}.(c), which reports the average data rate in Mbps under the same conditions. MRLMN not only sustains better coverage but also delivers higher throughput across various environment sizes.
For these two metrics, GVis and GA2C exhibit similar performance, both relying on GNN-based strategies for UAV coordination. However, MRLMN incorporates scenario-specific grouping and reward mechanisms, allowing for more adaptive decision-making. Additionally, its knowledge distillation method is designed for large-scale environments, further enhancing overall performance.
Moreover, it is observed that the objectives of connectivity coverage and communication data rate are correlated. Adjustments in UAV networking that improve network connectivity also tend to increase communication rates, reflecting the overall performance gains achieved through coordinated multi-hop networking.
In Figure \ref{compare}.(e), all methods exhibit a reduction in the number of available UAVs, highlighting the increasing challenge of sustaining multi-hop connectivity over larger regions.
Nonetheless, MRLMN maintains a higher number of available UAVs and lower variance compared to other methods, demonstrating its enhanced stability and robustness in large-scale networking scenarios. In smaller environments, MRLMN achieves nearly 100\% availability of UAVs, while in larger environments, it outperforms the best alternative by approximately 17\%.
This advantage stems from MRLMN's relay-oriented reward design and connectivity-preserving behavioral constraints, which enable more effective multi-hop link management.

\subsubsection{Impact of the number of UAVs}
Figure \ref{compare}.(b), (d), (f) illustrate the performance of different methods as the number of UAVs increases from 12 to 24, maintaining a 3.5 km $\times$ 3.5 km environment. As the number of UAVs increases, the performance of all methods improves, as a large-scale environment requires more UAVs to establish a stable network.
Considering these three metrics, MRLMN consistently outperforms all baseline methods, demonstrating superior networking performance even under limited UAV deployment. Specifically, MRLMN achieves an average of 23\% higher UE coverage, 52\% higher data rate, and a 19\% higher UAV availability ratio compared to other methods across all three metrics.
As the number of UAVs increases, MRLMN consistently maintains its advantage over other methods, highlighting its scalability. This is largely attributed to the proposed grouping strategy and the design of the group-based reward and behavioral constraint mechanisms, which provide more precise environmental feedback and offer clearer guidance for agent training.
Meanwhile, the knowledge distillation module utilizes the matching mechanism to decompose the LLM's decisions and deliver them to individual agents, providing clear supervisory signals to guide their policy learning.

\subsection{Ablation Study}

To investigate the contribution of each key component in our proposed framework, we conduct an ablation study by testing the system under three different configurations where specific modules are removed:
\begin{itemize}
    \item \textbf{Without Agent Grouping and Reward Decomposition (NR)}: In this configuration, the grouping and reward decomposition mechanism
    is omitted. As a result, agents are trained solely with a global reward signal, which obscures the evaluation of individual contributions. This lack of localized feedback makes it more challenging for agents to adjust their own policies effectively, thereby impeding coordinated behavior.

    \item \textbf{Without Knowledge Distillation and LLM Agent (NL)}: In this configuration, the knowledge distillation module that leverages the LLM agents is omitted. Consequently, agents rely solely on random initialization for their policy parameters during the early stages of training, which leads to extensive unproductive exploration and difficulties in handling the large exploration space. This negatively impacts training efficiency and convergence, ultimately degrading overall performance.

    \item \textbf{Without Behavioral Constraint (NC)}:
    This configuration removes the behavioral constraint module designed to ensure the connectivity stability of relay UAVs. As a result, relay UAVs are more prone to disconnections from BSs, increasing the risk of network disconnections. This disruption destabilizes the overall algorithm, leading to greater performance fluctuations and reduced reliability in dynamic environments.
\end{itemize}
Each configuration is evaluated in the same simulation environment to quantify the individual impact of these components on performance metrics, i.e. network coverage, quality of network service, and algorithm stability.

\begin{figure}
\centering
\subfloat[]{\includegraphics[width=.49\columnwidth]{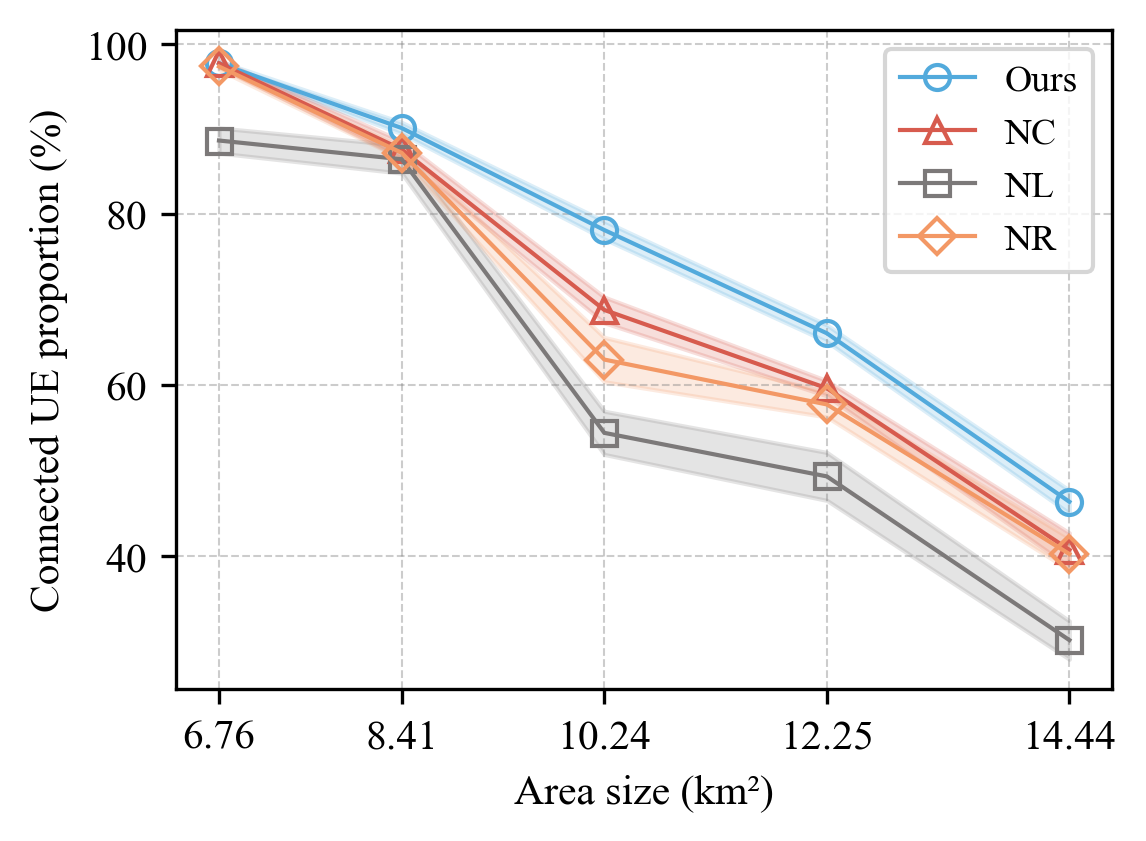}}
\subfloat[]{\includegraphics[width=.49\columnwidth]{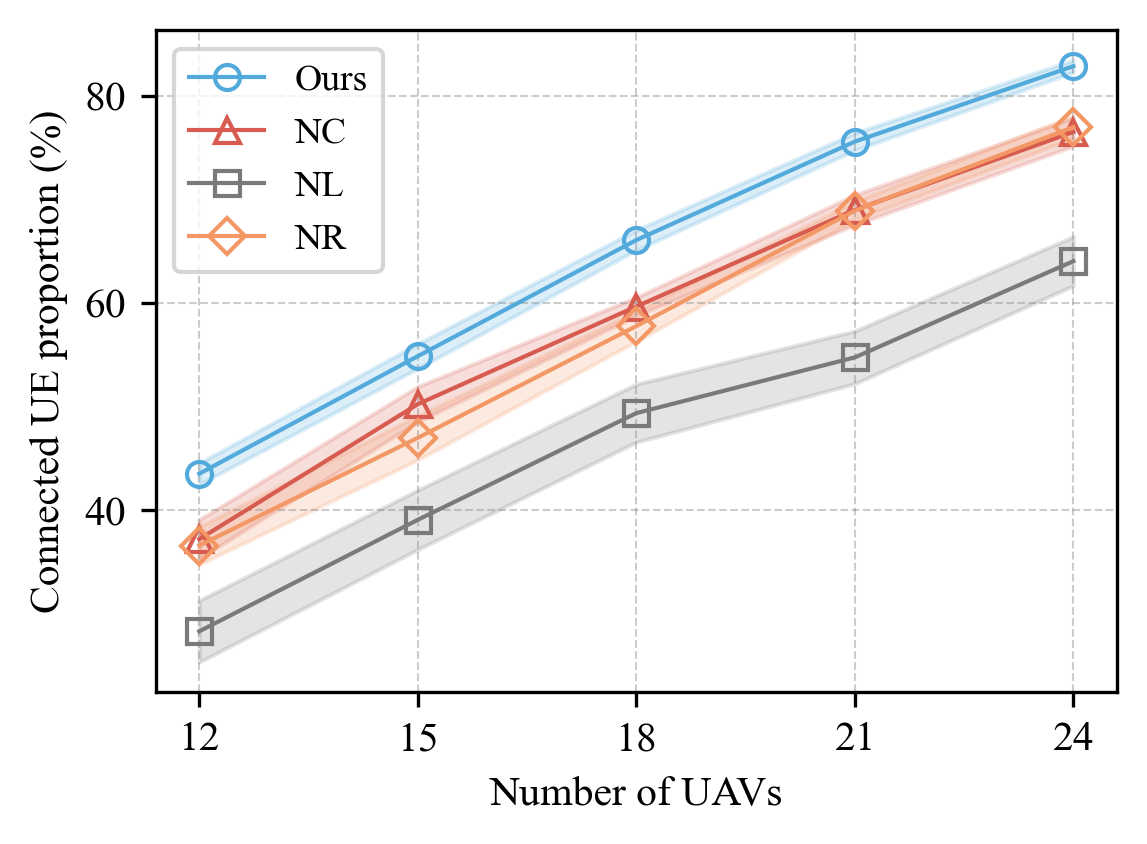}}\\
\vspace{-0.2cm}
\subfloat[]{\includegraphics[width=.49\columnwidth]{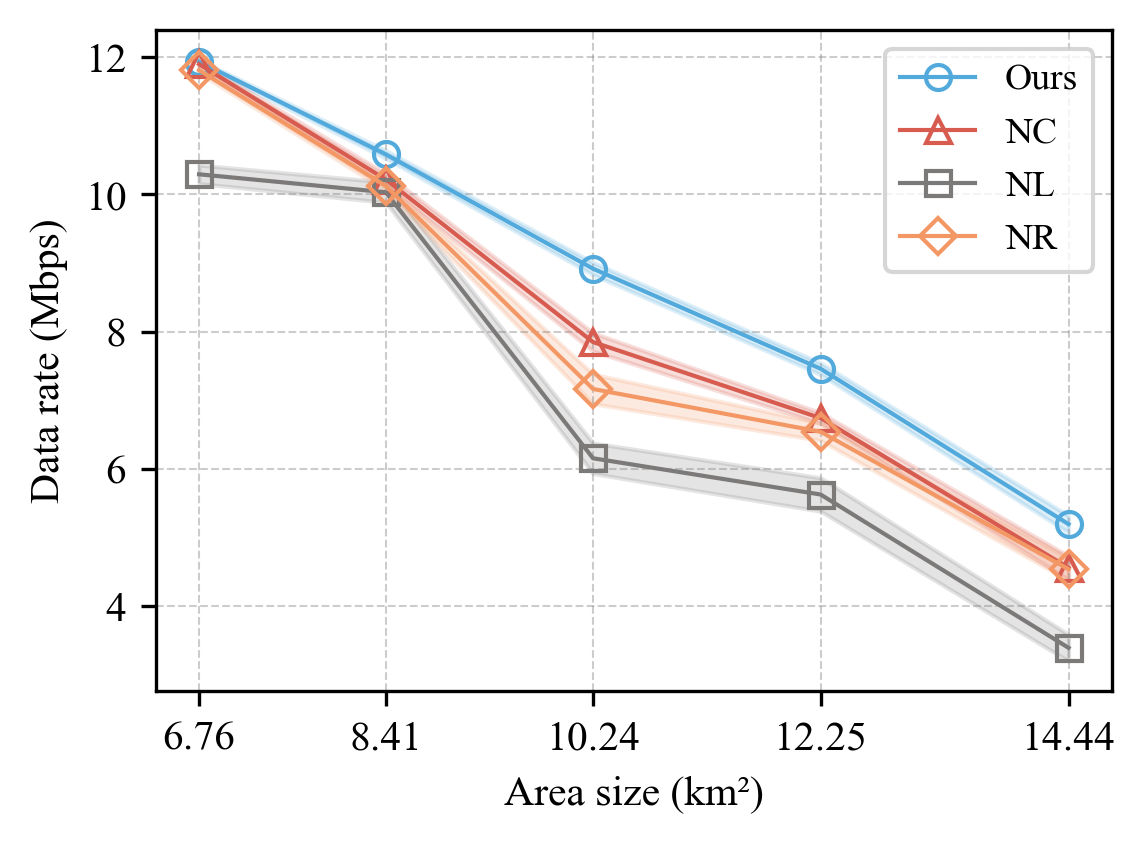}}
\subfloat[]{\includegraphics[width=.49\columnwidth]{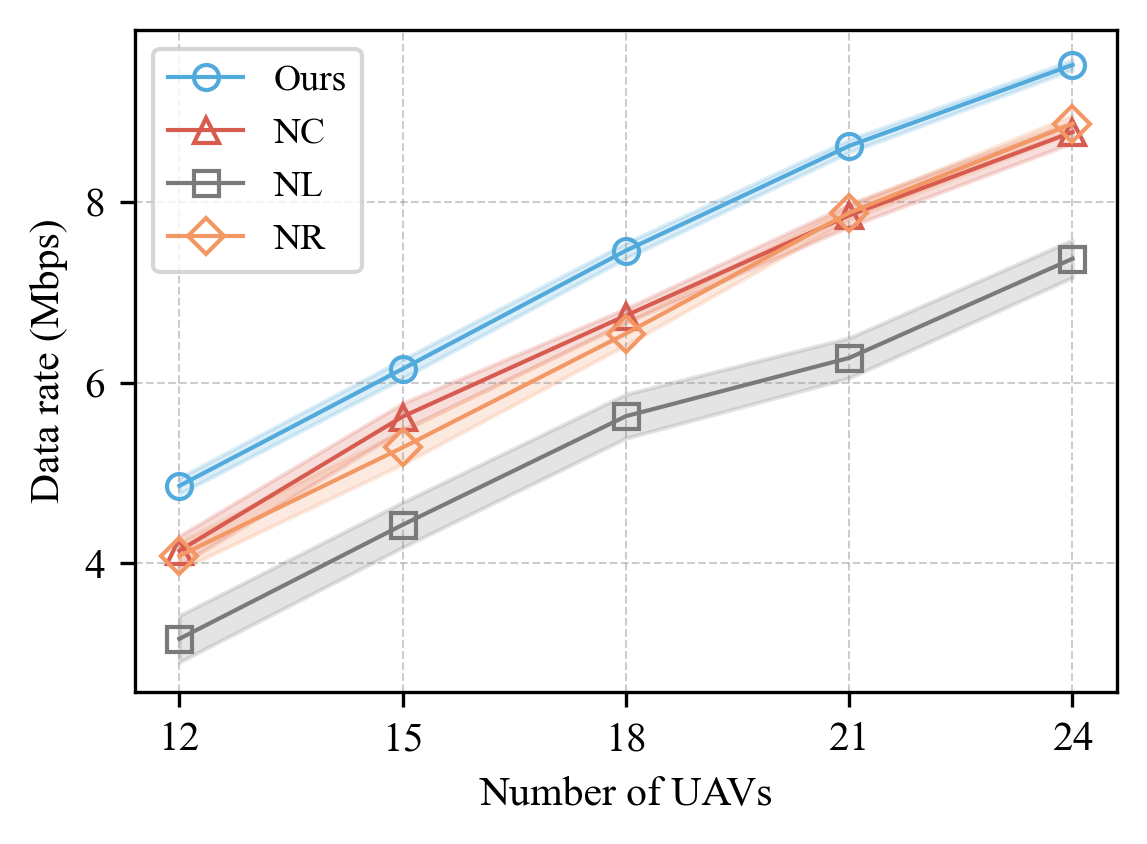}}\\
\vspace{-0.2cm}
\subfloat[]{\includegraphics[width=.49\columnwidth]{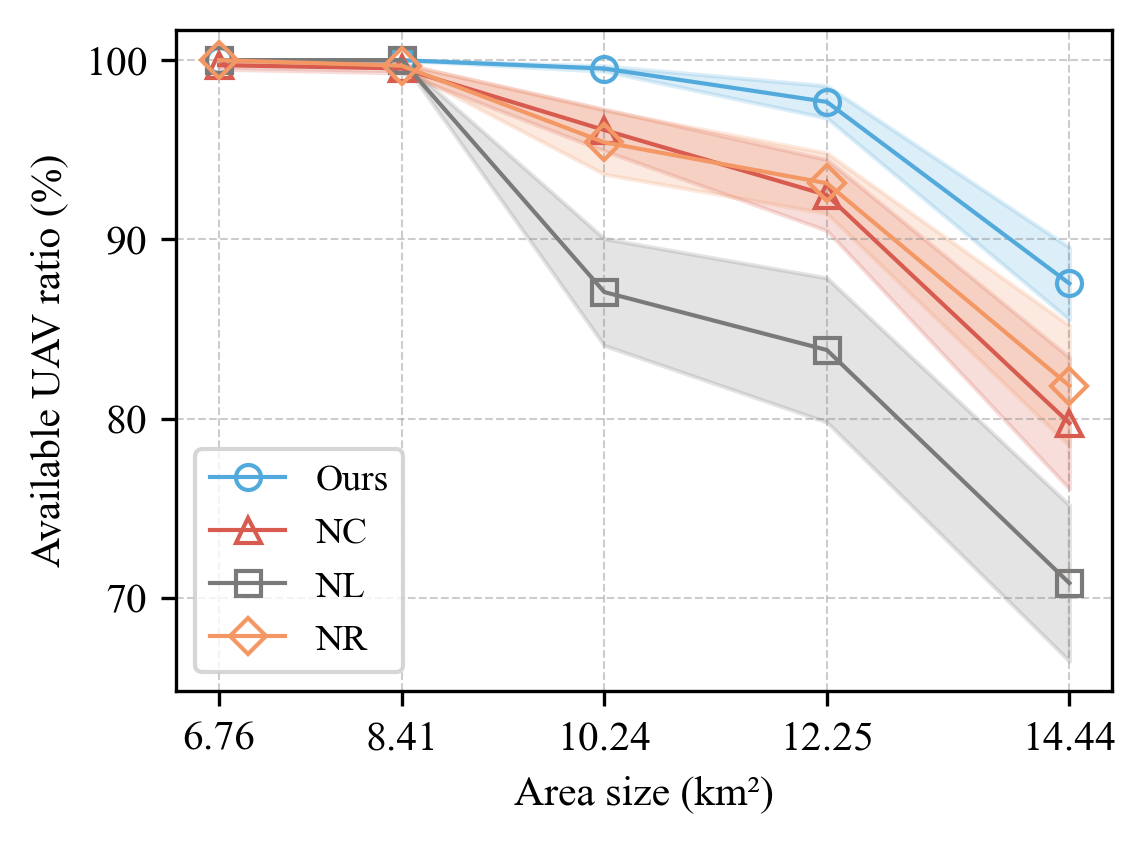}}
\subfloat[]{\includegraphics[width=.49\columnwidth]{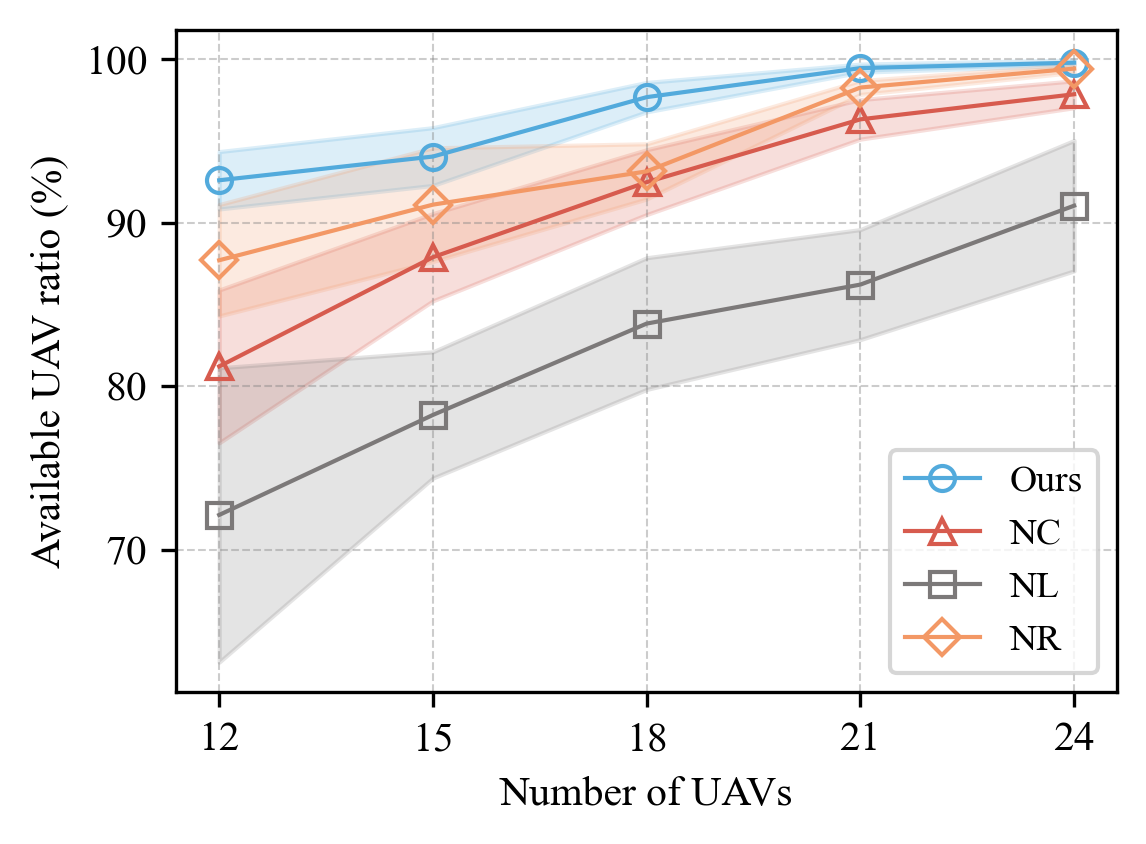}}\\
\vspace{-0.1cm}
\caption{Ablation study on model performance across different area sizes and the number of UAVs deployed. Shaded regions represent the standard deviation.}
\vspace{-0.3cm}
\label{ablation}
\end{figure}

Figures \ref{ablation}.(a), (c) and (e) illustrate the experimental results as the environmental size increases with a fixed deployment of 18 UAVs. The proposed method outperforms the methods NC, NL, and NR across all environmental configurations.
As the environment size expands to 14.44 km$^2$, the proposed method continues to deliver superior performance, achieving a connected UE proportion of 46\%, an available UAV ratio of 88\%, and a data rate of 5.2 Mbps. In contrast, the MRLMN shows a performance decline when specific modules are removed. For instance, the results of NR are notably lower, with a connected UE proportion of 40\%, an available UAV ratio of 82\%, and a data rate of 4.5 Mbps.
Moreover, Figures \ref{ablation}.(b), (d) and (f) demonstrate the performance of the four methods across key metrics as the number of UAVs increases from 12 to 24 in a fixed square environment with a 3.5 km side length.
The proposed algorithm exhibits noticeable performance degradation when any of the key modules is removed, indicating the importance of each component in addressing the challenges of multi-hop UAV networking. Specifically, removing any single module results in an average decline of at least 6\% in UE coverage and 10\% in data rate.
In terms of the proportion of UAVs connected to the BS, removing any module also leads to a noticeable performance drop, particularly in large-scale environments with a limited number of deployable UAVs.

These results indicate that the removal of certain modules negatively impacts the networking performance of the proposed algorithm.
The performance of NL suggests that the absence of the knowledge distillation module lowers MRLMN performance, emphasizing its critical role.
By constraining the exploration space, the distillation loss from LLM agents mitigates inherent exploration challenges in MARL. This facilitates faster convergence and enhances training effectiveness.
Additionally, the reduced available UAV ratio observed in NC underscores the role of behavioral constraints in stabilizing the model and preserving network connectivity.
The poorer performance of NR in terms of UE coverage and communication rate also highlights the importance of agent grouping and reward decomposition in improving the algorithm's overall performance.
The integration of both methods facilitates efficient training by providing clear and structured environmental feedback, thereby improving performance.

Collectively, these designed components establish a well-integrated framework that surpasses conventional approaches in maintaining connectivity, ensuring UAV availability, and optimizing data transmission rates.  The results indicate that the proposed method offers a highly effective solution for UAV-assisted communication systems, particularly in dynamic and large-scale networking scenarios.

\subsection{Discussion: Parameter Sharing within Agent Groups}
\label{pareto discussion}

\begin{figure}
	\centering
	\subfloat[]{\includegraphics[width=.49\columnwidth]{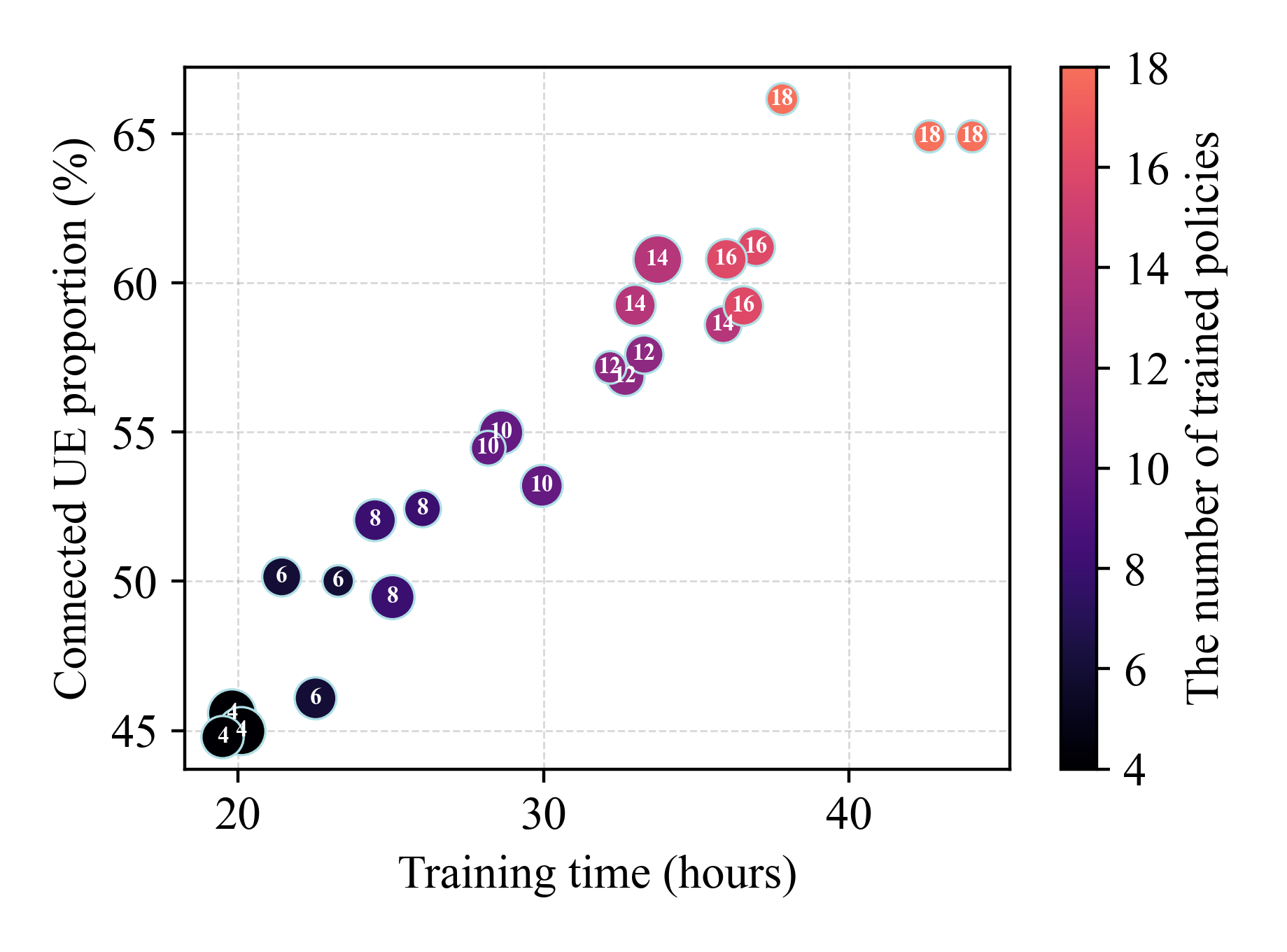}}
	\subfloat[]{\includegraphics[width=.49\columnwidth]{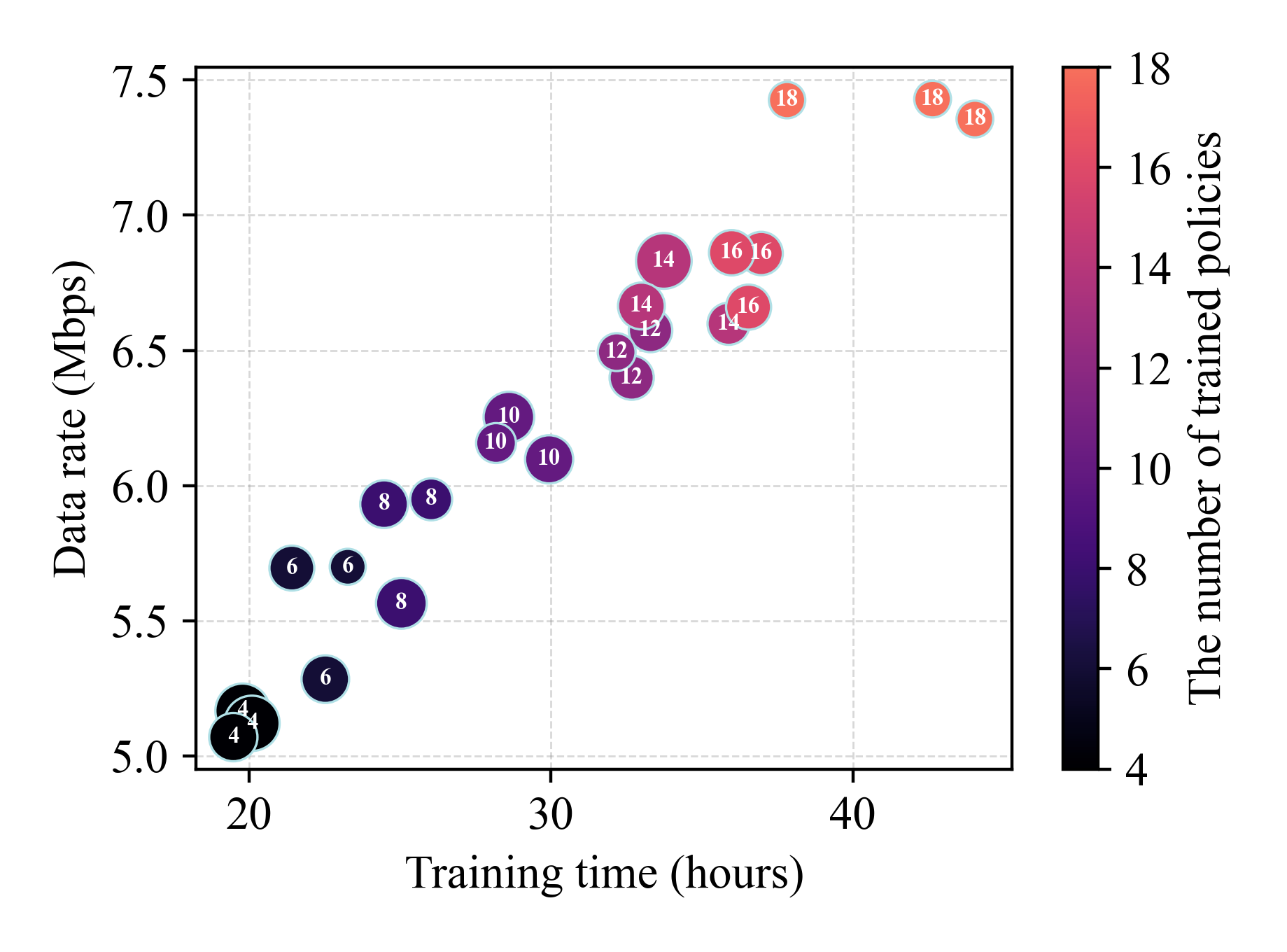}}\\
	\vspace{-0.2cm}
	\subfloat[]{\includegraphics[width=.49\columnwidth]{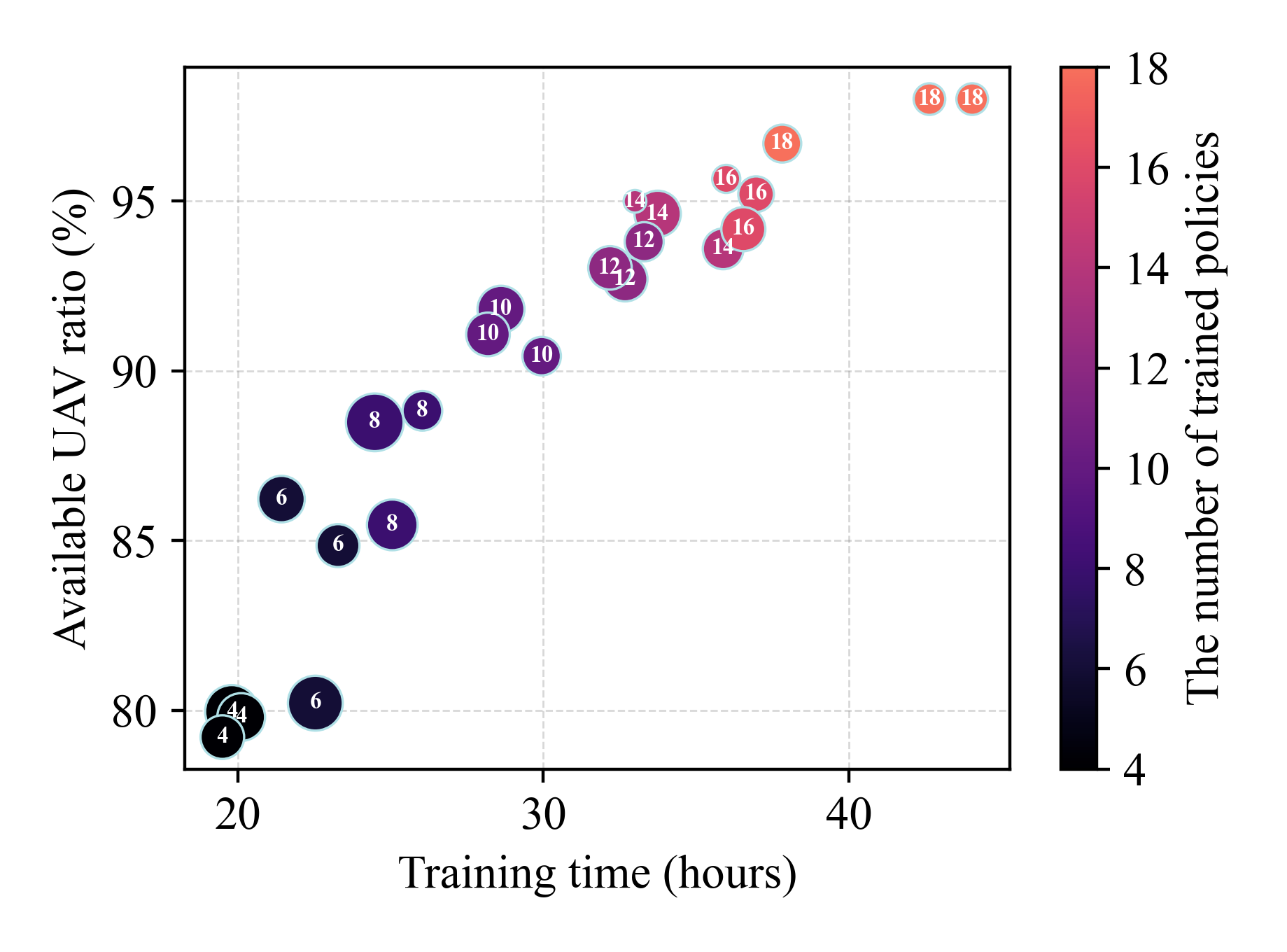}}\\
	\vspace{-0.1cm}
	\caption{Experimental analysis of the impact of policy sharing on algorithm training performance. Each dot represents an individual experimental result, with the circle size reflecting the relative magnitude of the standard deviation. }
        \vspace{-0.3cm}
	\label{pareto}
\end{figure}

In this section, the impact of incorporating intra-group strategy sharing within the agent grouping scheme on training performance and training time is discussed.
As discussed in Section \ref{Grouping and reward}, UAV agent grouping is established based on the responsibilities assigned to each UAV, which forms the foundation for parameter sharing. It is evident that agents within the same group share similar responsibilities, which makes network parameter sharing within these groups feasible.
By reducing the number of parameters that need to be individually trained, parameter sharing can significantly enhance the efficiency of the training process. This approach leads to a decrease in the total number of trainable parameters, resulting in faster convergence and reduced training time under a fixed number of training steps.
Specifically, in the experiment, a different number of agents within the UAV group are randomly selected to share their network parameters. This sharing results in varying numbers of networks being trained.

As shown in Figure \ref{pareto}, the experimental outcomes under different sharing configurations are presented. The experiments are conducted in a 3.5 km by 3.5 km environment with 18 UAVs deployed. In the experiments, agents are divided into four groups, with partial policy sharing applied within each group. As a result, the number of policies to be trained in the MARL framework gradually decreases from 18 (no sharing) to 4 (full sharing within each group).
To ensure statistical reliability, each sharing configuration is evaluated through three independent experimental runs.
The results clearly demonstrate that an increase in the number of policies leads to a longer training duration, yet yields a marked improvement in overall system performance.
As the number of trained policies increases from 4 to 18, the system exhibits consistent performance gains.  UE coverage improves from 45\% to 65\%, while the average data rate rises from 5.1 Mbps to 7.4 Mbps.  The proportion of available UAVs also grows steadily, reaching approximately 98\% with 18 policies.
However, as the number of networks rises from 4 to 18, the increasing volume of trainable network parameters leads to a longer training duration, extending from 20 hours to 40 hours.

Overall, the experimental results reveal a consistent trend across the three metrics: as the number of training strategies increases, both the training duration and performance improve significantly. This indicates that policy sharing within groups leads to a reduction in the diversity of UAV strategies. In scenarios involving large exploration spaces and dynamic UAV roles, maintaining strategy diversity proves crucial for enhancing the networking performance of the algorithm.
It is clear that a balance between training efficiency and algorithm performance needs to be considered during the learning process.

\subsection{Simulation Results}
\label{Simulation Results}

\begin{figure}
    \subfloat[t=1]{\includegraphics[width=.4\columnwidth]{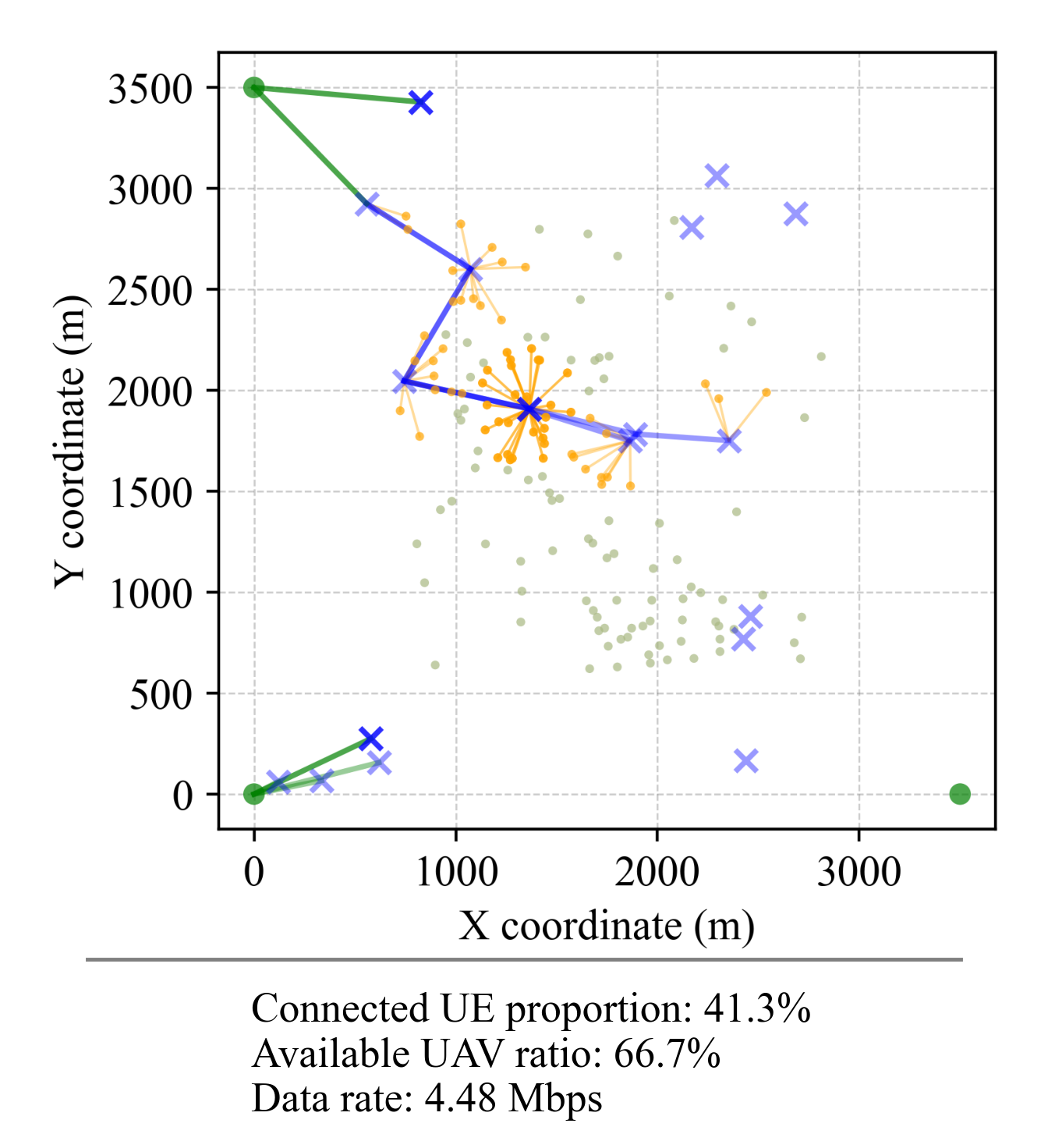}}
    \subfloat[t=100]{\includegraphics[width=.4\columnwidth]{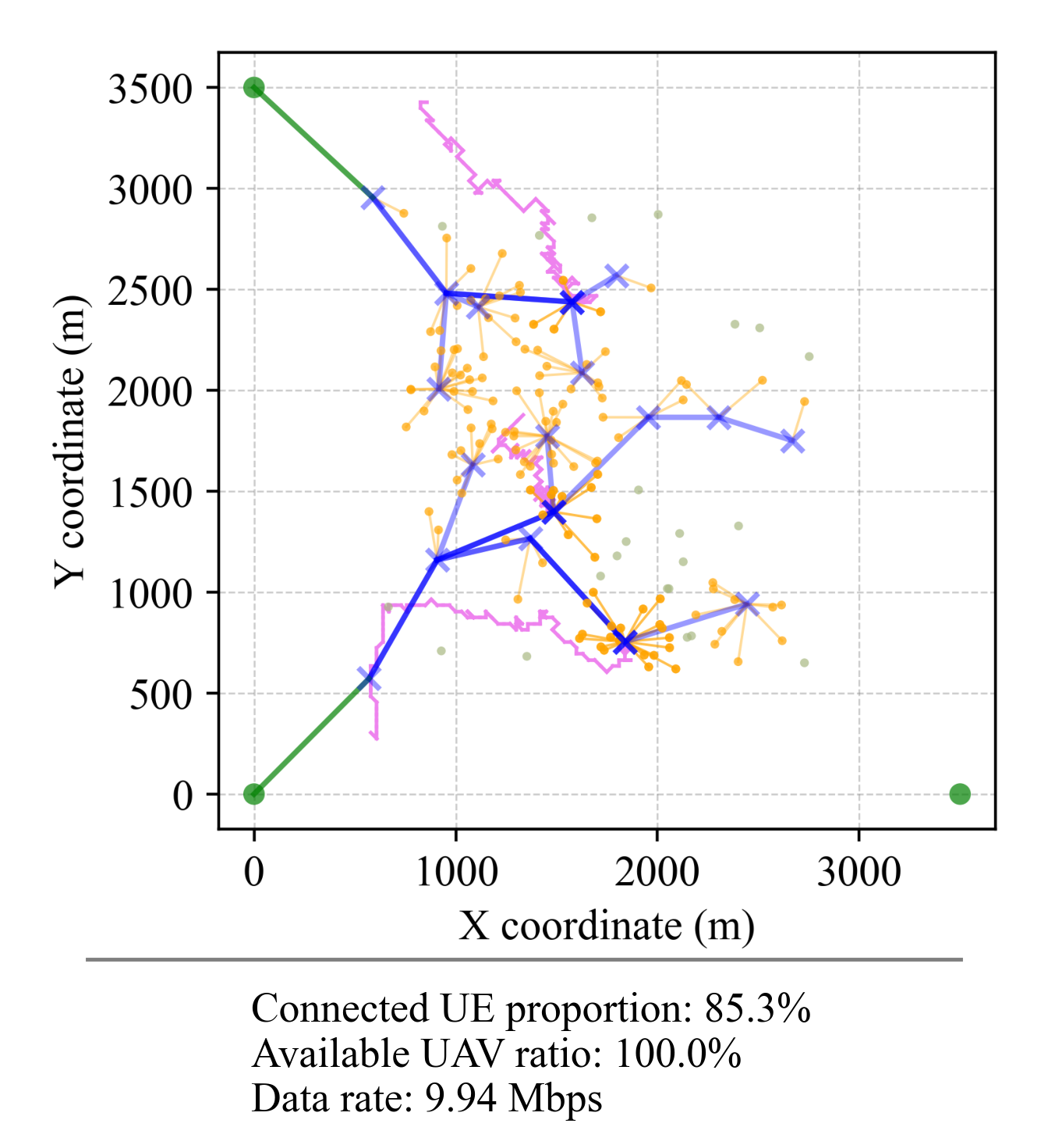}}\\
    \vspace{-0.1cm}
    \subfloat[t=200]{\includegraphics[width=.4\columnwidth]{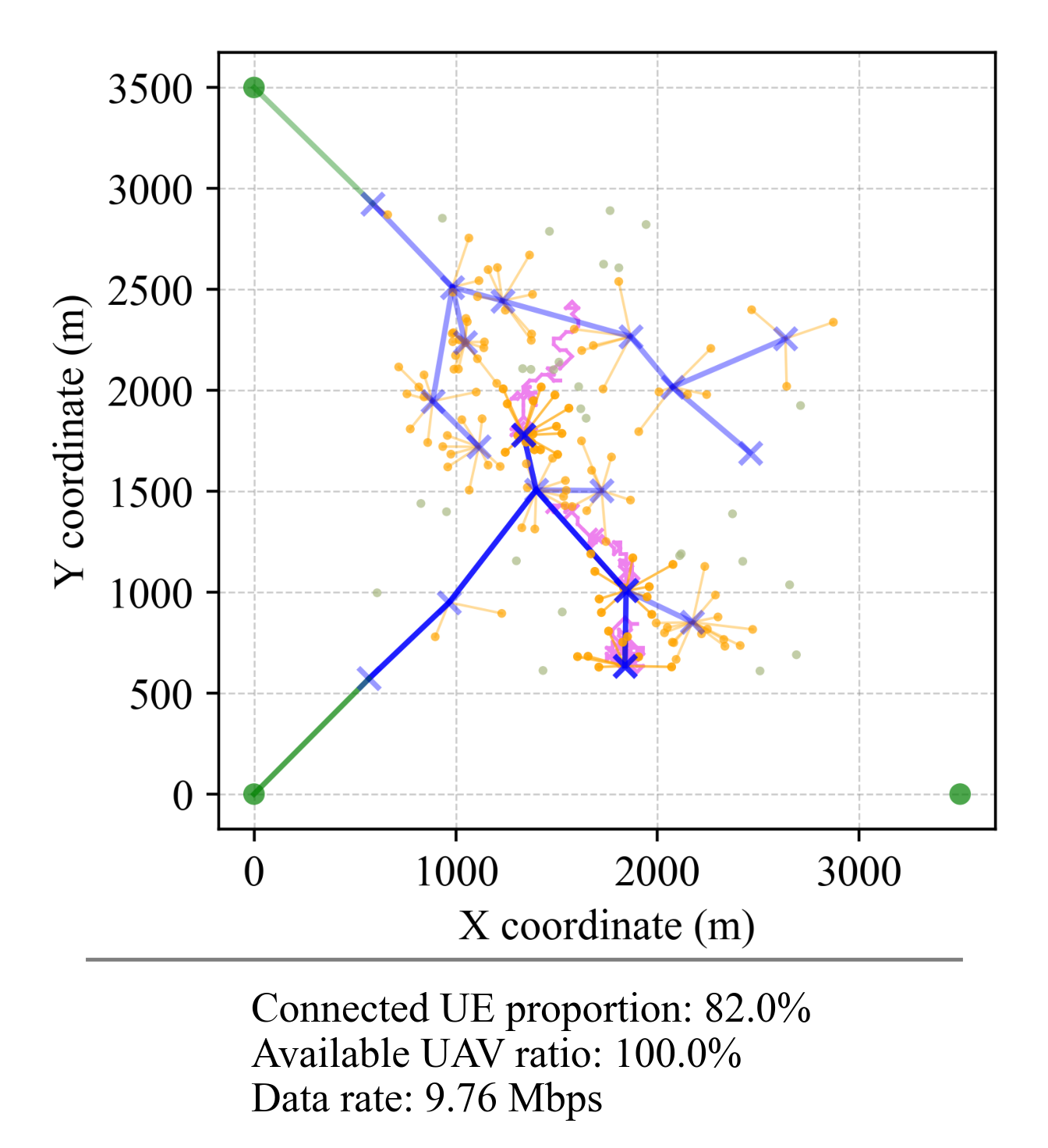}}
    \subfloat[t=400]{\includegraphics[width=.4\columnwidth]{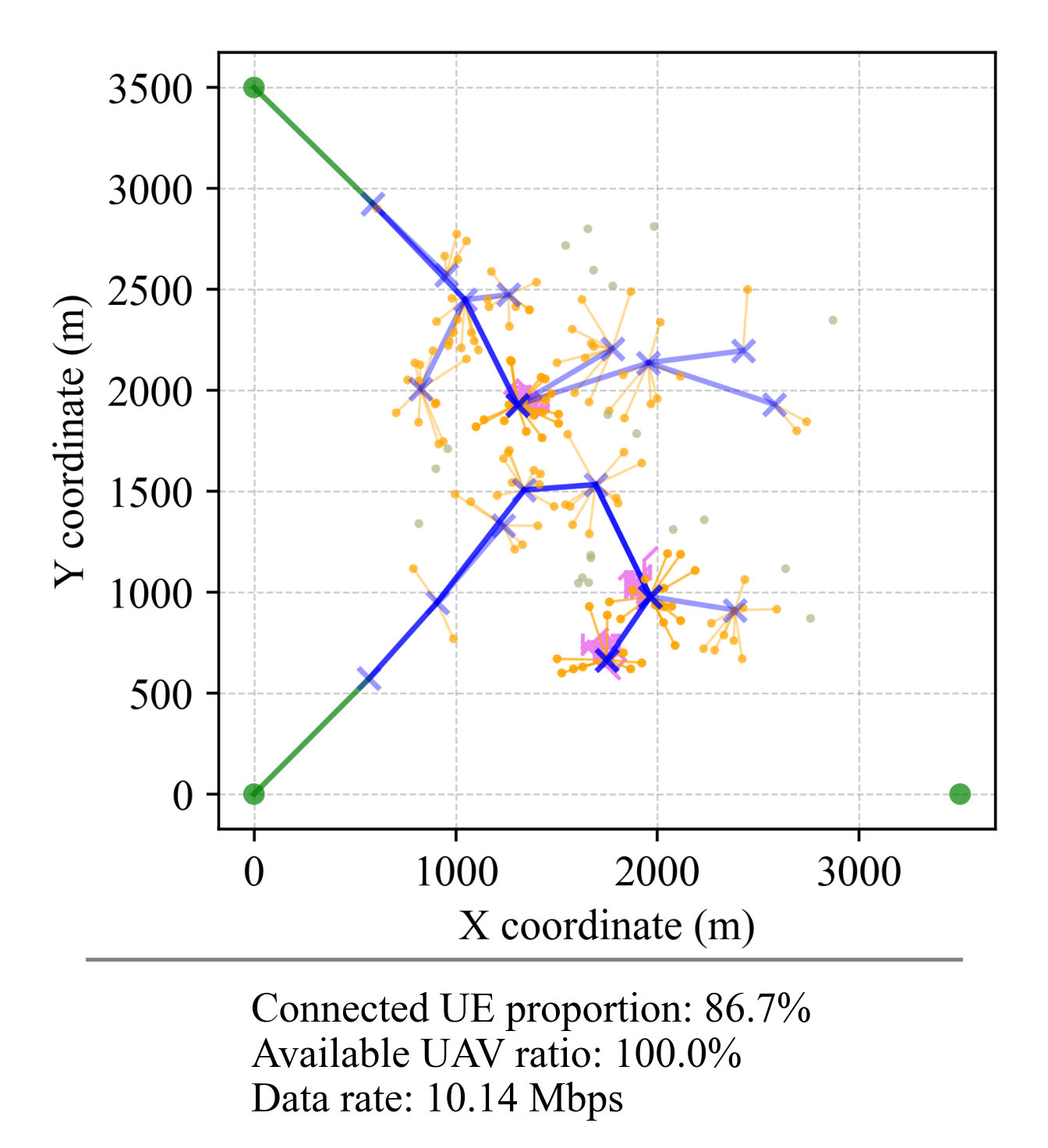}}
    \raisebox{0.4cm}{
    \subfloat{\includegraphics[width=.18\columnwidth]{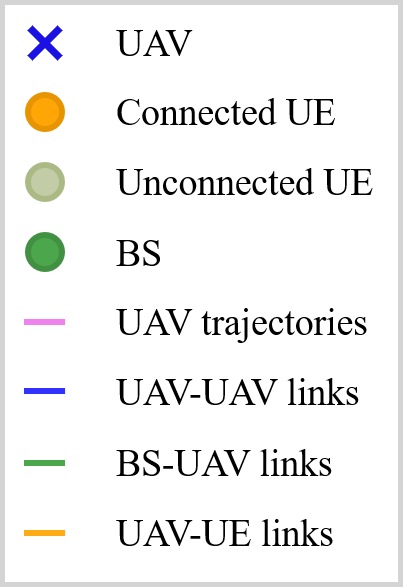}}}\\
\caption{Simulation results of the proposed algorithm within a single episode. Each figure shows the states of the nodes and the UAV network topology, with the evaluation metrics at the current time step displayed below. Three representative UAVs are highlighted using a darker shade, and their trajectories over 100 time steps are depicted by pink paths.}
\vspace{-0.308cm}
\label{traj}
\end{figure}

This section presents the experimental results of the proposed MRLMN algorithm within a simulated environment. In Figure \ref{traj}, a total of 18 UAVs are deployed.
Figure \ref{traj}.(a) illustrates the initial configuration at the start of the simulation. As the simulation progresses, UAV trajectories are dynamically planned by the MARL-trained policies, while UEs move randomly within the environment. Gradually, a multi-hop UAV network emerges and evolves, exhibiting enhanced connectivity over time. By the final stage, shown in Figure \ref{traj}.(d), the network establishes robust communication links with the majority of UEs. These results demonstrate the MRLMN algorithm's strong capability for multi-hop networking, ensuring both high connectivity and stability in dynamic UAV networking scenarios.

\section{Conclusion}
\label{Conclusion}
This paper introduces the MRLMN framework, which integrates MARL and LLMs to optimize UAV networking in disaster response scenarios.  To address the scalability of the networking problem and enhance coordination among UAVs, the framework incorporates agent grouping and reward decomposition modules.  Behavioral constraints based on the grouping mechanism are further applied to ensure robust and stable network formation.  Additionally, a knowledge distillation approach enables the transfer of high-level decision-making capabilities from LLMs to MARL agents, accelerating training and improving exploration efficiency.  Simulation results demonstrate substantial performance gains in large-scale dynamic environments, confirming the adaptability and effectiveness of the proposed framework across diverse configurations.
Future research directions include incorporating practical constraints such as UAV energy consumption, network load balancing, and UAV replacement mechanisms to ensure continuous and reliable operations in large-scale deployments. Enhancing communication reliability through interference management and physical-layer optimization can further support the efficient implementation of multi-hop aerial networks. At the same time, deeper integration of LLMs with MARL, for example, by replacing conventional MARL policies with more expressive decision models, has the potential to improve scalability and adaptability in complex environments. Complementing these developments with validation in real-world deployments would provide valuable insights into the operational feasibility and robustness of the proposed framework.

\bibliographystyle{IEEEtran}
\bibliography{ref}

@ARTICLE{11ML1,
  author={Wu, Qingqing and Zeng, Yong and Zhang, Rui},
  journal={IEEE Transactions on Wireless Communications}, 
  title={Joint Trajectory and Communication Design for Multi-UAV Enabled Wireless Networks}, 
  year={2018},
  volume={17},
  number={3},
  pages={2109-2121},
  doi={10.1109/TWC.2017.2789293}}

@ARTICLE{12ML2,
  author={Fazel, Fahimeh and Abouei, Jamshid and Jaseemuddin, Muhammad and Anpalagan, Alagan and Plataniotis, Konstantinos N.},
  journal={IEEE Internet of Things Journal}, 
  title={Secure Throughput Optimization for Cache-Enabled Multi-UAVs Networks}, 
  year={2022},
  volume={9},
  number={10},
  pages={7783-7801},
  doi={10.1109/JIOT.2021.3114086}}

@ARTICLE{7ML3,
  author={Tran, Dinh-Hieu and Vu, Thang X. and Chatzinotas, Symeon and ShahbazPanahi, Shahram and Ottersten, Björn},
  journal={IEEE Transactions on Vehicular Technology}, 
  title={Coarse Trajectory Design for Energy Minimization in UAV-Enabled}, 
  year={2020},
  volume={69},
  number={9},
  pages={9483-9496},
  doi={10.1109/TVT.2020.3001403}}

@ARTICLE{15ML7,
  author={Yao, Jingjing and Ansari, Nirwan},
  journal={IEEE Transactions on Vehicular Technology}, 
  title={QoS-Aware Power Control in Internet of Drones for Data Collection Service}, 
  year={2019},
  volume={68},
  number={7},
  pages={6649-6656},
  doi={10.1109/TVT.2019.2915270}}

@ARTICLE{18ML9,
  author={Samir, Moataz and Sharafeddine, Sanaa and Assi, Chadi M. and Nguyen, Tri Minh and Ghrayeb, Ali},
  journal={IEEE Transactions on Wireless Communications}, 
  title={UAV Trajectory Planning for Data Collection from Time-Constrained IoT Devices}, 
  year={2020},
  volume={19},
  number={1},
  pages={34-46},
  doi={10.1109/TWC.2019.2940447}}

@INPROCEEDINGS{13ML11,
  author={Wang, Chen and Zhai, Daosen and Zhang, Ruonan and Kaddoum, Georges and Singh, Satinder},
  booktitle={ICC 2023 - IEEE International Conference on Communications}, 
  title={Energy Consumption Minimization in Dynamic UAV-assisted Mobile Edge Computing Networks}, 
  year={2023},
  volume={},
  number={},
  pages={4671-4676},
  doi={10.1109/ICC45041.2023.10279245}}

@ARTICLE{20RL2,
  author={Wang, Liang and Wang, Kezhi and Pan, Cunhua and Xu, Wei and Aslam, Nauman and Hanzo, Lajos},
  journal={IEEE Transactions on Cognitive Communications and Networking}, 
  title={Multi-Agent Deep Reinforcement Learning-Based Trajectory Planning for Multi-UAV Assisted Mobile Edge Computing}, 
  year={2021},
  volume={7},
  number={1},
  pages={73-84},
  doi={10.1109/TCCN.2020.3027695}}

@article{5RLintro1,
author = {Vinyals, Oriol and Babuschkin, Igor and Czarnecki, Wojciech and Mathieu, Michaël and Dudzik, Andrew and Chung, Junyoung and Choi, David and Powell, Richard and Ewalds, Timo and Georgiev, Petko and Oh, Junhyuk and Horgan, Dan and Kroiss, Manuel and Danihelka, Ivo and Huang, Aja and Sifre, Laurent and Cai, Trevor and Agapiou, John and Jaderberg, Max and Silver, David},
year = {2019},
month = {11},
pages = {},
title = {Grandmaster level in StarCraft II using multi-agent reinforcement learning},
volume = {575},
journal = {Nature},
doi = {10.1038/s41586-019-1724-z}
}

@ARTICLE{21RL3,
  author={Zhang, Wenqi and Wang, Qiang and Liu, Xiao and Liu, Yuanwei and Chen, Yue},
  journal={IEEE Transactions on Vehicular Technology}, 
  title={Three-Dimension Trajectory Design for Multi-UAV Wireless Network With Deep Reinforcement Learning}, 
  year={2021},
  volume={70},
  number={1},
  pages={600-612},
  doi={10.1109/TVT.2020.3047800}}

@INPROCEEDINGS{22RL4,
  author={Catté, Esteban and Sana, Mohamed and Maman, Mickael},
  booktitle={ICC 2023 - IEEE International Conference on Communications}, 
  title={Dual-Attention Deep Reinforcement Learning for Multi-MAP 3D Trajectory Optimization in Dynamic 5G Networks}, 
  year={2023},
  volume={},
  number={},
  pages={6417-6422},
  doi={10.1109/ICC45041.2023.10278657}}

@ARTICLE{31RL6,
  author={Zhang, Xiaochen and Zhao, Haitao and Wei, Jibo and Yan, Chao and Xiong, Jun and Liu, Xiaoran},
  journal={IEEE Transactions on Wireless Communications}, 
  title={Cooperative Trajectory Design of Multiple UAV Base Stations With Heterogeneous Graph Neural Networks}, 
  year={2023},
  volume={22},
  number={3},
  pages={1495-1509},
  doi={10.1109/TWC.2022.3204794}}

@ARTICLE{29RL8,
  author={He, Yejun and Gan, Youhui and Cui, Haixia and Guizani, Mohsen},
  journal={IEEE Internet of Things Journal}, 
  title={Fairness-Based 3-D Multi-UAV Trajectory Optimization in Multi-UAV-Assisted MEC System}, 
  year={2023},
  volume={10},
  number={13},
  pages={11383-11395},
  doi={10.1109/JIOT.2023.3241087}}

@ARTICLE{30RL12,
  author={Qin, Yunhui and Zhang, Zhongshan and Li, Xulong and Huangfu, Wei and Zhang, Haijun},
  journal={IEEE Transactions on Wireless Communications}, 
  title={Deep Reinforcement Learning Based Resource Allocation and Trajectory Planning in Integrated Sensing and Communications UAV Network}, 
  year={2023},
  volume={22},
  number={11},
  pages={8158-8169},
  doi={10.1109/TWC.2023.3260304}}

@ARTICLE{26RL13,
  author={Song, Fuhong and Xing, Huanlai and Wang, Xinhan and Luo, Shouxi and Dai, Penglin and Xiao, Zhiwen and Zhao, Bowen},
  journal={IEEE Transactions on Mobile Computing}, 
  title={Evolutionary Multi-Objective Reinforcement Learning Based Trajectory Control and Task Offloading in UAV-Assisted Mobile Edge Computing}, 
  year={2023},
  volume={22},
  number={12},
  pages={7387-7405},
  doi={10.1109/TMC.2022.3208457}}

@InProceedings{27MARL1,
  title = 	 {Scaling Multi-Agent Reinforcement Learning with Selective Parameter Sharing},
  author =       {Christianos, Filippos and Papoudakis, Georgios and Rahman, Muhammad A and Albrecht, Stefano V},
  booktitle = 	 {Proceedings of the 38th International Conference on Machine Learning},
  pages = 	 {1989--1998},
  year = 	 {2021},
  editor = 	 {Meila, Marina and Zhang, Tong},
  volume = 	 {139},
  series = 	 {Proceedings of Machine Learning Research},
  month = 	 {18--24 Jul},
  publisher =    {PMLR},
  url = 	 {https://proceedings.mlr.press/v139/christianos21a.html}
}

@inproceedings{28MARL2,
  title={The Surprising Effectiveness of PPO in Cooperative Multi-Agent Games},
  author={Chao Yu and Akash Velu and Eugene Vinitsky and Yu Wang and Alexandre M. Bayen and Yi Wu},
  booktitle={Neural Information Processing Systems},
  year={2021},
  url={https://api.semanticscholar.org/CorpusID:232092445}
}

@inproceedings{32GNN1,
author = {Ren, Jiyuan and Xu, Yanggang and Li, Zuxin and Hong, Chaopeng and Zhang, Xiao-Ping and Chen, Xinlei},
title = {Scheduling UAV Swarm with Attention-Based Graph Reinforcement Learning for Ground-to-Air Heterogeneous Data Communication},
year = {2023},
isbn = {9798400702006},
publisher = {Association for Computing Machinery},
address = {New York, NY, USA},
url = {https://doi.org/10.1145/3594739.3612905},
doi = {10.1145/3594739.3612905},
booktitle = {Adjunct Proceedings of the 2023 ACM International Joint Conference on Pervasive and Ubiquitous Computing \& the 2023 ACM International Symposium on Wearable Computing},
pages = {670–675},
numpages = {6},
location = {<conf-loc>, <city>Cancun, Quintana Roo</city>, <country>Mexico</country>, </conf-loc>},
series = {UbiComp/ISWC '23 Adjunct}
}

@ARTICLE{1intro2,
  author={Gupta, Lav and Jain, Raj and Vaszkun, Gabor},
  journal={IEEE Communications Surveys \& Tutorials}, 
  title={Survey of Important Issues in UAV Communication Networks}, 
  year={2016},
  volume={18},
  number={2},
  pages={1123-1152},
  doi={10.1109/COMST.2015.2495297}}

@ARTICLE{3LoS,
  author={Li, Bin and Fei, Zesong and Zhang, Yan},
  journal={IEEE Internet of Things Journal}, 
  title={UAV Communications for 5G and Beyond: Recent Advances and Future Trends}, 
  year={2019},
  volume={6},
  number={2},
  pages={2241-2263},
  doi={10.1109/JIOT.2018.2887086}}

@article{PPO,
  author       = {John Schulman and
                  Filip Wolski and
                  Prafulla Dhariwal and
                  Alec Radford and
                  Oleg Klimov},
  title        = {Proximal Policy Optimization Algorithms},
  journal      = {CoRR},
  volume       = {abs/1707.06347},
  year         = {2017},
  url          = {http://arxiv.org/abs/1707.06347},
  eprinttype    = {arXiv},
  eprint       = {1707.06347},
  bibsource    = {dblp computer science bibliography, https://dblp.org}
}

@article{IPPO,
  author       = {Christian Schr{\"{o}}der de Witt and
                  Tarun Gupta and
                  Denys Makoviichuk and
                  Viktor Makoviychuk and
                  Philip H. S. Torr and
                  Mingfei Sun and
                  Shimon Whiteson},
  title        = {Is Independent Learning All You Need in the StarCraft Multi-Agent
                  Challenge?},
  journal      = {CoRR},
  volume       = {abs/2011.09533},
  year         = {2020},
  url          = {https://arxiv.org/abs/2011.09533},
  eprinttype    = {arXiv},
  eprint       = {2011.09533},
  bibsource    = {dblp computer science bibliography, https://dblp.org}
}

@article{voyager,
  title   = {Voyager: An Open-Ended Embodied Agent with Large Language Models},
  author  = {Guanzhi Wang and Yuqi Xie and Yunfan Jiang and Ajay Mandlekar and Chaowei Xiao and Yuke Zhu and Linxi Fan and Anima Anandkumar},
  year    = {2023},
  journal = {arXiv preprint arXiv: Arxiv-2305.16291}
}

@InProceedings{Guiding,
  title = 	 {Guiding Pretraining in Reinforcement Learning with Large Language Models},
  author =       {Du, Yuqing and Watkins, Olivia and Wang, Zihan and Colas, C\'{e}dric and Darrell, Trevor and Abbeel, Pieter and Gupta, Abhishek and Andreas, Jacob},
  booktitle = 	 {Proceedings of the 40th International Conference on Machine Learning},
  pages = 	 {8657--8677},
  year = 	 {2023},
  editor = 	 {Krause, Andreas and Brunskill, Emma and Cho, Kyunghyun and Engelhardt, Barbara and Sabato, Sivan and Scarlett, Jonathan},
  volume = 	 {202},
  series = 	 {Proceedings of Machine Learning Research},
  month = 	 {23--29 Jul},
  publisher =    {PMLR},
  url = 	 {https://proceedings.mlr.press/v202/du23f.html}
}

@misc{react,
      title={ReAct: Synergizing Reasoning and Acting in Language Models}, 
      author={Shunyu Yao and Jeffrey Zhao and Dian Yu and Nan Du and Izhak Shafran and Karthik Narasimhan and Yuan Cao},
      year={2023},
      eprint={2210.03629},
      archivePrefix={arXiv},
      primaryClass={cs.CL}
}

@article{CoT,
  author       = {Jason Wei and
                  Xuezhi Wang and
                  Dale Schuurmans and
                  Maarten Bosma and
                  Ed H. Chi and
                  Quoc Le and
                  Denny Zhou},
  title        = {Chain of Thought Prompting Elicits Reasoning in Large Language Models},
  journal      = {CoRR},
  volume       = {abs/2201.11903},
  year         = {2022},
  url          = {https://arxiv.org/abs/2201.11903},
  eprinttype    = {arXiv},
  eprint       = {2201.11903},
  bibsource    = {dblp computer science bibliography, https://dblp.org}
}

@article{DETR,
  author       = {Nicolas Carion and
                  Francisco Massa and
                  Gabriel Synnaeve and
                  Nicolas Usunier and
                  Alexander Kirillov and
                  Sergey Zagoruyko},
  title        = {End-to-End Object Detection with Transformers},
  journal      = {CoRR},
  volume       = {abs/2005.12872},
  year         = {2020},
  url          = {https://arxiv.org/abs/2005.12872},
  eprinttype    = {arXiv},
  eprint       = {2005.12872},
  bibsource    = {dblp computer science bibliography, https://dblp.org}
}

@inproceedings{KD,title	= {Distilling the Knowledge in a Neural Network},author	= {Geoffrey Hinton and Oriol Vinyals and Jeffrey Dean},year	= {2015},URL	= {http://arxiv.org/abs/1503.02531},booktitle	= {NIPS Deep Learning and Representation Learning Workshop}}

@ARTICLE{intro_research1,
  author={Dai, Chen and Zhu, Kun and Hossain, Ekram},
  journal={IEEE Transactions on Mobile Computing}, 
  title={Multi-Agent Deep Reinforcement Learning for Joint Decoupled User Association and Trajectory Design in Full-Duplex Multi-UAV Networks}, 
  year={2023},
  volume={22},
  number={10},
  pages={6056-6070},
  doi={10.1109/TMC.2022.3188473}}

@ARTICLE{intro_research2,
  author={Khairy, Sami and Balaprakash, Prasanna and Cai, Lin X. and Cheng, Yu},
  journal={IEEE Journal on Selected Areas in Communications}, 
  title={Constrained Deep Reinforcement Learning for Energy Sustainable Multi-UAV Based Random Access IoT Networks With NOMA}, 
  year={2021},
  volume={39},
  number={4},
  pages={1101-1115},
  doi={10.1109/JSAC.2020.3018804}}

@ARTICLE{intro_research3,
  author={Cheng, Sike and Lin, Xiangbo and Li, Xuanheng and Wang, Jingjing},
  journal={IEEE Transactions on Wireless Communications}, 
  title={Joint UAV Trajectory and RadCom Task Schedule for IVNs: A Game-Embedding Multi-Agent Deep Reinforcement Learning Approach}, 
  year={2025},
  volume={24},
  number={1},
  pages={181-196},
  doi={10.1109/TWC.2024.3489624}}

@ARTICLE{intro_research4,
  author={Liu, Chi Harold and Chen, Zheyu and Zhan, Yufeng},
  journal={IEEE Journal on Selected Areas in Communications}, 
  title={Energy-Efficient Distributed Mobile Crowd Sensing: A Deep Learning Approach}, 
  year={2019},
  volume={37},
  number={6},
  pages={1262-1276},
  doi={10.1109/JSAC.2019.2904353}}

@ARTICLE{intro_research7,
  author={Ji, Jiequ and Zhu, Kun and Cai, Lin},
  journal={IEEE Transactions on Mobile Computing}, 
  title={Trajectory and Communication Design for Cache- Enabled UAVs in Cellular Networks: A Deep Reinforcement Learning Approach}, 
  year={2023},
  volume={22},
  number={10},
  pages={6190-6204},
  doi={10.1109/TMC.2022.3181308}}

@INPROCEEDINGS{llm_uav,
  author={Xiang, Xiancai and Xue, Jian and Zhao, Lin and Lei, Yuan and Yue, Chao and Lu, Ke},
  booktitle={2024 International Joint Conference on Neural Networks (IJCNN)}, 
  title={Real-time Integration of Fine-tuned Large Language Model for Improved Decision-Making in Reinforcement Learning}, 
  year={2024},
  volume={},
  number={},
  pages={1-8},
  doi={10.1109/IJCNN60899.2024.10650538}}

@InProceedings{llm_state,
  title = 	 {{LLM}-Empowered State Representation for Reinforcement Learning},
  author =       {Wang, Boyuan and Qu, Yun and Jiang, Yuhang and Shao, Jianzhun and Liu, Chang and Yang, Wenming and Ji, Xiangyang},
  booktitle = 	 {Proceedings of the 41st International Conference on Machine Learning},
  pages = 	 {51348--51375},
  year = 	 {2024},
  editor = 	 {Salakhutdinov, Ruslan and Kolter, Zico and Heller, Katherine and Weller, Adrian and Oliver, Nuria and Scarlett, Jonathan and Berkenkamp, Felix},
  volume = 	 {235},
  series = 	 {Proceedings of Machine Learning Research},
  month = 	 {21--27 Jul},
  publisher =    {PMLR},
  url = 	 {https://proceedings.mlr.press/v235/wang24bh.html}
}

@misc{llm_action,
      title={SRLM: Human-in-Loop Interactive Social Robot Navigation with Large Language Model and Deep Reinforcement Learning}, 
      author={Weizheng Wang and Ike Obi and Byung-Cheol Min},
      year={2024},
      eprint={2403.15648},
      archivePrefix={arXiv},
      primaryClass={cs.RO},
      url={https://arxiv.org/abs/2403.15648}, 
}

@ARTICLE{compare2,
  author={Li, Kai and Ni, Wei and Yuan, Xin and Noor, Alam and Jamalipour, Abbas},
  journal={IEEE Internet of Things Journal}, 
  title={Deep-Graph-Based Reinforcement Learning for Joint Cruise Control and Task Offloading for Aerial Edge Internet of Things (EdgeIoT)}, 
  year={2022},
  volume={9},
  number={21},
  pages={21676-21686},
  doi={10.1109/JIOT.2022.3182119}}

@inproceedings{p1,
	author = {Xu, Yanggang and Zha, Jirong and Ren, Jiyuan and Jiang, Xintao and Zhang, Hongfei and Chen, Xinlei},
	title = {Scalable Multi-Agent Reinforcement Learning for Effective UAV Scheduling in Multi-Hop Emergency Networks},
	year = {2024},
	isbn = {9798400704895},
	publisher = {Association for Computing Machinery},
	address = {New York, NY, USA},
	url = {https://doi.org/10.1145/3636534.3694730},
	doi = {10.1145/3636534.3694730},
	booktitle = {Proceedings of the 30th Annual International Conference on Mobile Computing and Networking},
	pages = {2028–2033},
	numpages = {6},
	location = {Washington D.C., DC, USA},
	series = {ACM MobiCom '24}
}

@online{NOAA2025,
  author       = {{NOAA National Centers for Environmental Information (NCEI)}},
  title        = {U.S. Billion-Dollar Weather and Climate Disasters},
  year         = {2025},
  url          = {https://www.ncei.noaa.gov/access/billions/},
  doi          = {10.25921/stkw-7w73}
}

@inproceedings{zhou2024large,
  title={Large Language Model as a Policy Teacher for Training Reinforcement Learning Agents},
  author={Zhou, Zihao and Hu, Bin and Zhao, Chenyang and Zhang, Pu and Liu, Bin},
  booktitle={The 33rd International Joint Conference on Artificial Intelligence (IJCAI 2024)},
  year={2024}
}

@InProceedings{pmlr-v235-sun24e,
  title = 	 {{DFA}-{RAG}: Conversational Semantic Router for Large Language Model with Definite Finite Automaton},
  author =       {Sun, Yiyou and Hu, Junjie and Cheng, Wei and Chen, Haifeng},
  booktitle = 	 {Proceedings of the 41st International Conference on Machine Learning},
  pages = 	 {47033--47055},
  year = 	 {2024},
  editor = 	 {Salakhutdinov, Ruslan and Kolter, Zico and Heller, Katherine and Weller, Adrian and Oliver, Nuria and Scarlett, Jonathan and Berkenkamp, Felix},
  volume = 	 {235},
  series = 	 {Proceedings of Machine Learning Research},
  month = 	 {21--27 Jul},
  publisher =    {PMLR},
  url = 	 {https://proceedings.mlr.press/v235/sun24e.html}
}

@inproceedings{10.5555/3692070.3693945,
author = {Song, Weixi and Li, Zuchao and Zhang, Lefei and Zhao, Hai and Du, Bo},
title = {Sparse is enough in fine-tuning pre-trained large language models},
year = {2024},
publisher = {JMLR.org},
booktitle = {Proceedings of the 41st International Conference on Machine Learning},
articleno = {1875},
numpages = {15},
location = {Vienna, Austria},
series = {ICML'24}
}

@misc{openai_gpt4o,
  title        = {GPT-4o System Card},
  author       = {Openai (2024)},
  year         = {2024},
  howpublished = {\url{https://openai.com/research/gpt-4o-system-card}},
  note         = {Accessed: 2025-12-06}
}

@article{zhao2025flight,
  title={Flight Dynamics to Sensing Modalities: Exploiting Drone Ground Effect for Accurate Edge Detection},
  author={Zhao, Chenyu and Xu, Jingao and Ruan, Ciyu and Wang, Haoyang and Wang, Shengbo and Li, Jiaqi and Zha, Jirong and Hong, Weijie and Yang, Zheng and Liu, Yunhao and others},
  journal={arXiv preprint arXiv:2509.21085},
  year={2025}
}

@article{zha2024diffusion,
  title={Diffusion-based filter for fast and accurate collaborative tracking with low data transmission},
  author={Zha, Jirong and Zhou, Nan and Liu, Zhenyu and Sun, Tao and Chen, Xinlei},
  journal={Authorea Preprints},
  year={2024},
  publisher={Authorea}
}

@article{zha2025aircopbench,
  title={AirCopBench: A Benchmark for Multi-drone Collaborative Embodied Perception and Reasoning},
  author={Zha, Jirong and Fan, Yuxuan and Zhang, Tianyu and Chen, Geng and Chen, Yingfeng and Gao, Chen and Chen, Xinlei},
  journal={arXiv preprint arXiv:2511.11025},
  year={2025}
}

@inproceedings{cheng2024multi,
  title={Multi-Agent Target Pursuit Using Perception Uncertainty-Aware Reinforcement Learning},
  author={Cheng, Yuhan and Zha, Jirong and Yang, Renjue and Sun, Zhi and Xu, Susu and Chen, Xinlei},
  booktitle={Proceedings of the 30th Annual International Conference on Mobile Computing and Networking},
  pages={1992--1997},
  year={2024}
}

@article{chen2025survey,
  title={A Survey on Reinforcement Learning Methods for UAV Systems},
  author={Chen, Hengsheng and Lin, Yuanguo and Fu, Mingjian and Yao, Lina and Sheng, Michael},
  journal={ACM Computing Surveys},
  volume={58},
  number={4},
  pages={1--37},
  year={2025},
  publisher={ACM New York, NY}
}

@article{hussain2024computing,
  title={Computing Challenges of UAV Networks: A Comprehensive Survey.},
  author={Hussain, Altaf and Li, Shuaiyong and Hussain, Tariq and Lin, Xianxuan and Ali, Farman and AlZubi, Ahmad Ali},
  journal={Computers, Materials \& Continua},
  volume={81},
  number={2},
  year={2024}
}

@inproceedings{mnih2016asynchronous,
  title={Asynchronous methods for deep reinforcement learning},
  author={Mnih, Volodymyr and Badia, Adria Puigdomenech and Mirza, Mehdi and Graves, Alex and Lillicrap, Timothy and Harley, Tim and Silver, David and Kavukcuoglu, Koray},
  booktitle={International Conference on Machine Learning (ICML)},
  pages={1928--1937},
  year={2016},
  organization={PMLR}
}

@inproceedings{xu2024emergency,
  title={Emergency networking using UAVs: A reinforcement learning approach with large language model},
  author={Xu, Yanggang and Jian, Zhuozhu and Zha, Jirong and Chen, Xinlei},
  booktitle={2024 23rd ACM/IEEE International Conference on Information Processing in Sensor Networks (IPSN)},
  pages={281--282},
  year={2024},
  organization={IEEE}
}

@article{chen2024soscheduler,
  title={Soscheduler: Toward proactive and adaptive wildfire suppression via multi-uav collaborative scheduling},
  author={Chen, Xuecheng and Xiao, Zijian and Cheng, Yuhan and Hsia, Chen-Chun and Wang, Haoyang and Xu, Jingao and Xu, Susu and Dang, Fan and Zhang, Xiao-Ping and Liu, Yunhao and others},
  journal={IEEE Internet of Things Journal},
  volume={11},
  number={14},
  pages={24858--24871},
  year={2024},
  publisher={IEEE}
}

@article{wang2025novel,
  title={A Novel Integrated Sensing and Communication Scheme in UAVs-Enabled Vehicular Networks with MARL-Driven Adaptive Control},
  author={Wang, Ziyuan and Zhang, Xiao-Ping and Ding, Wenbo and Dong, Yuhan and Chen, Xinlei},
  journal={IEEE Transactions on Mobile Computing},
  year={2025},
  publisher={IEEE}
}

@article{wang2025aerial,
  title={Aerial shepherds: Enabling hierarchical localization in heterogeneous mav swarms},
  author={Wang, Haoyang and Xu, Jingao and Zhao, Chenyu and Cheng, Yuhan and Chen, Xuecheng and Hong, Chaopeng and Zhang, Xiao-Ping and Liu, Yunhao and Chen, Xinlei},
  journal={IEEE Transactions on Mobile Computing},
  year={2025},
  publisher={IEEE}
}

@article{jian2024lvcp,
  title={Lvcp: Lidar-vision tightly coupled collaborative real-time relative positioning},
  author={Jian, Zhuozhu and Li, Qixuan and Zheng, Shengtao and Wang, Xueqian and Chen, Xinlei},
  journal={arXiv preprint arXiv:2407.10782},
  year={2024}
}

\vspace{-1cm}

\begin{IEEEbiography}[{\includegraphics[width=1in,height=1.25in,clip,keepaspectratio]{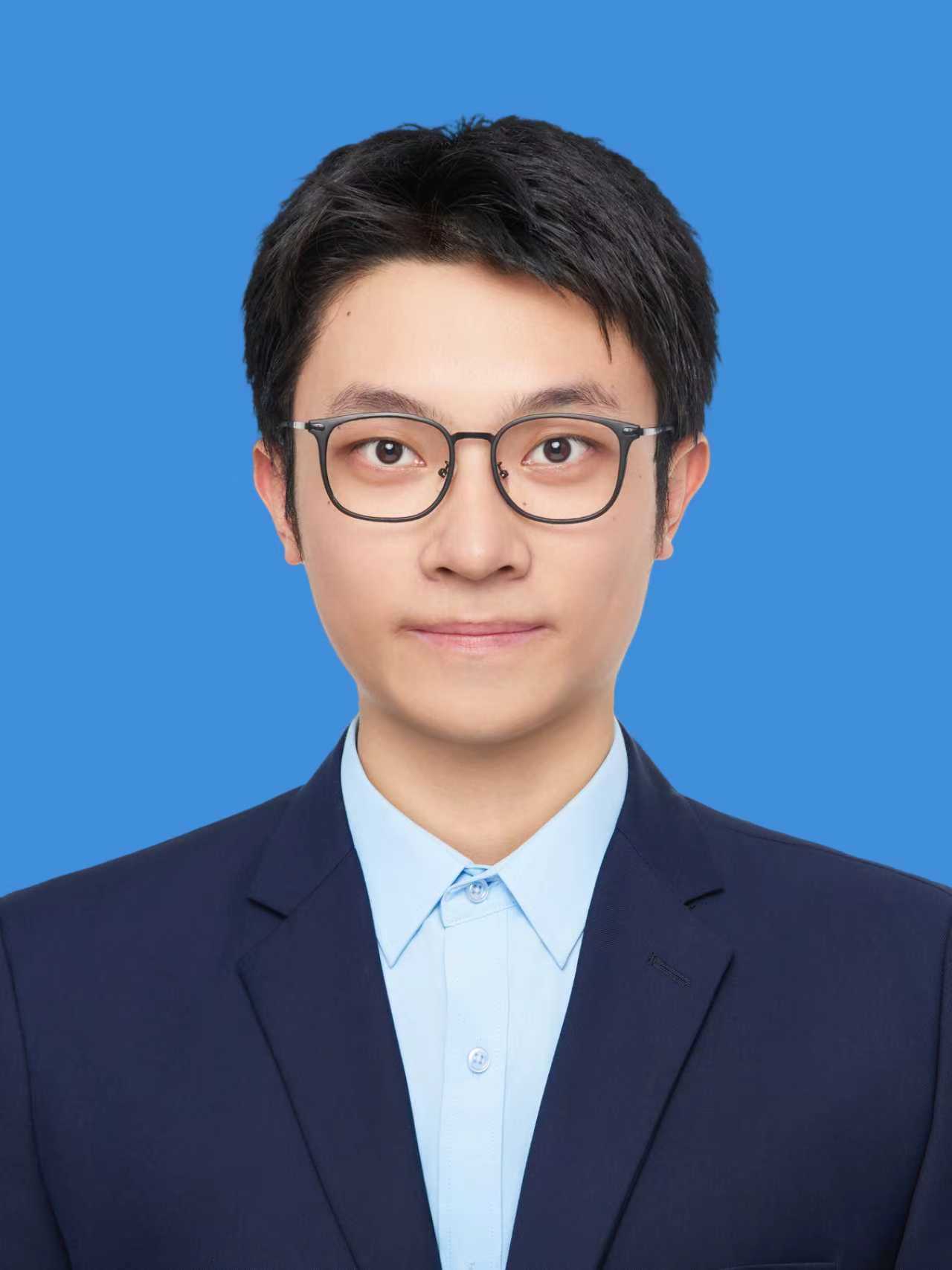}}]{Yanggang Xu}
was born in Sichuan, China, in 1999. He received the B.Eng. degree in communication engineering from the University of Electronic Science and Technology of China (UESTC), Chengdu, China, and the B.Eng. degree in electronics and electrical engineering from the University of Glasgow, Glasgow, U.K., in 2022. He received the M.S. degree in data science and information technology from Tsinghua University, Shenzhen, China, in 2025. He is currently working at Tencent, Shenzhen, China. His research interests include reinforcement learning, large language models, multi-agent systems, robotics, and artificial intelligence.
\end{IEEEbiography}

\begin{IEEEbiography}
[{\includegraphics[width=1in,height=1.25in,clip,keepaspectratio]{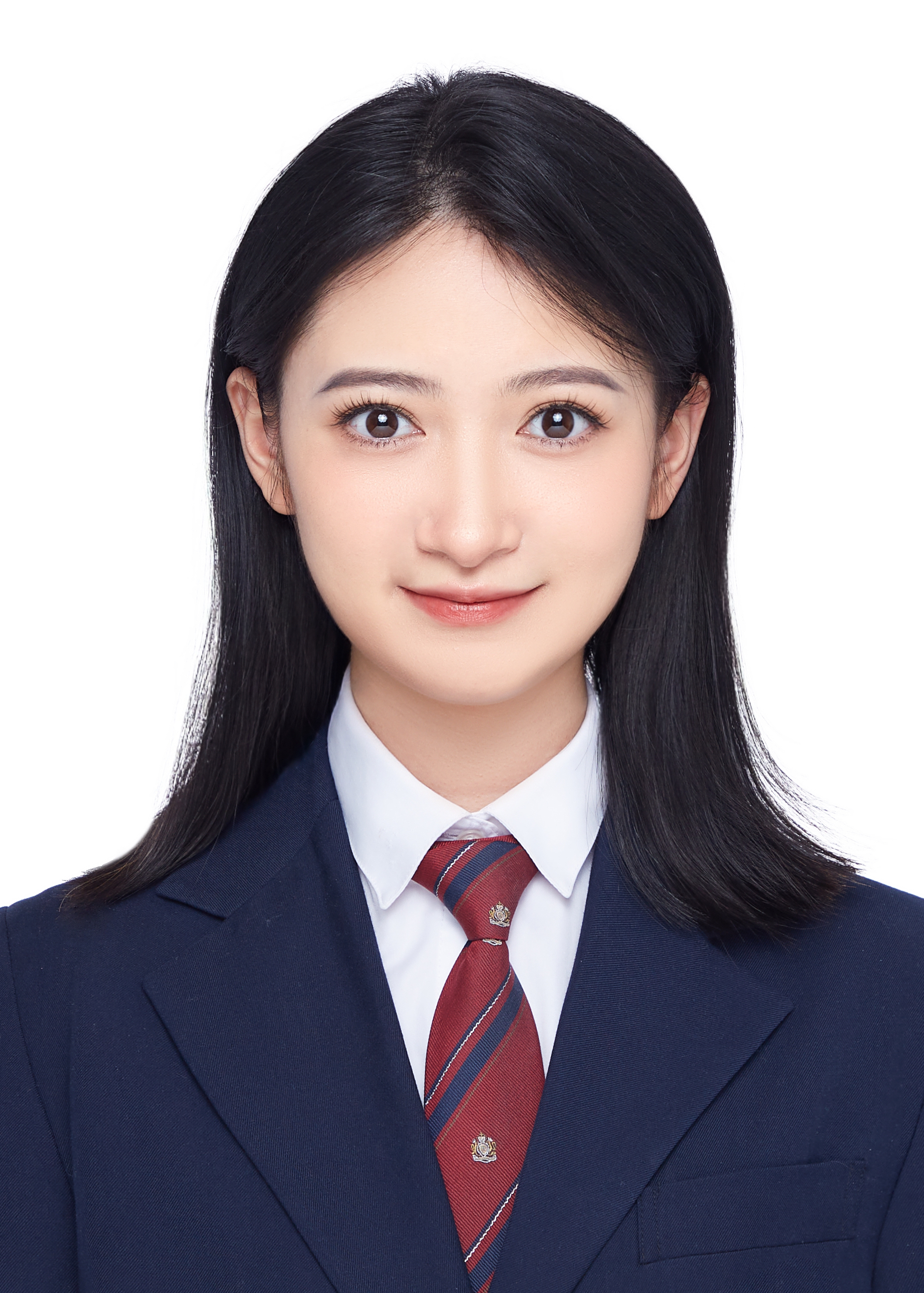}}]
{Jirong Zha}
is currently pursuing her Ph.D. degree in data science and information technology from Tsinghua University, China. She
received the M.S. and B.S. degrees from Beihang University, China, in 2023 and 2020, respectively. Her research interests include collaborative perception, large language models, multi-agent systems, and distributed state estimation.
\end{IEEEbiography}

\begin{IEEEbiography}
[{\includegraphics[width=1in,height=1.25in,clip,keepaspectratio]{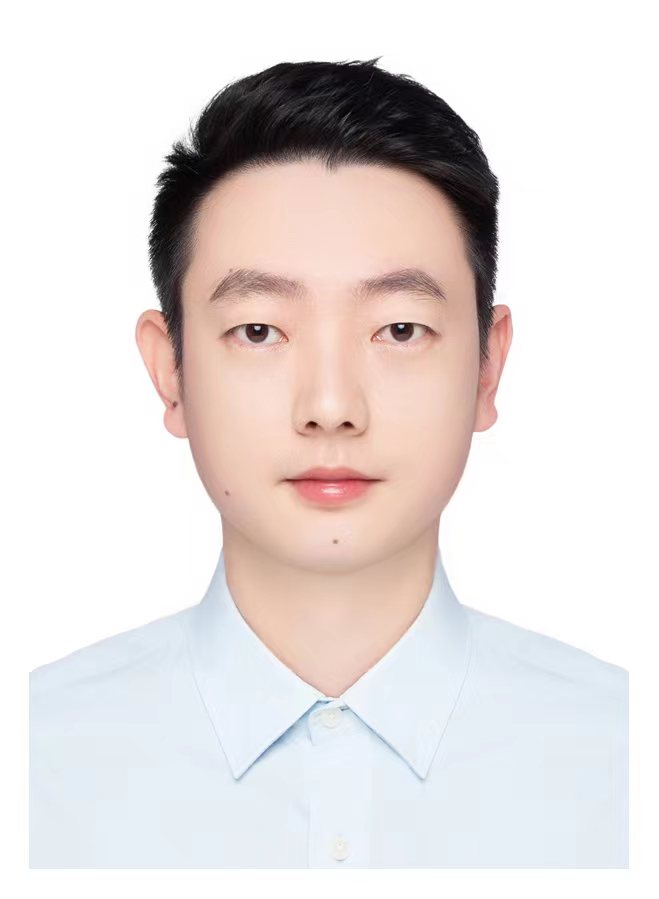}}]
{Weijie Hong} (Member, IEEE) received  a B.S. degree in information and computing science from South China University of Technology in 2009 and is currently pursuing a master’s degree at Tsinghua University. He currently serves as CTO of Shenzhen Smart City Communications Co., Ltd., and Director of Guangdong Engineering Technology Research Center for Multimodal Fusion Communication and IoT Sensing. His research interests include AI-IoT, computing power networks, Optical networks, and Low-Altitude Communications.

\end{IEEEbiography}

\begin{IEEEbiography} 
[{\includegraphics[width=1in,height=1.25in,clip,keepaspectratio]{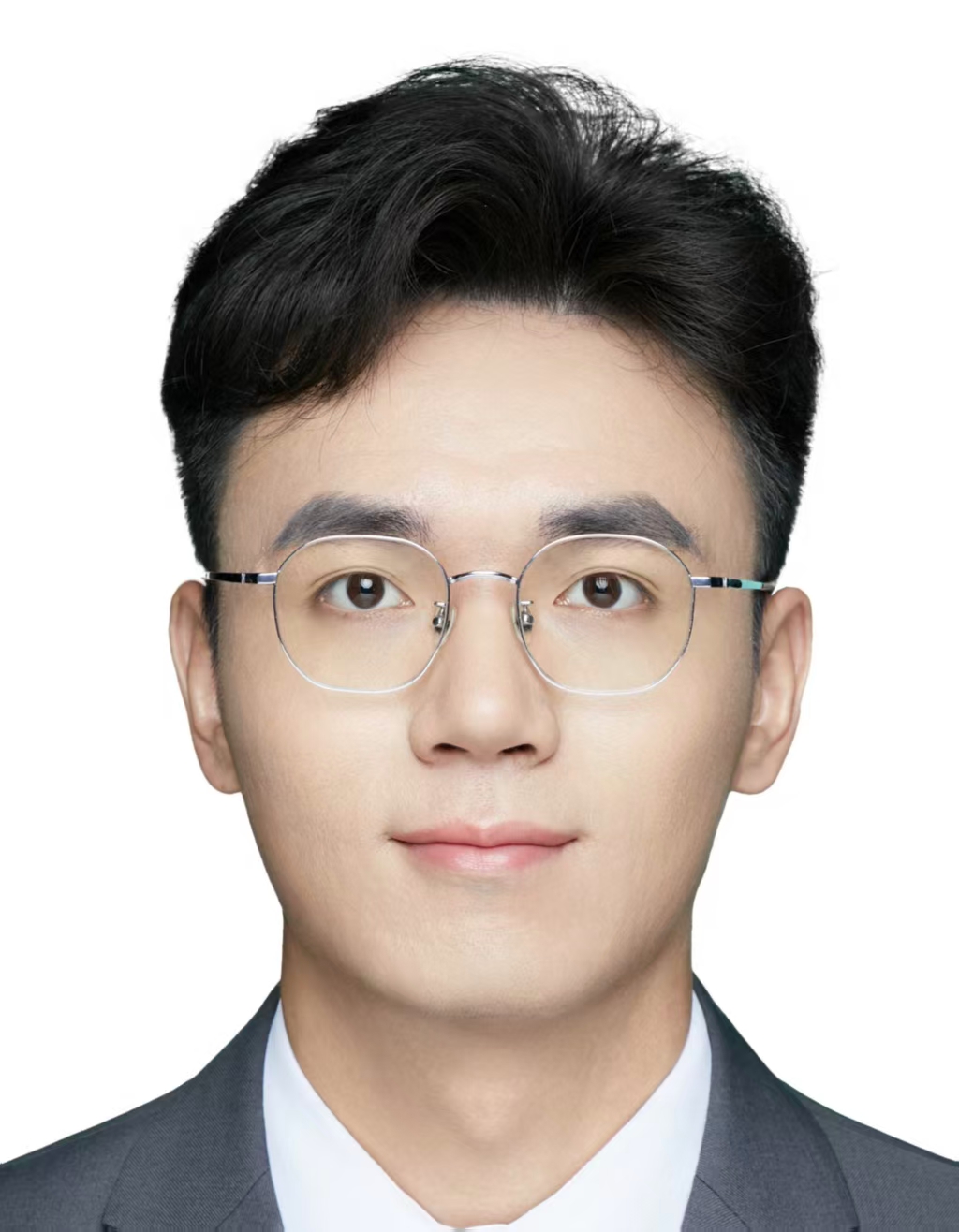}}]
{Xiangmin Yi}
received the BE degree from the School of Artificial Intelligence, Beijing University of Posts and Telecommunications, Beijing, China, in 2025. He is currently pursuing the MS degree with the Tsinghua Shenzhen International Graduate School, Tsinghua University, Shenzhen, China. His research interests include artificial intelligence, large language models, and reinforcement learning.
\end{IEEEbiography}

\begin{IEEEbiography} 
[{\includegraphics[width=1in,height=1.25in,clip,keepaspectratio]{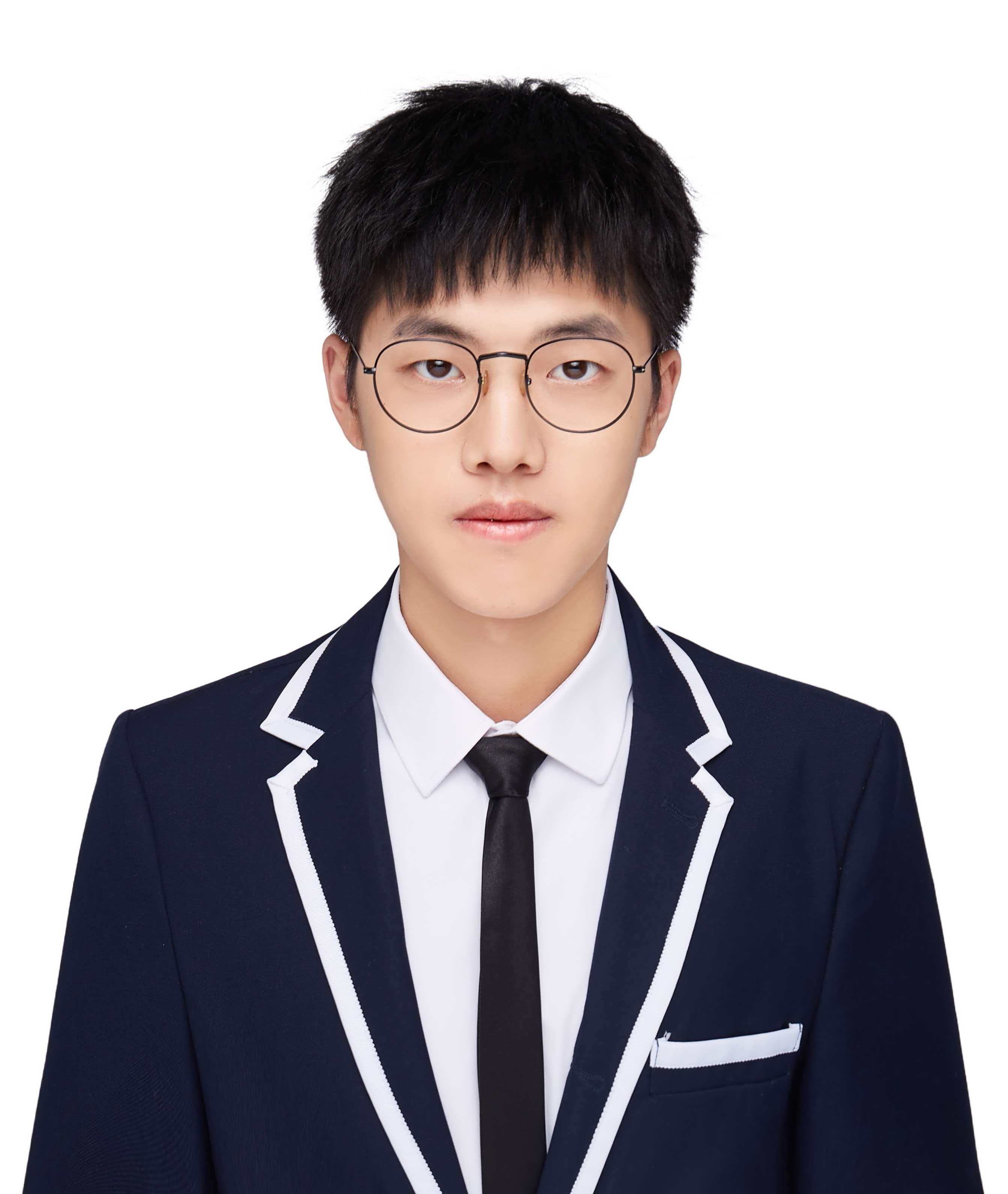}}]
{Geng Chen}
is currently an undergraduate student majoring in software engineering at Jilin University (2022–2026). His research interests include reinforcement learning and multimodal large language models.
 
\end{IEEEbiography}

\noindent
\begin{minipage}{\linewidth}
\begin{IEEEbiography} 
[{\includegraphics[width=1in,height=1.25in,clip,keepaspectratio]{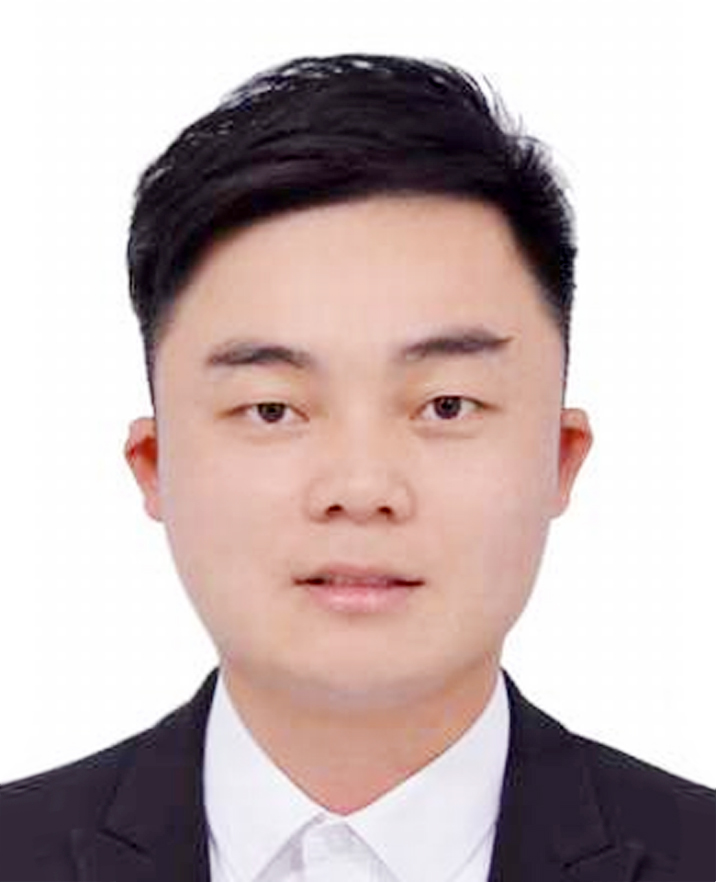}}]
{Jianfeng Zheng}
received the B.E. degree in art and design from Guangxi Minzu University, China, in 2015. He is currently a Project Manager with Shenzhen Smart City Communication Co., Ltd., Shenzhen, China. His research interests include smart city systems, intelligent transportation, IoT-based sensing, and large-scale system integration.
 
\end{IEEEbiography}

\begin{IEEEbiography} 
[{\includegraphics[width=1in,height=1.25in,clip,keepaspectratio]{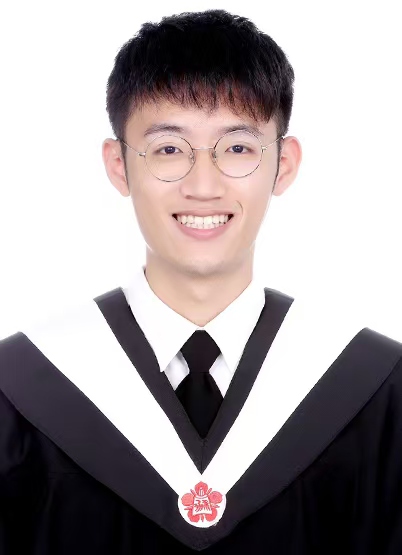}}]
{Chen-Chun Hsia}
received the B.E. degree from National Cheng Kung University, Tainan, Taiwan, in 2021, and the M.S. degree from the Shenzhen International Graduate School, Tsinghua University, Shenzhen, China, in 2024. He is currently an ML Engineer at Ant Group (Alipay), working on large-scale recommendation retrieval and merchant intelligence systems for payment growth. His interests include representation learning, multimodal real-time generative systems, and end-to-end ML deployment.

\end{IEEEbiography}

\begin{IEEEbiography}
[{\includegraphics[width=1in,height=1.25in,clip,keepaspectratio]{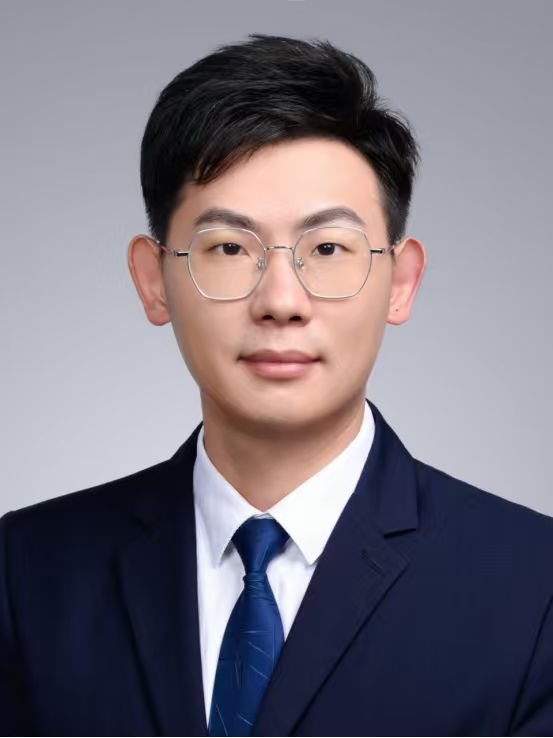}}]
{Xinlei Chen}
is an associate professor at Shenzhen International Graduate School, Tsinghua University. His research interests lie in mobile sensing, embodied AI and etc. Dr. Chen has won several awards from top-tier conference and been selected in Elsevier's Global Top 2\% Scientists List in the past three years.
\end{IEEEbiography}
\end{minipage}

\end{document}